# Cosmological and High-Energy Physics Implications from Gravitational-Wave Background Searches in LIGO-Virgo-KAGRA's O1–O4a Runs

A. G. Abac *et al.*[*]

(LIGO Scientific Collaboration, Virgo Collaboration, and KAGRA Collaboration)

(Received 10 November 2025; accepted 21 July 2026; published 22 October 2026)

We search for gravitational-wave background signals produced by various early Universe processes in the Advanced LIGO O4a dataset, combined with the data from the earlier O1, O2, and O3 (LIGO-Virgo) runs. The absence of detectable signals enables powerful constraints on fundamental physics. We derive gravitational-wave background energy density upper limits from the O1–O4a data to constrain parameters associated with various possible processes in the early Universe: first-order phase transitions, cosmic strings, domain walls, stiff equation of state, axion inflation, second-order scalar perturbations, primordial black hole binaries, and parity violation. In our analyses, the presence of an astrophysical background produced by compact (black hole and neutron star) binary coalescences throughout the Universe is also considered. We address the implications for various cosmological and high-energy physics models based on the obtained parameter constraints. We conclude that LIGO-Virgo data already yield significant constraints on numerous early Universe scenarios.

## I. INTRODUCTION

Advanced LIGO [1], Advanced Virgo [2], and KAGRA [3] have completed three observational runs. LIGO, Virgo, and KAGRA are now in their fourth observational run, O4, which began in May 2023. The O4a part of the run started on May 24, 2023, going until January 16, 2024 [4]. Data acquired in these observation runs have resulted in a series of novel scientific pursuits. This includes discovery of over 200 compact binary (black hole and neutron star) mergers [5], increasingly stringent tests of general relativity [6], multimessenger measurements of the Hubble constant [7], and measurements of the neutron star equation of state [8].

One of the primary targets of these observations is the gravitational-wave background produced by a superposition of a large number of uncorrelated gravitational-wave signals [9]. Observations of stellar mass compact binary mergers by Advanced LIGO and Advanced Virgo imply that a gravitational-wave background of astrophysical origin [10–15] should exist and that it may be detectable by the LIGO-Virgo-KAGRA network in the near future [16–19]. Furthermore, a gravitational-wave background could be of cosmological origin, generated in a variety of processes in early phases of the Universe. Consequently, gravitational-wave background searches can be used to probe high-energy physics models at energy scales beyond the ones reached at the Large Hadron Collider [20] and to explore early Universe cosmological scenarios [21–24].

In what follows, we present searches for a gravitational-wave background produced by various cosmological models, using the LIGO O1 [25], O2, O3 [26], and O4a [27,28] data, plus Virgo O3 [29] data, and we report the resulting constraints on their parameters. Motivations for considering sources of a cosmological gravitational-wave background descend from open questions of fundamental physics. Indeed, despite the extraordinary success of the Standard Model of particle physics and cosmology, our understanding of basic aspects of fundamental physics is still incomplete. The nature of dark matter, the origin of the matter-antimatter asymmetry, the explanation of the neutrino masses, and the realization of inflation remain as important open questions. In order to solve these issues, scenarios of physics beyond the Standard Model are investigated and are under scrutiny in astroparticle experiments, from colliders to telescopes. Many of these beyond the Standard Model scenarios also imply novel phenomena happening in the very early stages of the Universe, potentially leaving footprints in the form of a gravitational-wave background. The models for which we conduct gravitational-wave background searches are first-order phase transitions, cosmic strings, domain walls, stiff equation of state, axion inflation, second-order scalar perturbations, primordial black holes, and parity violation.

[*]Full author list given at the end of the article.

The first three models (first-order phase transitions, cosmic strings, and domain walls) we consider are related to cosmological phase transitions, common in theories with spontaneously broken symmetries [30–32]. If the phase transition is first order, it can generate a gravitational-wave background [33]. Phase transitions followed by spontaneously broken symmetries can lead to topological defects, such as cosmic strings [34,35] and domain walls [36–42], extended objects in the Universe that can produce a gravitational-wave background.

The next three models (stiff equation of state, axion inflation, and second-order scalar perturbations) we study lead to a gravitational-wave background generated during inflation. While single-field slow-roll inflation within the ΛCDM cosmological model predicts a gravitational-wave background that is too weak to be observed with current detectors, other inflationary models may produce a detectable gravitational-wave background. For an exotic early Universe cosmological model with a stiff equation of state [43–51], we explore the gravitational-wave background generated during inflation [50–53]. For the axion inflation model, we consider a coupling between the scalar field driving inflation and an SU(2) gauge field and calculate the produced gravitational-wave background [54–57]. For second-order scalar perturbations [58], we estimate the gravitational-wave background induced by primordial curvature perturbations [59–63].

Next, we study a gravitational-wave background generated by mergers of primordial black holes [64] that could have formed by a variety of mechanisms in the early Universe. Finally, we study a gravitational-wave background that exhibits chiral polarization. We consider a model-independent parametrization of such a background, as well as a particular case where parity violation originates from axion inflation [65].

For each of these cosmological models, the absence of a gravitational-wave background constrains their parameters. The results of our analysis show that the LIGO-Virgo O1–O4a data can already be used to successfully derive new constraints on a wide range of theories beyond the Standard Model.

*Formalism and methodology*—The gravitational-wave background is quantified in terms of its energy density per logarithmic frequency interval and compared to the critical energy density of the Universe. Specifically,

$$\Omega_{\rm GW}(f) = \frac{f}{\rho_c}\frac{{\rm d}\rho_{\rm GW}}{{\rm d}f}, \tag{1}$$

where $\rho_{\rm GW}$ is the energy density of gravitational waves, the critical energy density of the Universe is $\rho_c = 3c^2H_0^2/(8\pi G) \approx 7.7\times10^{-9}\ {\rm erg\,cm^{-3}}$, $H_0 = 100\ h\,{\rm km/s/Mpc}$ is the Hubble constant with $h = 0.68$ from the Planck measurements [66], $c$ is the speed of light, and $G$ is Newton's constant. The gravitational-wave background spectrum is often approximated by the power-law (PL) form with the spectral index $\alpha$:

$$\Omega^{\rm PL}_{\rm GW}(f) = \Omega_{\rm ref}\left(\frac{f}{f_{\rm ref}}\right)^{\alpha}, \tag{2}$$

where $\Omega_{\rm ref}$ is the gravitational-wave energy density at the reference frequency $f_{\rm ref}$. A background produced by compact binary mergers from population I or II stars can be approximated as having $\alpha = 2/3$ [67]. Cosmological backgrounds can also be typically approximated as power laws, or as broken power laws, as we show in the following.

For the first three observing runs, the LIGO-Virgo-KAGRA Collaboration reported an upper limit on the strength of an isotropic gravitational-wave background of $\Omega_{\rm GW} \le 5.8\times10^{-9}$ for $\alpha = 0$ and 95% credible level [68] with 99% of the sensitivity coming from the band (20–76.6) Hz. For $\alpha = 2/3$, the limit was $\Omega_{\rm GW}(25\ {\rm Hz}) \le 3.4\times10^{-9}$ in the band (20–90.6) Hz. This upper bound was derived from the LIGO data acquired during O1, O2, and O3 observing runs, as well as Virgo data of O3. There have also been searches for backgrounds with nonstandard polarizations, such as scalar and vector (in addition to tensor) [69,70]. Anisotropic backgrounds have been also explored [15,71].

The O4a data, combined with the data from O1, O2, and O3, do not provide evidence for the detection of a gravitational-wave background [19]. As such, an upper limit of $\Omega_{\rm GW} \le 2.8\times10^{-9}$ is set for $\alpha = 0$ and 95% confidence, with 99% of the sensitivity coming from the band (20–58.2) Hz. For $\alpha = 2/3$, the limit is $\Omega_{\rm GW}(25\ {\rm Hz}) \le 2.0\times10^{-9}$ in the band (20–86.8) Hz [19]. The LIGO-Virgo-KAGRA Collaboration has also searched for an anisotropic gravitational-wave background [71].

To derive constraints on the cosmological model parameters, we perform a Bayesian analysis [72] using the data from the LIGO-Virgo O1–O4a observing runs, following the methods developed in Ref. [73]. Assuming the cross-correlation estimator $\hat{C}_{IJ}(f)$ [68] is Gaussian distributed, we write the following likelihood function:

$$p[\hat{C}_{IJ}(f)|\boldsymbol{\theta},\lambda] \propto \exp\left[-\frac{1}{2}\sum_f \frac{[\hat{C}_{IJ}(f) - \lambda\Omega_{\rm GW}(f,\boldsymbol{\theta})]^2}{\sigma^2_{IJ}(f)}\right], \tag{3}$$

using data from detectors $I$ and $J$, while $\sigma^2_{IJ}(f)$ is the variance. Both $\hat{C}_{IJ}(f)$ and $\sigma^2_{IJ}(f)$ are data products from the LIGO-Virgo-KAGRA Collaboration isotropic gravitational-wave background search analysis [19], where they are calculated from the LIGO-Virgo data. It is assumed that such an isotropic search does not have correlated noise between detectors $I$ and $J$, for example, from correlated magnetic noise [74]. Hence, a standard Gaussian noise model is preferred [68], and the potential contribution from correlated magnetic noise (Schumann resonances) [75] can be neglected [19]. The function $\Omega_{\rm GW}(f,\boldsymbol{\theta})$ corresponds to the model considered, described by the set of parameters $\boldsymbol{\theta}$, while the parameter $\lambda$, which we marginalize over, accounts for the detectors' calibration uncertainties [19]. A minimum of two detectors are needed in order to conduct this analysis. The two LIGO detectors contribute the most to the correlation due to the smallest distance separation in the network, their optimal alignment [76,77], and their sensitivities [68].

In our analyses, we take into account the contribution from an isotropic astrophysical background of compact binary coalescences (CBC) $\Omega_{\rm CBC}(f,\boldsymbol{\theta}_{\rm CBC}) = \Omega_{\rm CBC}(f,\Omega_{\rm ref},\alpha)$, which we model by Eq. (2) where $f_{\rm ref}= 25$ Hz [68]. The cosmologically produced gravitational-wave background is $\Omega_{\rm Cosmo}(f,\boldsymbol{\theta}_{\rm Cosmo})$. The total gravitational-wave background is

$$\Omega_{\rm GW}(f,\boldsymbol{\theta}) = \Omega_{\rm CBC}(f,\Omega_{\rm ref},\alpha) + \Omega_{\rm Cosmo}(f,\boldsymbol{\theta}_{\rm Cosmo}). \tag{4}$$

This publication is organized as follows: We present limits on various cosmological models in the following sections; first-order phase transitions in Sec. II; cosmic strings in Sec. III; domain walls in Sec. IV; stiff equation of state in Sec. V; axion inflation in Sec. VI; second-order scalar perturbations in Sec. VII; primordial black holes in Sec. VIII; and parity violation in Sec. IX. For each of the eight scenarios, we discuss the motivation, present the model considered, and give the constraints to the model parameters using O1–O4a LIGO-Virgo data. The models analyzed in this work span a wide range of physical processes. While we aim for a uniform presentation, the specific statistical products (such as two-dimensional corner plots) vary between models. This is due to the inherent differences in parameter dimensionality and the nature of the constraints in the LIGO-Virgo frequency band. For models where certain parameters are fixed or where the data constrain only a single combination of parameters (e.g., cosmic strings), two-dimensional plots are omitted in favor of more descriptive one-dimensional upper limits or exclusion regions. Conclusions are given in Sec. X. The data supporting the findings presented in this article are openly available at the Gravitational Wave Open Science Center (GWOSC) [4,78].

## II. FIRST-ORDER PHASE TRANSITIONS

### A. Motivation

Cosmological phase transitions are among the most well-motivated early Universe phenomena we anticipate (see, e.g., Refs. [30–32]). They are a common feature of particle physics models that exhibit symmetry breaking. The phase transition is first order when the effective potential of the theory develops a new minimum (true vacuum) with a free energy density lower than that of the minimum at high temperature (false vacuum), and both are separated by a potential barrier. The Universe then violently transitions from the symmetric higher-energy false vacuum to the broken lower-energy true vacuum. A first-order phase transition could be a powerful source of a gravitational-wave background.

Given our knowledge of the cosmological history and particle physics, such transitions could have occurred within the first one-trillionth of a second after the Big Bang, at energies higher than those accessible in present-day particle accelerators. Their gravitational-wave imprints would be an important key to determining the correct theory beyond the Standard Model realized in nature.

The Standard Model itself undergoes two phase transitions: the electroweak phase transition due to the breaking of electroweak symmetry and the confinement-deconfinement phase transition due to chiral symmetry breaking in quantum chromodynamics (QCD). Neither of these Standard Model phase transitions is first order (or generates stable topological defects); thus, no corresponding gravitational-wave signatures are expected.

However, many extensions of the Standard Model with enlarged symmetry structures at high energies necessarily undergo spontaneous symmetry breaking (for sufficiently high reheating temperature). Such a first-order phase transition corresponds to nucleation of bubbles of true vacuum in various points in the Universe. Those bubbles then expand, collide with each other, and eventually fill out the entire space. During this process, gravitational waves are generated from processes such as bubble collisions [79,80], sound waves propagating in the early Universe plasma [81,82], and magnetohydrodynamic turbulence [83], with the first two contributions being typically dominant (see, e.g., Ref. [33] and references therein).

The strong connection of the resulting primordial gravitational-wave background to particle physics provides gravitational-wave astronomy with a unique opportunity to probe regions of parameter space of physics models completely inaccessible in any other types of experiments. This includes various extensions of the Standard Model, e.g., models with an extended electroweak gauge sector [84–86], theories with dark sectors [87–89], axion models [90–92], unification models [93,94], supersymmetric theories [95–97], or theories with extra dimensions [98]. In turn, the information about physics at energies beyond the electroweak scale may provide insight into solutions to problems such as the nature of dark matter or the origin of the matter-antimatter asymmetry of the Universe.

As has recently been shown based on the LIGO-Virgo observing runs O1–O3, already current data can be used to provide meaningful constraints on the parameters of first-order phase transitions [99]. This method was successfully applied to particle physics models in the context of supercooled transitions (see, e.g., Refs. [100,101]), leading to novel constraints on beyond Standard Model theories [102]. In the following, we utilize the LIGO-Virgo O1–O4a data to derive new and improved bounds on the parameters of early Universe first-order phase transitions.

### B. Model

In this scenario, the cosmological component $\Omega_{\rm Cosmo}$ of the gravitational-wave background in Eq. (4) depends on the parameters describing the phase transition, i.e., on the details of the particle physics model. From an effective theory point of view, a first-order phase transition can be fully described by just several parameters: $v_w$, the bubble wall velocity (given in units of the speed of light); $T_{\rm PT}$, the temperature of the phase transition; $\alpha_{\rm PT}$, the strength of the phase transition, which is equal to the density of the energy released divided by the energy density of radiation; $\kappa$, the fraction of the energy corresponding to a given source; $\beta$, the inverse time duration of the transition; and $g_*$, the number of effective degrees of freedom (equal to 106.75 in the Standard Model at high temperatures).

In our analysis, we first discuss the sound wave contribution, which is typically dominant in case of thermal phase transitions (where friction from the Standard Model plasma is relevant), and then focus on the bubble collision contribution, which is the leading source of gravitational waves for vacuum phase transitions (where friction is negligible). We disregard magnetohydrodynamic turbulence, since its effects are typically subdominant and their characterization is the subject of ongoing research. As for the gravitational-wave background spectral shape for these two types of contributions, we first select two representative forms while acknowledging the fact that new results and simulations are continuously proposed in the literature [see discussion after Eq. (10)]. In light of this, to maintain a model-independent approach, we then conduct a more generic search using the broken-power-law (BPL) model.

#### *1. Sound wave contribution*

We first address the leading source of gravitational waves for a thermal first-order phase transition which comes from sound waves in the primordial plasma and is caused by the coupling between the scalar field undergoing the phase transition and the thermal bath [81,82,103]. A fruitful description of the physics of this process is provided by the sound shell model [104–106], although its accuracy has been challenged [103,107]. We employ the gravitational-wave spectrum as summarized in Ref. [108] based on the numerical simulations performed in Refs. [82,103]. This yields (upon setting $v_w = 1$)

$$h^2\Omega_{\rm SW}(f) \approx 0.4\times10^{-5}\left(\frac{H_{\rm PT}}{\beta}\right)\left(\frac{\alpha_{\rm PT}\kappa_{\rm SW}}{\alpha_{\rm PT}+1}\right)^2 \times\left(\frac{100}{g_*}\right)^{1/3}\frac{(f/f_{\rm SW})^3}{[1+0.75(f/f_{\rm SW})^2]^{7/2}}\Upsilon, \tag{5}$$

where the peak frequency $f_{\rm SW}$ is

$$f_{\rm SW} = 0.089\ {\rm Hz}\left(\frac{\beta}{H_{\rm PT}}\right)\left(\frac{T_{\rm PT}}{1\ {\rm PeV}}\right)\left(\frac{g_*}{100}\right)^{1/6}. \tag{6}$$

In the equations above, $H_{\rm PT}$ is the Hubble constant at the phase transition, $\kappa_{\rm sw}$ is the fraction of the latent heat transformed into the bulk motion of the plasma [109]

$$\kappa_{\rm SW} = \frac{\alpha_{\rm PT}}{0.73+0.083\sqrt{\alpha_{\rm PT}}+\alpha_{\rm PT}}, \tag{7}$$

and the suppression factor $\Upsilon$ due to the finite lifetime of sound waves is [106,110]

$$\Upsilon = 1-\frac{1}{[1+8\pi^{1/3}v_w(\frac{H_{\rm PT}}{\beta})(\frac{\alpha_{\rm PT}+1}{3\alpha_{\rm PT}\kappa_{\rm SW}})^{1/2}]^{1/2}}, \tag{8}$$

derived assuming a lifetime on the order of the timescale for the onset of turbulence.

#### *2. Bubble collision contribution*

In some cases, e.g., when the first-order phase transition occurs in the vacuum of a dark sector without sizable interactions with the Standard Model, the sound wave contribution is suppressed and the bubble collision part becomes dominant. The corresponding spectrum is obtained within the envelope approximation by assuming a zero width for the bubble wall and neglecting contributions from overlapping bubble segments [80,111–113]. We consider the gravitational-wave spectrum from bubble collisions as reported in Ref. [114], and based on the numerical simulations of Ref. [115], given by (assuming $v_w = 1$)

$$h^2\Omega_{\rm BC}(f) \approx 1.3\times10^{-5}\left(\frac{H_{\rm PT}}{\beta}\right)^2\left(\frac{\alpha_{\rm PT}}{\alpha_{\rm PT}+1}\right)^2 \times\left(\frac{100}{g_*}\right)^{1/3}\frac{(f/f_{\rm BC})^{2.4}}{[1+(f/f_{\rm BC})^{1.2}]^4}, \tag{9}$$

where the peak frequency $f_{\rm BC}$ is

$$f_{\rm BC} = 0.018\ {\rm Hz}\left(\frac{\beta}{H_{\rm PT}}\right)\left(\frac{T_{\rm PT}}{1\ {\rm PeV}}\right)\left(\frac{g_*}{100}\right)^{1/6}. \tag{10}$$

The shape of the spectrum at low frequencies $\sim f^{2.4}$, close to the expected $\sim f^3$ from causality, whereas the high-frequency behavior is $\sim 1/f^{2.4}$.

While we use Eq. (9) in our analysis, we note that the precise shape of the gravitational-wave spectrum for bubble collisions is not fully settled. It was reported in Ref. [116] that simulations beyond the envelope approximation yield at high frequencies $\sim 1/f^{1.5}$. In Ref. [117], a dependence on wall thickness was found to change the high-frequency spectrum from $\sim 1/f^{1.4}$ to $\sim 1/f^{2.3}$ with increasing thickness. Further variations of the spectrum were discussed in Refs. [118–121].

#### 3. Broken power law

Given the continuous developments in the modeling of the gravitational-wave signals from phase transitions through numerical simulations and semianalytical methods, and the resulting modification in the predicted spectra, we now proceed with a generic BPL model search.

We consider the following parametrization for a BPL gravitational-wave background:

$$\Omega_{\rm BPL}(f) = \Omega_* \left(\frac{f}{f_*}\right)^3 \left[1 + \left(\frac{f}{f_*}\right)^\Delta\right]^{(n_2-3)/\Delta}, \qquad (11)$$

where the $f^3$ dependence arises from causality, while, e.g., $n_2 = -4$ corresponds to sound waves and $n_2 = -1$ to bubble collisions, as considered above. In our search, we scan over the values $n_2 \in [-8, 0]$, thus including both of those cases and possible generalizations. We perform the analysis and present our results for two representative values of the parameter $\Delta$: the value $\Delta = 2$ for signals with a smooth peak and $\Delta = 4$ for spectra with a more abrupt transition at the peak.

### C. Constraints using O1–O4a LIGO-Virgo data

Here, we determine the 95% confidence level (CL) upper limits on the strength of the gravitational-wave signal from sound waves $\Omega_{\rm SW}$, bubble collisions $\Omega_{\rm BC}$, and the generic BPL model $\Omega_{\rm BPL}$, arising from the LIGO-Virgo observing runs O1–O4a data. The previous analyses of this type in Refs. [99,102] were based only on the O1–O3 dataset.

To this end, we perform a Bayesian analysis for these cases (using PYGWB [122]) assuming the priors on the parameters and the CBC background specified in Table I for sound waves and bubble collisions and in Table II for the BPL model. For the CBC background, we fix the power-law index to $\alpha = 2/3$ and vary the amplitude $\Omega_{\rm ref}$.

TABLE I. Prior distributions assumed for the parameters of the model and the CBC background. For $\alpha_{\rm PT} \ll 1$ the gravitational-wave signal would be suppressed, while for $\alpha_{\rm PT} \gg 1$ the signal shape becomes independent from $\alpha_{\rm PT}$. The parameter $\beta/H_{\rm PT}$ cannot be less than 1 for consistency (see explanation in the text), while larger values suppress the gravitational-wave signal beyond detectability. The parameter $T_{\rm PT}$ is chosen such that the broken-power-law spectrum peak is at frequencies close to the ones accessible with LIGO-Virgo-KAGRA.

| Parameter | Prior |
|---|---|
| $\alpha_{\rm PT}$ | LogUniform$[10^{-2}, 10^2]$ |
| $\beta/H_{\rm PT}$ | LogUniform$[1, 10^2]$ |
| $T_{\rm PT}$/GeV | LogUniform$[10^5, 10^{10}]$ |
| $\Omega_{\rm ref}$ | LogUniform$[10^{-13}, 10^{-6}]$ |
| $v_w$ | Fixed at 1 |

TABLE II. Prior distributions adopted for the parameters of the BPL spectrum and the CBC background.

| Parameter | Prior |
|---|---|
| $\Omega_*$ | LogUniform$[10^{-9}, 10^{-4}]$ |
| $f_*$ | Uniform[0, 256 Hz] |
| $n_2$ | Uniform$[-8, 0]$ |
| $\Omega_{\rm ref}$ | LogUniform$[10^{-13}, 10^{-6}]$ |
| $\Delta$ | $\Delta = 2$ or $\Delta = 4$ |

#### 1. Results for sound waves and bubble collisions

Our analysis for the case of a dominant sound wave contribution, using the spectrum given by Eq. (5), yields the Bayes factor $\ln \mathcal{B}^{\rm CBC+SW}_{\rm noise} = -0.571$, so we do not find any evidence of a signal in the data. Figure 1 shows the posterior distributions for the combined CBC and first-order phase transition search in this case. The 95% and 68% CL exclusion contours are highlighted. In the posterior distributions, we also show with a gray dashed line the LogUniform priors.

In the case when bubble collisions dominate, we get $\ln \mathcal{B}^{\rm CBC+BC}_{\rm noise} = -0.631$ for the spectrum in Eq. (9). Figure 2 presents the corresponding constraints from the combined CBC and first-order phase transition search. As Fig. 2 demonstrates, with the priors listed in Table I, the data exclude at 95% CL a small portion of the parameter space of the bubble-collision-dominated phase transitions, especially the region of $T_{\rm PT} \gtrsim 10^8$ GeV and $\beta/H_{\rm PT} \lesssim 3$.

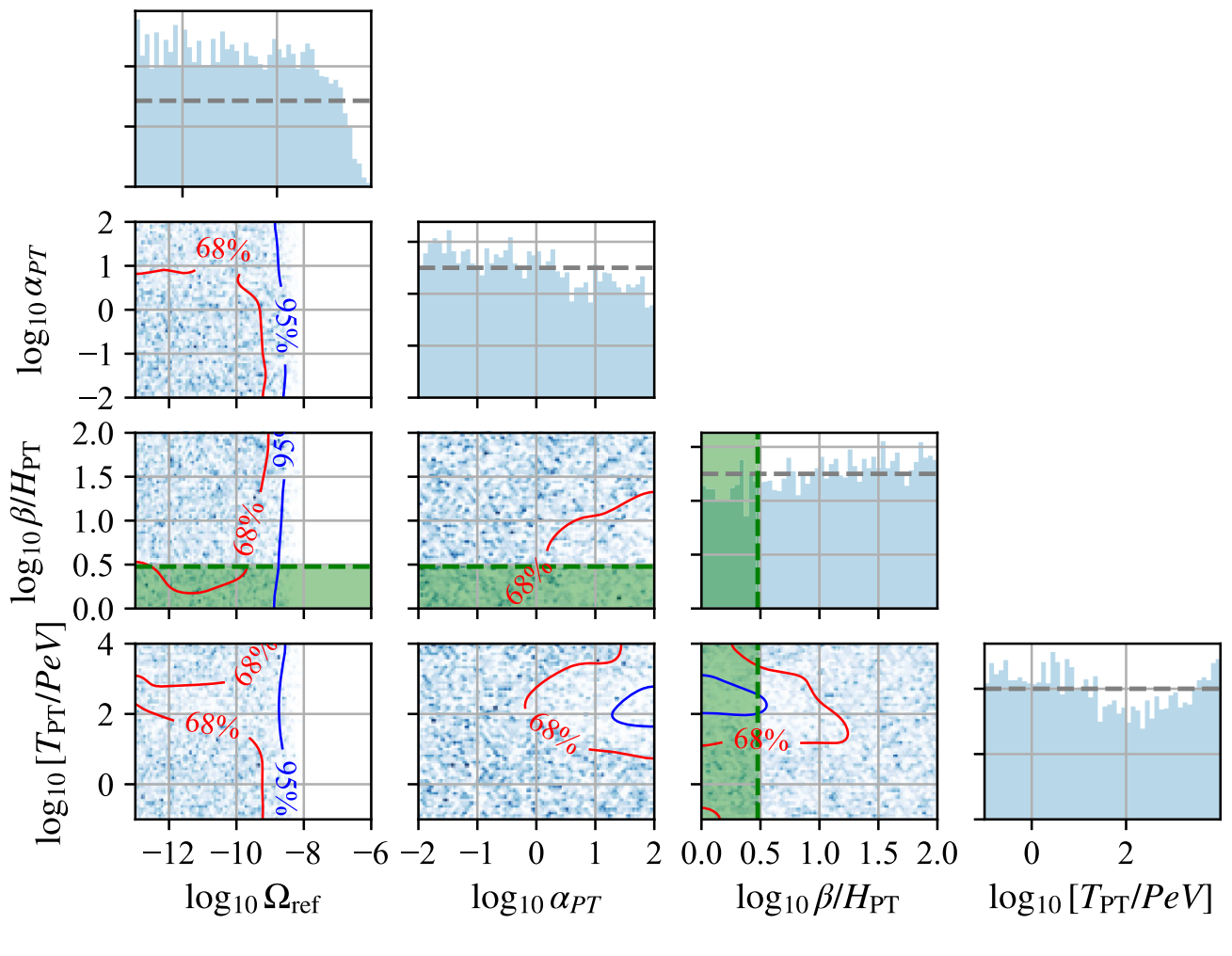

FIG. 1. Constraints from the LIGO-Virgo observing runs O1–O4a data on the first-order phase transition parameters $\alpha_{\rm PT}$, $\beta/H_{\rm PT}$, and $T_{\rm PT}$, assuming a dominant sound wave contribution in Eq. (5) and taking into account the CBC background. The priors selected are shown in Table I. The 95% and 68% confidence level exclusion contours are shown in blue and red, respectively. Clearly, less conservative choices of priors could lead to stronger constraints. The green shaded region corresponds to $\beta/H_{\rm PT} < (8\pi)^{1/3}$.

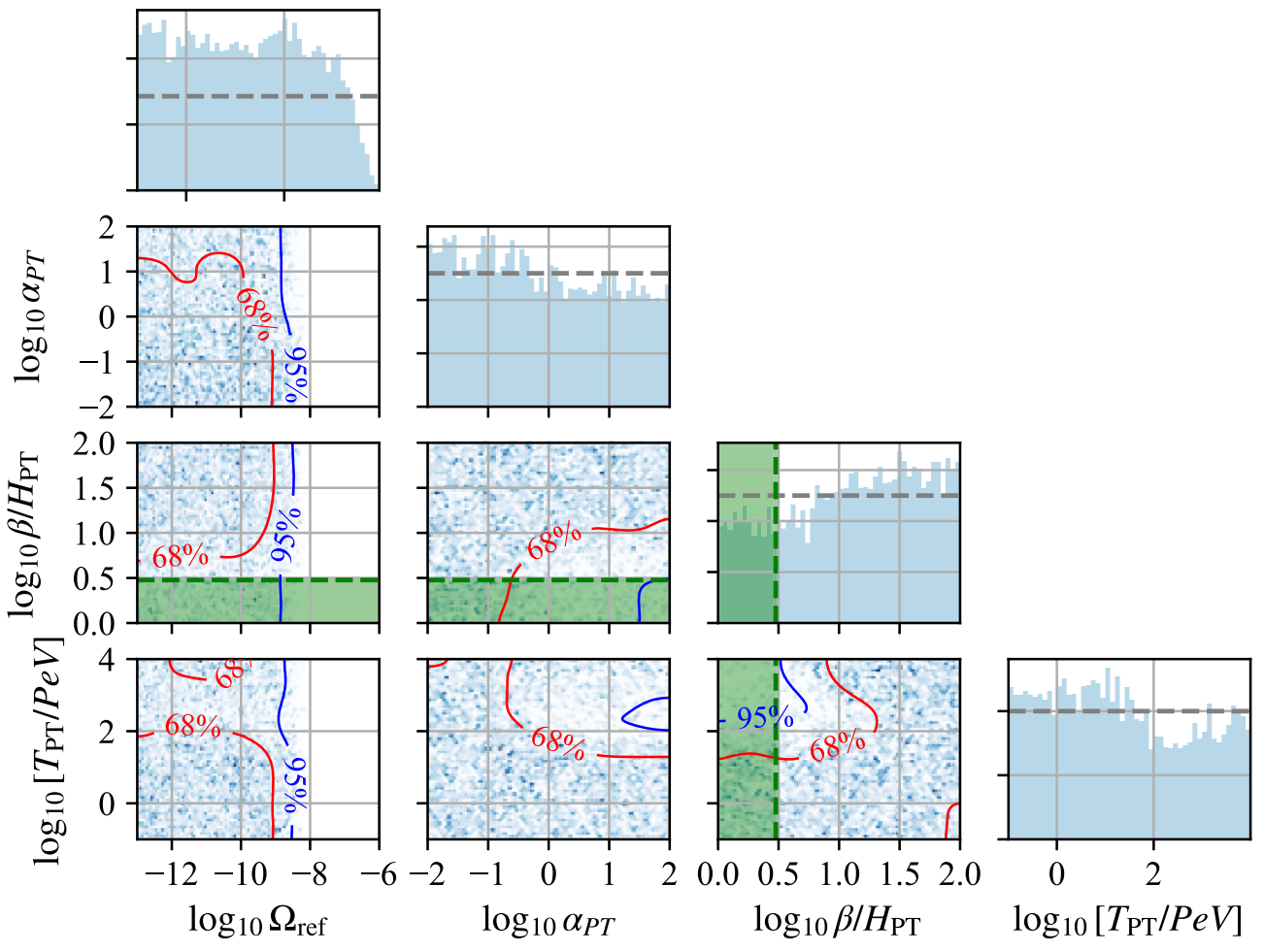

FIG. 2. Similarly as in Fig. 1, the constraints from the LIGO-Virgo observing runs O1–O4a data on the first-order phase transition parameters including the CBC background but for a dominant bubble collision contribution in Eq. (9).

It is important to remark that small values of $\beta$ are at the edge of the consistency with the description of the phase transition. The bubble size is related to the $\beta$ parameter as $R_*^{-3} \sim (1/8\pi)(\beta/v_w)^3$ [123]. We indicated with a green dashed line in Figs. 1 and 2 the value of $\beta$ for which the size of the bubbles becomes comparable to the Hubble volume (see, e.g., Refs. [124–126] for recent studies of this regime), shading the region in which it is greater.

Finally, we note that our analysis sets a 95% CL upper limit on the amplitude of the CBC background, $\Omega_{\rm ref}$, of $2 \times 10^{-9}$ for both the sound wave and bubble collision cases. These values are compatible with the ones reported in Ref. [19].

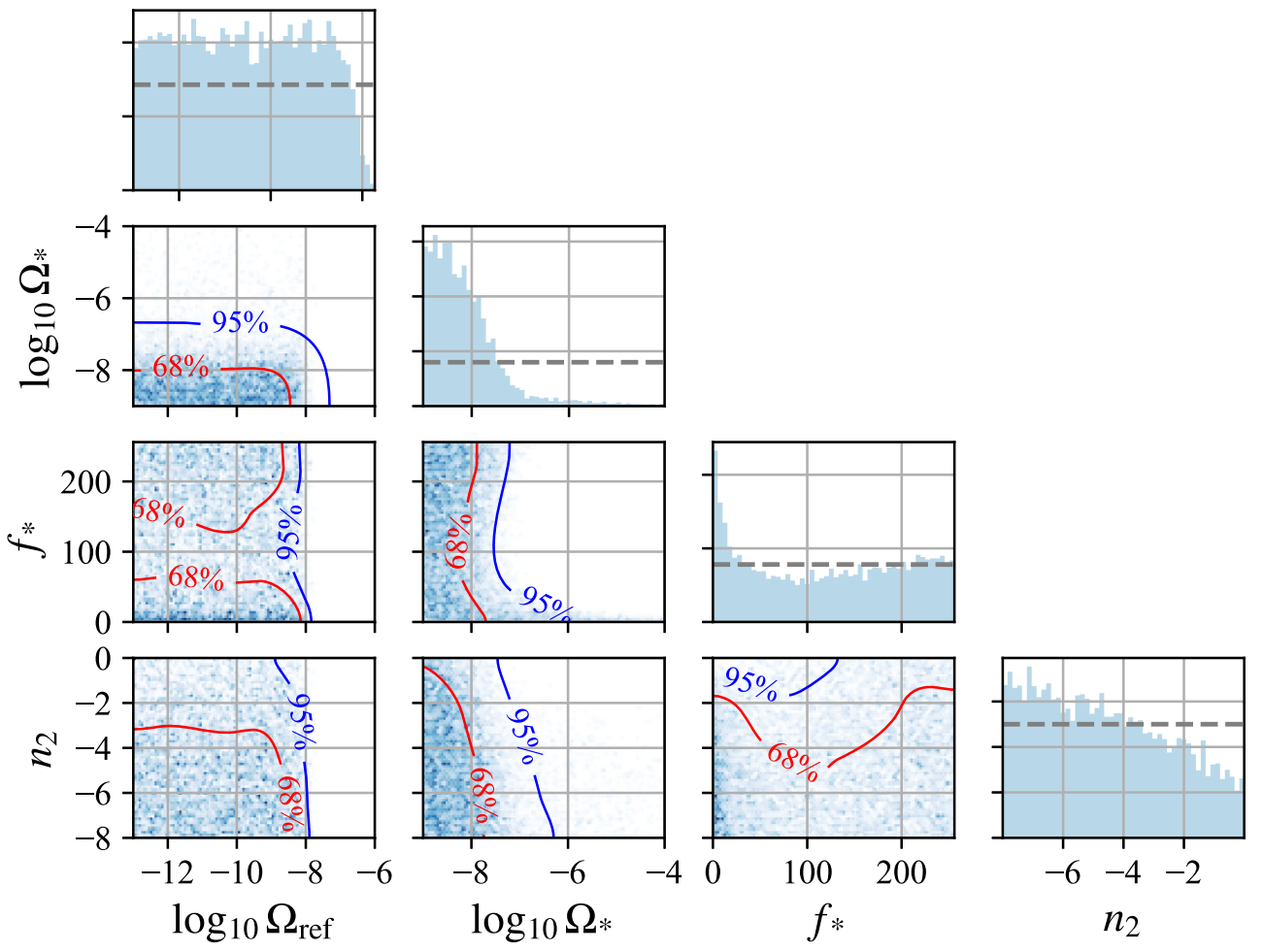

FIG. 3. Constraints from the LIGO-Virgo observing runs O1–O4a data on the first-order phase transition parameters including the CBC background using the BPL distribution in Eq. (11) with $\Delta = 2$.

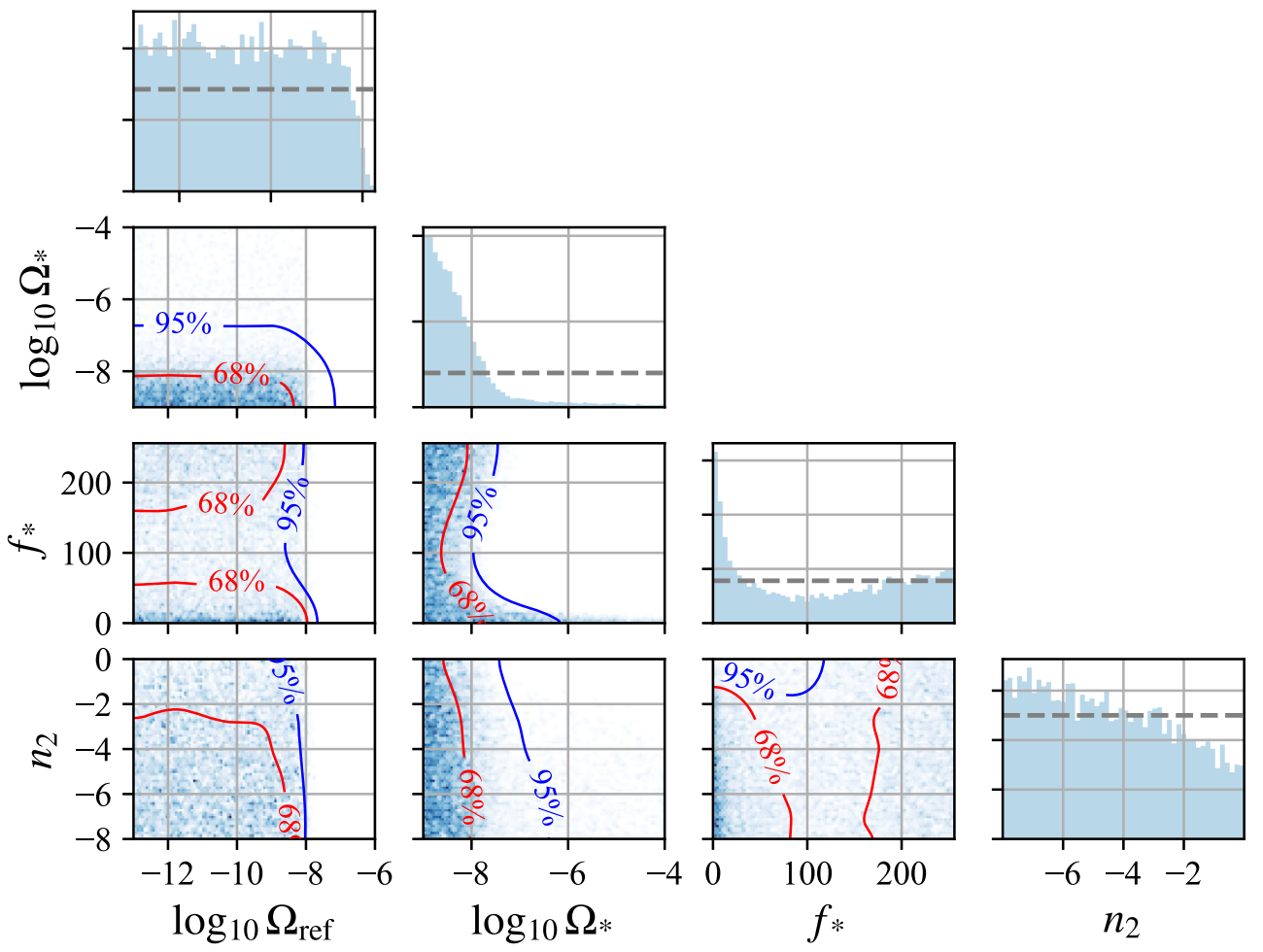

FIG. 4. The same as Fig. 3 but taking $\Delta = 4$ in Eq. (11).

### 2. *Results for the broken-power-law spectrum*

We now report the results for the generic BPL model described in Eq. (11). For $\Delta = 2$, our analysis yields the Bayes factor $\ln \mathcal{B}^{\rm CBC+BPL(\Delta=2)}_{\rm noise} = -1.757$, whereas for $\Delta = 4$, we get $\ln \mathcal{B}^{\rm CBC+BPL(\Delta=4)}_{\rm noise} = -2.093$. Figures 3 and 4 show the 95% and 68% CL exclusion contours for the two cases, respectively. For $\Delta = 2$, the 95% CL upper limits are $\Omega_{\rm ref} = 4.5 \times 10^{-9}$ and $\Omega_* = 2.1 \times 10^{-7}$, whereas in the $\Delta = 4$ case those limits are $\Omega_{\rm ref} = 4.5 \times 10^{-9}$ and $\Omega_* = 3.8 \times 10^{-7}$. In both cases, we are able to set significant constraints on the peak amplitude of a BPL signal. The limits on the CBC background are compatible with the ones obtained in Ref. [19].

The best sensitivity of the search corresponds to the peak frequency close to $f_* \approx 100$ Hz. The 95% CL upper limits on $\Omega_*$ are largely insensitive to the value of $n_2$; the exclusion for lower $n_2$ values is only slightly stronger than for larger $n_2$ values.

Although the results show no evidence for a combined first-order phase transition signal plus CBC background in the LIGO-Virgo O1–O4a data, our analysis demonstrates that particle physics models predicting such phase transitions are indeed testable with the LIGO-Virgo-KAGRA data.

## III. COSMIC STRINGS

### A. Motivation

Cosmic strings are topological defects in the Universe that can be generated from spontaneous symmetry breaking of a global or gauge symmetry which has nontrivial winding of the vacuum manifold during cosmological phase transitions [34] via the Kibble-Zurek mechanism [35,127–129]. The width of the cosmic strings is inversely proportional to the energy scale of the symmetry breaking and is, thus, generally tiny, making these strings linelike. Formed mainly as superhorizon objects, these long strings intercommute and intersect to form a network of string loops. String loops oscillate due to their tension and shrink due to the emission of gravitational waves, Nambu-Goldstone bosons, or gauge bosons, depending on which coupling of the radiated particle to the string world sheet is dominant.

One of the simplest string models is the axion string model, resulting from a spontaneous symmetry breaking of U(1) global symmetry. The QCD axion string, as a global string, is one of the well-motivated axion string models, since QCD axions [130–138] could contribute to the dark matter relic abundance. Axion strings predominantly radiate Nambu-Goldstone bosons (axions), and, thus, gravitational waves radiated by axion strings are subdominant. A recent study [139] embeds the QCD axion string into the gauged global string model resulting from two subsequent spontaneous symmetry breakings of global U(1) and gauge U(1) symmetries. The model contains both global strings and gauge strings, and the gauge string as a bound state of two types of global strings can radiate either gravitational waves or axions depending on whether the gauge coupling is significantly smaller than the gravitational coupling. Thus, it enriches the radiation channels from gauge strings. Without further assumption on the string model, in what follows, we consider only Nambu-Goto gauge strings, for which gravitational waves are the dominant radiation channel.

The evolution of the string loops in an expanding Universe eventually results in a scaling distribution, with loop sizes proportional to the cosmic time or the Hubble radius. At high frequencies, the gravitational-wave production is dominated by cusps, kinks, and kink-kink collisions. The dimensionless decay constant that characterizes the radiation power of gravitational waves from cusps, kinks, and kink-kink collisions can be estimated by

$$\Gamma_{\rm d} \equiv \frac{P_{\rm GW}}{G\mu^2} = \sum_i \frac{P_{{\rm GW},i}}{G\mu^2}, \qquad (12)$$

where $G\mu$ is the string tension and $i = \{{\rm c, k, kk}\}$ denotes cusp, kink, and kink-kink collision cases, respectively. Incoherent superpositions of these gravitational waves lead to a gravitational-wave background, the detection of which can be used to infer the energy scale of the symmetry breaking and is, thus, an important target for gravitational-wave detectors.

This target has been previously searched for with LIGO's O1 [140] and O2 [141] data and, more recently, with LIGO and Virgo's O3 data combined with previous O1 and O2 data [142]. It has also been searched for by pulsar timing array experiments [143–145] and remains an important source of gravitational-wave background for future space-based gravitational-wave detectors [146–148] and atomic interferometers [149].

### B. Model

The gravitational-wave spectrum is

$$\Omega_{\rm CS}(f) = \frac{4\pi^2}{3H_0^2} f^3 \sum_i \int dz \int dl h_i^2 \frac{d^2R_i}{dzdl}, \qquad (13)$$

where the index $i$ runs over cusps, kinks, and kink-kink collisions, $l$ denotes the invariant loop length, $z$ stands for the redshift, and $h_i = A_i(l,z) f^{-q_i}$, with $A_i = g_{1,i} G\mu l^{2-q_i} / [(1+z)^{q_i - 1} r(z)]$, $r(z)$ the comoving distance of the loop, $q = 4/3, 5/3$, 2, respectively, for cusps, kinks, and kink-kink collisions, and $g_{1,i} \approx 0.85$, 0.29, 0.10 correspondingly. For each type $i$, the burst rate per redshift and loop size is

$$\frac{d^2R_i}{dzdl} = \frac{\varphi_V(z)}{H_0^3(1+z)} \frac{2N_i}{l} n(l,t)\Delta_i, \qquad (14)$$

where $\Delta_i = (\theta_m/2)^{3(2-q_i)}$, with $\theta_m \equiv [g_2 f(1+z)l]^{-1/3}$ and $g_2 = \sqrt{3}/4$, denotes the fraction of burst events that can be detected, $n(l,t)$ is the loop distribution function (the number of loops of size $l$ at time $t$ per loop size per volume), $N_i$ is the number of burst events per loop oscillation time, and $\varphi_V(z) = H_0^3 dV(z)/dz$, with $V(z)$ the proper volume at redshift $z$.

The spectrum above includes only the contribution from subhorizon string loops, though long strings can also emit gravitational waves. As long strings intercommute, they are building a small-scale structure, resulting in the emission of radiation [150,151]. This additional contribution is generally subdominant as compared with that from string loops and, hence, usually neglected.

The gravitational-wave background sourced from cosmic string loops is very sensitive to the scaling loop distribution and the string microstructure as well as the number of cusps, kinks, and kink-kink collisions. The loop distribution is subject to several theoretical uncertainties, and numerical results for the matter era loop distribution are inconsistent [152–154]. In our analysis, as in O3 studies [142], we consider three typical models of the string loop population $n(\gamma, z)$ in a scaling regime within a Friedmann-Lemaître-Robertson-Walker metric, where $\gamma = \ell/t$ is the dimensionless loop size, and derive constraints on each of them. Models $A$ [153,154] and $B$ [155] (called model 2 and 3, respectively, in the O1 study) are based on numerical simulations of Nambu-Goto string networks of Nambu-Goto string networks (zero thickness strings with intercommutation probability equal to unity) but differ in their interpretation of the results. The analytical modeling [152] of model $B$ considers also the effect of gravitational-wave backreaction on the loops. Model $B$ leads to a higher number of small loops than model $A$, leading to important consequences in the rate of gravitational-wave events we can detect and on the amplitude of the gravitational-wave background. Model $C$ [156] is constructed to incorporate features of both models $A$ and $B$. It assumes that the scaling loop distribution is a power law but leaves its slope unspecified. As in the O3 study, we consider two different examples of model $C$ by choosing parameters to reproduce $A$ and $B$ in radiation and matter eras. Model $C1$ (respectively, $C2$) reproduces qualitatively the loop production function of model $A$ (respectively, $B$) in the radiation-dominated era and the loop production of model $B$ (respectively, $A$) in the matter-dominated era. We investigate the effect of the number of kinks, cusps, and kink-kink collisions.

Following the same approach as in the O1 LVK study [140], the parameters for model $B$ are obtained from Ref. [155]. Similarly, for models $C1$ and $C2$, the parameters are obtained from Ref. [156], as in the O3 analysis [142]. See Appendix A in Ref. [142] for the equations governing the models and Sec. II in Ref. [142] for the parameter values.

### C. Constraints using O1–O4a LIGO-Virgo data

Following the O3 study [142], we carry out a Bayesian analysis with the same posterior as previously:

$$p(G\mu|N_k) \propto \mathcal{L}(\hat{C}_a^{IJ}|G\mu, N_k)p(G\mu|I, N_k), \quad (15)$$

where $\mathcal{L}(\hat{C}_a^{IJ}|G\mu, N_k)$ is the likelihood

$$\ln \mathcal{L} = -\frac{1}{2}\sum_{IJ,a} \frac{[\hat{C}_a^{IJ} - \Omega_{\rm CS}(f_a; G\mu, N_k)]^2}{\sigma_{IJ}^2(f_a)}, \quad (16)$$

$\hat{C}_a^{IJ} \equiv \hat{C}^{IJ}(f_a)$ and $\sigma_{IJ}$ are, respectively, the cross-correlation estimator and the variance for the detector pair $IJ$, running over LIGO-Livingston and LIGO-Hanford, LIGO-Hanford and Virgo, and LIGO-Livingston and Virgo [19]. The data used encompass those used in O3 analysis, i.e., from O1, O2, and O3 runs, and, in addition, the data of LIGO-Livingston and LIGO-Hanford from the O4a period, while Virgo was not running during this period. In addition, $p(G\mu|I)$ is the prior on $G\mu$, and we impose a log-uniform prior for $G\mu$ in the range $10^{-18} \leq G\mu \leq 10^{-6}$ and do the analysis for each value of $N_k$ while fixing $N_c = 1$.

Since there is no detection, we show in Fig. 5 the exclusion region at the 95% CL on the parameters $G\mu$ versus $N_k$ for the four loop distribution models, with O1, O2, O3, and O4a data. In the frequency band of LIGO-Virgo, the main contribution to the gravitational-wave background comes from loops in the radiation era; therefore, models $B$ and $C2$ lead to similar findings. Model $C1$ presents more subtle features. Larger values of the string tension do not necessarily imply signals with larger amplitude, leading to features in the corresponding plot. As pointed out [157], the peculiar feature observed in the exclusion plot for model $C1$ must be interpreted as exclusion for a set of $G\mu$ intervals. As discussed earlier, model $C1$ reproduces qualitatively the loop production function of model $A$ in the radiation-dominated era and the loop production of model $B$ in the matter-dominated era. Model $B$ leads to an additional population of small loops, as compared to model $A$, which modifies the gravitational-wave spectrum at frequencies relative for LIGO-Virgo. This extra population of small loops during the matter era leads to a peak in the spectrum around a given frequency characterized by the gravitational backreaction scale. This peak enters the detectability band at a certain range of string tension, hence excluding another $G\mu$ interval [157].

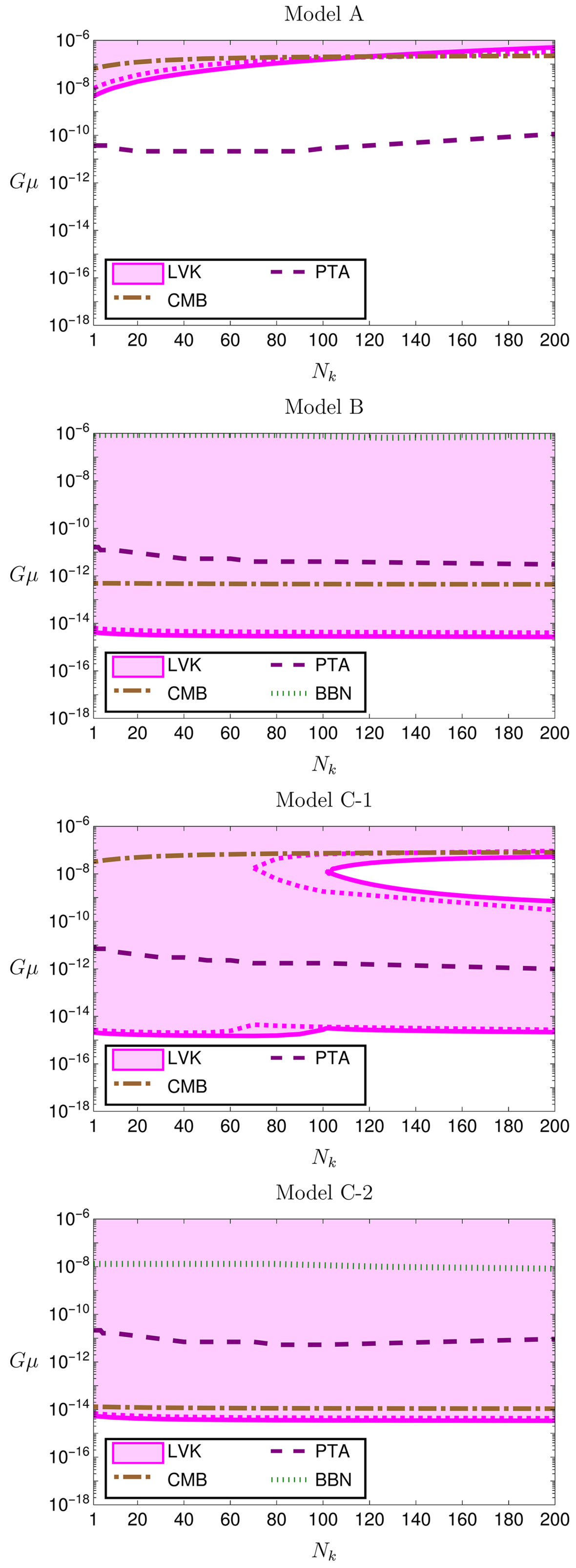

FIG. 5. Exclusion regions at the 95% CL on the cosmic string parameter space $(G\mu, N_k)$.

Compared with previous results, the limits generally become stringent. For model $A$, the region $G\mu \gtrsim (4.5 \times 10^{-9}$–$5.1 \times 10^{-7})$ is excluded for $1 \leq N_k \leq 200$, with the strongest constraint achieved at $N_k = 1$. This result improves over that of O1–O3 for a wide range of $N_k$. It becomes worse only for unrealistically high values of $N_k \geq 120$. More precisely, for $N_k < 120$ the constraint on $G\mu$ is better at most by a factor of about 0.47, while for $N_k \geq 120$ it is worse by at most 1.6.

For model $B$, the region $G\mu \gtrsim (2.7$–$4.2) \times 10^{-15}$ is excluded, and this improves over that of O3 by a factor of 0.66 over the range of $N_k$ considered here. For model $C1$, the region $G\mu \gtrsim (1.5$–$3.2) \times 10^{-15}$ is excluded, and this improves on the previous result in all the range of $N_k$ by a factor of (0.33–0.79). Because of the features of the spectrum for this model, there is a region that cannot be excluded for higher values of $G\mu$. However, this region shrinks after the O4a data are included. For model $C2$, the region $G\mu \gtrsim (3.4$–$5.5) \times 10^{-15}$ is excluded, which improves by a factor of 0.77 compared to O3.

It should be noted that the results presented here treat each choice of $N_k$ as a separate model (with $N_c = 1$) in the Bayesian analysis. Increasing $N_c$ has a similar effect as increasing $N_k$, as both lead to enhanced power of gravitational-wave emission, while the resulting changes to the constraints are different for the three models, with model $A$ weakened and models $B$ and $C$ less sensitive. Ideally, a joint distribution of $N_k$ together with $N_c$ should be used and a marginalization over these two parameters should be performed to obtain the constraint on $G\mu$. Because of a lack of simulations to get this information, we have adopted this approach. It should also be noted that the choice $\Gamma_d = 50$ is commonly used, according to simulation results. Enforcing this power emission corresponds to setting $(N_c, N_k) \approx (1, 9)$ or (0,18). The constraints for $(N_c = 0, N_k = 18)$ are slightly more stringent for most models except for $C1$. This is due to the smaller value of $\Gamma_d$, despite the absence of gravitational-wave emission from cusps. More specifically, the excluded regions are $G\mu \gtrsim 9.2 \times 10^{-9}$ for model $A$, $G\mu \gtrsim 3.0 \times 10^{-15}$ for model $B$, $G\mu \gtrsim 2.6 \times 10^{-15}$ for model $C1$, and $G\mu \gtrsim 3.5 \times 10^{-15}$ for model $C2$.

In this analysis, the average number of cusps per oscillation on a loop has been set to 1. As has been already shown in the O3 analysis [142], a high number of cusps gives qualitatively the same result as increasing the number of kinks. More precisely, numerical simulations have shown that the constraints are weakened for model $A$, whereas the bounds are insensitive to $N_\mathrm{c}$ for models $B$ and $C$.

We include in Fig. 5 the corresponding constraints from pulsar timing arrays (PTA), cosmic microwave background (CMB) [158], and Big Bang nucleosynthesis (BBN) [159] obtained from the O1 [140] and O3 analysis [142]. Note that the PTA limits are obtained considering the nanohertz limit of the primordial background obtained from the Parkes Pulsar Timing Array [22] $\Omega_{\mathrm{CS}} < 2.3 \times 10^{-10}$ at a single frequency of $2.8 \times 10^{-9}$ Hz, which is comparable to the latest results of NANOGrav, EPTA, and ParkesParkes [143–145].

In our analysis, we assume that the local curvature of the cosmic string is small compared to its thickness; hence, their dynamics are described by the Nambu-Goto action. Such strings decay by emission of gravitational waves. However, considering cosmic strings as solitonic solutions of a classical field theory, another decay mechanism is energy released into excitations of their elementary constituents. Large-scale field theory simulations find that loops decay into field modes within a Hubble time, and, thus, they do not contribute to the gravitational-wave background [160–162]. Such a decay channel does not exist for Nambu-Goto strings, and this discrepancy remains under debate. More recent field theory simulations find that loops seem to behave according to the Nambu-Goto action, except in regions of high curvature where energy is emitted in the form of massive radiation [163]. Finally, an Abelian-Higgs string network was considered, allowing for both particle and gravitational-wave production, leading to multimessenger constraints from NANOGrav [164].

We briefly comment on the contribution from long strings [150,165,166]. Considering a kinky Nambu-Goto string network, recent works [167–169] suggest enhanced gravitational-wave production from long strings using a semianalytic approach and a modeling of the kink structure with sharpness [170]. This makes the long string scenario potentially detectable by the LIGO-Virgo-KAGRA network. Adopting the spectrum from Ref. [169], we find that the excluded region is $G\mu \gtrsim 2.06 \times 10^{-7}$, comparable to the results obtained from the loop distribution of model $A$.

## IV. DOMAIN WALLS

### A. Motivation

In a cosmological context, domain walls (DWs) are two-dimensional defects that arise when a discrete symmetry is spontaneously broken during the thermal history of the Universe [35,36]. Around the temperature of this symmetry-breaking phase transition, uncorrelated patches in space will select one among the possible disconnected degenerate vacua of the theory. DWs are then formed at the boundaries of those regions where the scalar field interpolates between different vacua. These field configurations are topologically stable owing to the underlying discrete symmetry. At the center of the DW, the field is trapped at the maximum of the scalar potential leading to a high-energy density localized within the wall width. This results in a large DW tension, which is effectively the wall mass per unit surface. The relativistic motion of the DWs, driven by their own tension force or by vacuum pressure, could act as a powerful source of gravitational waves that can be detected today.

Similarly to other topological defects such as cosmic strings, DWs in the early Universe are known to reach a scaling regime with a constant $\mathcal{O}(1)$ number of walls per Hubble volume [37–42]. Differently from the strings, this implies that the relative importance of the DW network in the energy budget of the Universe grows with time, potentially leading to a phase of DW domination. As this is inconsistent with the standard evolution of the Universe, DWs have been often regarded as a cosmological problem. Crucially, however, a DW network is not expected to be absolutely stable, as the underlying discrete symmetry needs not to be exact but only approximate. In this case, DWs can annihilate before dominating the expansion of the Universe, leading to a strong gravitational-wave signal and no contradiction with standard cosmology.

New physics scenarios involving the formation of DWs are characterized by the presence of (approximate) discrete symmetries that are spontaneously broken in the early Universe. Relevant examples include the QCD axion [130–135,171] and, more generally, axionlike particles, where a residual $\mathbb{Z}_N$ subgroup of the original U(1) Peccei-Quinn symmetry [we generally refer to U(1) Peccei-Quinn also for the case of axionlike particles that are not related to the strong *CP* problem] is left untouched by its chiral anomaly. This particular class of models implies the formation of cosmic strings at temperatures of the order of the axion decay constant, $f_a$, when the Peccei-Quinn symmetry is spontaneously broken and the axion is effectively massless. If this occurs after cosmic inflation, the strings and the corresponding inhomogeneous axion field will play an important role in the subsequent evolution of the system. In fact, as the axion mass increases while the Universe cools down, a network of axion DWs will ultimately form with each string attached to $N$ walls of tension $\sigma_{\rm DW} \sim m_a f_a^2$, where $m_a$ is the axion mass [172,173]. The temperature of DW formation in this case can be estimated as the moment when the axionlike particle mass overcomes the Hubble friction, $m_a \sim H$.

The following dynamics depends on the value of $N$, which is referred to as the DW number. For $N = 1$ the string-wall network collapses very quickly after DW formation, as the theory actually possesses a unique vacuum. On the other hand, for $N > 1$, axion DWs are topologically stable and can be long-lived depending on the quality of the underlying Peccei-Quinn symmetry. This latter scenario is the one relevant for a gravitational-wave signal from DWs, as the string-wall dynamics is mostly controlled by the walls in this case.

Minimal QCD axion models predict the formation of DWs at temperatures around the QCD scale, $\Lambda_{\rm QCD} \sim 150$ MeV, so that the corresponding gravitational waves would not overlap with the LIGO-Virgo-KAGRA observation band. Earlier formation of DWs leading to a detectable gravitational-wave signal is, however, possible for the so-called heavy QCD-axion models [174–176], which still solve the strong *CP* problem and ameliorate the issue with the quality of the U(1) Peccei-Quinn symmetry [177–182], or for general axionlike particles depending on the relevant scales [183].

Beyond the case of axions and axionlike particles, many other scenarios of new physics involve new discrete symmetries that can ultimately lead to the formation of a DW network. Well-motivated models include discrete flavor symmetries [184], left-right symmetric models [185], supersymmetry [186–190], grand unification [191–193], and discrete spacetime symmetries [194].

### B. Model

The dynamics of the DW network is controlled, on one hand, by the tension force, which tends to stretch the walls and reduce their surface to minimize the energy, and, on the other hand, by the Hubble expansion as well as the interaction of the walls with the primordial plasma. When particle friction can be neglected, DWs are known to approach a scaling regime in which the typical scale of the network, such as the average curvature and distance between the walls, is given by the Hubble radius $H^{-1}$, indicating the presence of $\mathcal{O}(1)$ DWs per Hubble volume at any time [37–42]. In this regime, the energy density of the network is given by

$$\rho_{\rm DW} = 2\mathcal{A}\sigma_{\rm DW}H, \tag{17}$$

where $\mathcal{A} = \mathcal{O}(1)$ and $\sigma_{\rm DW}$ is the DW tension or mass per unit surface. Equation (17) indicates that the energy density of the network decreases more slowly than matter or radiation, eventually leading to a DW-dominated epoch that is inconsistent with cosmological observations [36].

The temperature at which this would occur can be estimated by equating the energy density of the DWs to the critical density of the Universe, $\rho_c = 3H^2c^2/(8\pi G)$, yielding

$$T_{\mathrm{dom}} = \left(\frac{80G}{\pi c^4 g_*}\right)^{1/4}\sqrt{\sigma_{\mathrm{DW}}}, \tag{18}$$

where we have assumed radiation domination with $g_*$ being the number of relativistic degrees of freedom.

Crucially, DW domination can be avoided if the underlying discrete symmetry is only approximate and the different vacua of the theory are actually biased by a small energy difference $\Delta V$ such that there exists only a unique true vacuum state [173,195]. The microscopic origin of this bias will depend on the specific particle physics under consideration. However, according to the no global symmetry conjecture in quantum gravity [196–199], one expects the DW discrete symmetry to be ultimately broken at the Planck scale or earlier. In the case of axions and axionlike particles, the bias term can then descend from Planck-suppressed higher-dimensional operators that break the U(1) Peccei-Quinn symmetry as well as its $\mathbb{Z}_N$ subgroup relevant for DW formation.

The vacuum pressure resulting from the potential bias $\Delta V$ competes with the tension force trying to annihilate the DW network. The temperature at which the collapse initiates, $T_{\mathrm{ann}}$, can be estimated by equating the bias to the tension force or, equivalently, the DW energy density in the scaling regime, namely, $\Delta V \sim \rho_{\mathrm{DW}}$, leading to

$$T_{\mathrm{ann}} \sim 10^8\ \mathrm{GeV}\left(\frac{10^{11}\ \mathrm{GeV}}{\sigma_{\mathrm{DW}}^{1/3}}\right)^{3/2}\left(\frac{\Delta V^{1/4}}{10^8\ \mathrm{GeV}}\right)^2\left(\frac{100}{g_*}\right)^{1/4}. \tag{19}$$

For consistency, the annihilation temperature needs to be smaller than the temperature at which DWs form, which is at most as large as the DW tension $\sigma_{\mathrm{DW}}^{1/3}$ and parametrically suppressed for axion DWs. In the following, we hence restrict ourselves to $T_{\mathrm{ann}} \lesssim \sigma_{\mathrm{DW}}^{1/3}$.

During the lifetime of the DW network, gravitational waves are copiously produced by the relativistic motion of the walls [200–203]. As the energy density of the network actually increases with time compared to the critical density according to Eq. (17), the gravitational-wave emission is the strongest around the final time of DW annihilation. While the dynamics of the DW collapse itself can contribute to the emission of gravitational waves [204,205], we here consider only the gravitational-wave spectrum coming from the last period of scaling just before the collapse begins, namely, at $T = T_{\mathrm{ann}}$. Several numerical simulations of the gravitational waves emitted by a domain wall network in scaling have been performed [203,206–212], where it has been shown that the gravitational-wave spectrum is a broken power law. In the following, we employ the recent simulation [212], where the spectrum scales as [203]

$$\Omega_{\mathrm{DW}}(f) = \Omega_{\mathrm{DW}}^{\mathrm{peak}} \times \begin{cases} (f/f_{\mathrm{peak}})^3 & f < f_{\mathrm{peak}}, \\ (f/f_{\mathrm{peak}})^{-1.3} & f > f_{\mathrm{peak}}, \end{cases} \tag{20}$$

where the peak amplitude associated to gravitational-wave emission at $T_{\mathrm{ann}}$ and redshifted until today is given by

$$\Omega_{\mathrm{DW}}^{\mathrm{peak}} = 2.45\times 10^{-7}\left(\frac{\tilde{\epsilon}_{\mathrm{gw}}}{0.35}\right)\left(\frac{g_*}{100}\right)\left(\frac{g_{*s}}{100}\right)^{-4/3}\left(\frac{T_{\mathrm{dom}}}{T_{\mathrm{ann}}}\right)^4, \tag{21}$$

with $\tilde{\epsilon}_{\mathrm{gw}} \sim O(1)$ the efficiency of the DW network in emitting gravitational waves and $g_{*s}$ denotes the effective number of entropy degrees of freedom. The redshifted peak frequency is

$$f_{\mathrm{peak}} = 17\ \mathrm{Hz}\left(\frac{g_*}{100}\right)^{1/2}\left(\frac{g_{*s}}{100}\right)^{-1/3}\left(\frac{T_{\mathrm{ann}}}{10^8\ \mathrm{GeV}}\right). \tag{22}$$

In Fig. 6, we show a benchmark spectrum to highlight the effect of varying the DW tension, as well as the annihilation temperature. The power-law integrated sensitivity curves displayed here and in subsequent figures follow the method described in Ref. [213].

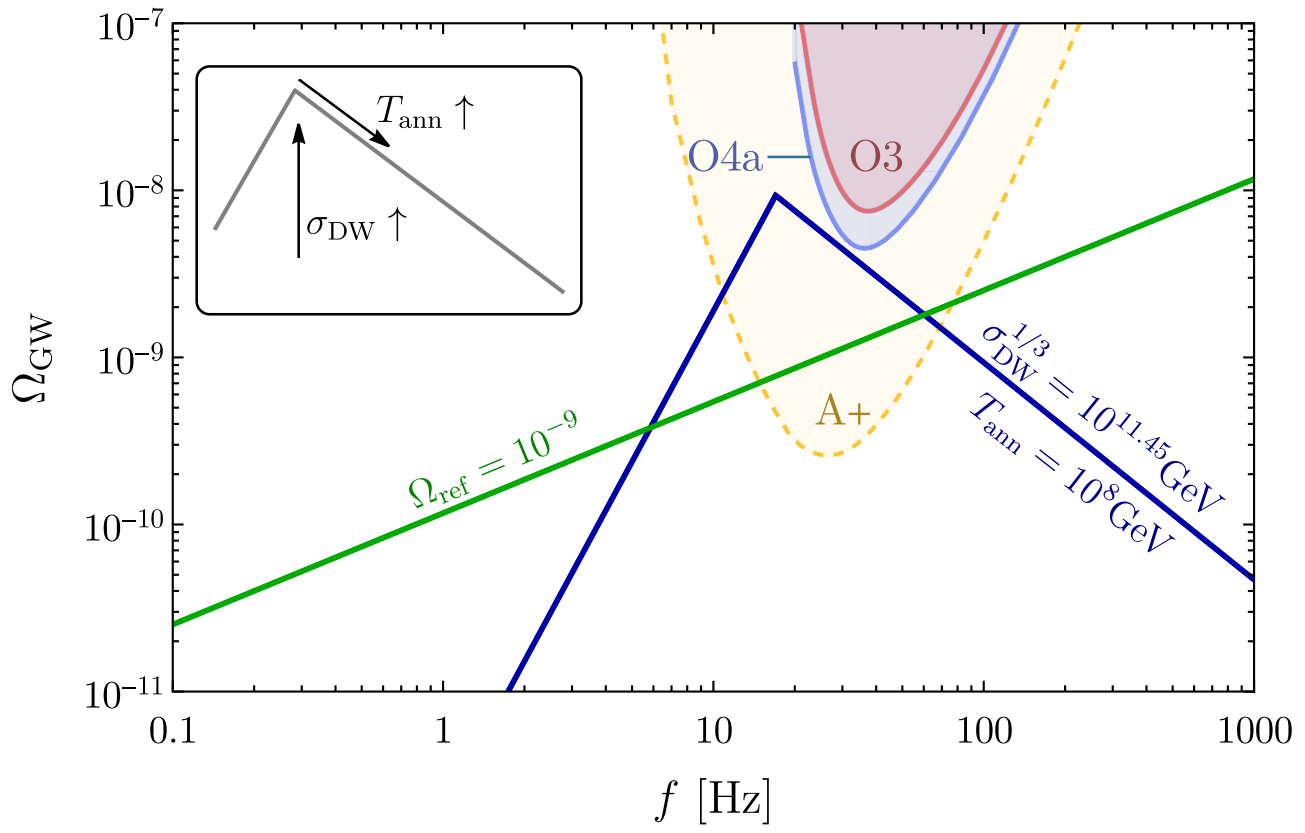

FIG. 6. The gravitational-wave spectrum of DW networks for a benchmark for the wall tension $\sigma_{\mathrm{DW}}$ and annihilation temperature $T_{\mathrm{ann}}$. The schematic in the top left box illustrates the impact of these parameters on the spectra: increasing the wall tension $\sigma_{\mathrm{DW}}$ enhances the peak amplitude, while a higher annihilation temperature $T_{\mathrm{ann}}$ reduces the amplitude and shifts the spectrum to higher frequencies. In addition, the assumed shape for the overlapping astrophysical gravitational-wave background is displayed in green, for a representative value of $\Omega_{\mathrm{ref}}$ (and with $\alpha = 2/3$). Finally, power-law integrated sensitivity curves [213] for LIGO-Virgo O3 run [68] as well as the ones of the O4a run [19] and LIGO A+ detector [214] are included.

TABLE III. Prior distributions assumed for the parameters of the DW model and the CBC background. The prior on $\Omega_{\mathrm{ref}}$ comes from estimates of the CBC background [17], whereas the range for the priors on the tension $\sigma_{\mathrm{DW}}$ and the annihilation temperature $T_{\mathrm{ann}}$ are chosen large enough to include region of parameter space that would lead to gravitational-wave signals within the LIGO-Virgo-KAGRA observational band. Values for $T_{\mathrm{ann}}$ larger than $\sigma_{\mathrm{DW}}^{1/3}$ are not considered, as previously discussed around Eq. (19).

| Parameter | Prior |
|---|---|
| $\Omega_{\mathrm{ref}}$ | LogUniform$[10^{-13}, 10^{-6}]$ |
| $\sigma_{\mathrm{DW}}^{1/3}/\mathrm{GeV}$ | LogUniform$[10^{10}, 10^{13}]$ |
| $T_{\mathrm{ann}}/\mathrm{GeV}$ | LogUniform$[10^{6}, 10^{10}]$ |

### C. Constraints using O1–O4a LIGO-Virgo data

We perform a Bayesian analysis following the approach described in Sec. I and using the PYGWB package [122]. For the gravitational-wave spectrum, we consider the contribution from a DW network, given by Eq. (20), in combination with the astrophysical gravitational-wave background from unresolved CBCs [defined as in Eq. (2) with $\alpha = 2/3$ and $f_{\mathrm{ref}} = 25$ Hz]:

$$\Omega_{\mathrm{GW}}(f|\theta) = \Omega_{\mathrm{DW}}(f) + \Omega_{\mathrm{CBC}}(f). \tag{23}$$

The parameters of interest are $\theta = (\Omega_{\mathrm{ref}}, \sigma_{\mathrm{DW}}, T_{\mathrm{ann}})$, with $\Omega_{\mathrm{ref}}$ the CBC background amplitude and $T_{\mathrm{ann}}$ and $\sigma_{\mathrm{DW}}^{1/3}$ the parameters determining the cosmological gravitational-wave signal from DWs. The prior distributions for these parameters are summarized in Table III.

The resulting posterior distributions are shown in Fig. 7, which displays contour regions corresponding to 1 to $2\sigma$ CL. From the posterior of the CBC background amplitude $\Omega_{\mathrm{ref}}$, we set an upper limit at the 95% CL, with a value of $2.82 \times 10^{-9}$.

For the parameters controlling the DW signal, the region excluded by the gravitational-wave data is visualized in white in the bottom-middle panel. In the same panel, we have shaded in gray the values of $T_{\mathrm{ann}}$ and $\sigma_{\mathrm{DW}}$ for which the DW system would have dominated the Universe, leading to inconsistent cosmology. The excluded white region is close to DW domination, as the gravitational-wave signal is strongest when $T_{\mathrm{ann}} \sim T_{\mathrm{dom}}$. Our analysis can rule out annihilation temperatures in the range $10^7$ GeV $< T_{\mathrm{ann}} < 10^9$ GeV for sufficiently large DW tension.

Considering the hypothesis of a combined gravitational-wave background signal from a DW network and CBCs versus noise, the analysis yields a Bayes factor of $\ln \mathcal{B}_{\mathrm{noise}}^{\mathrm{DW+CBC}} = -1.10$, indicating no evidence for such a background in the data. Similarly, for a CBC-only background, we find $\ln \mathcal{B}_{\mathrm{noise}}^{\mathrm{CBC}} = -0.52$, implying a preference for the CBC-only scenario with $\ln \mathcal{B}_{\mathrm{CBC}}^{\mathrm{DW+CBC}} = -0.58$.

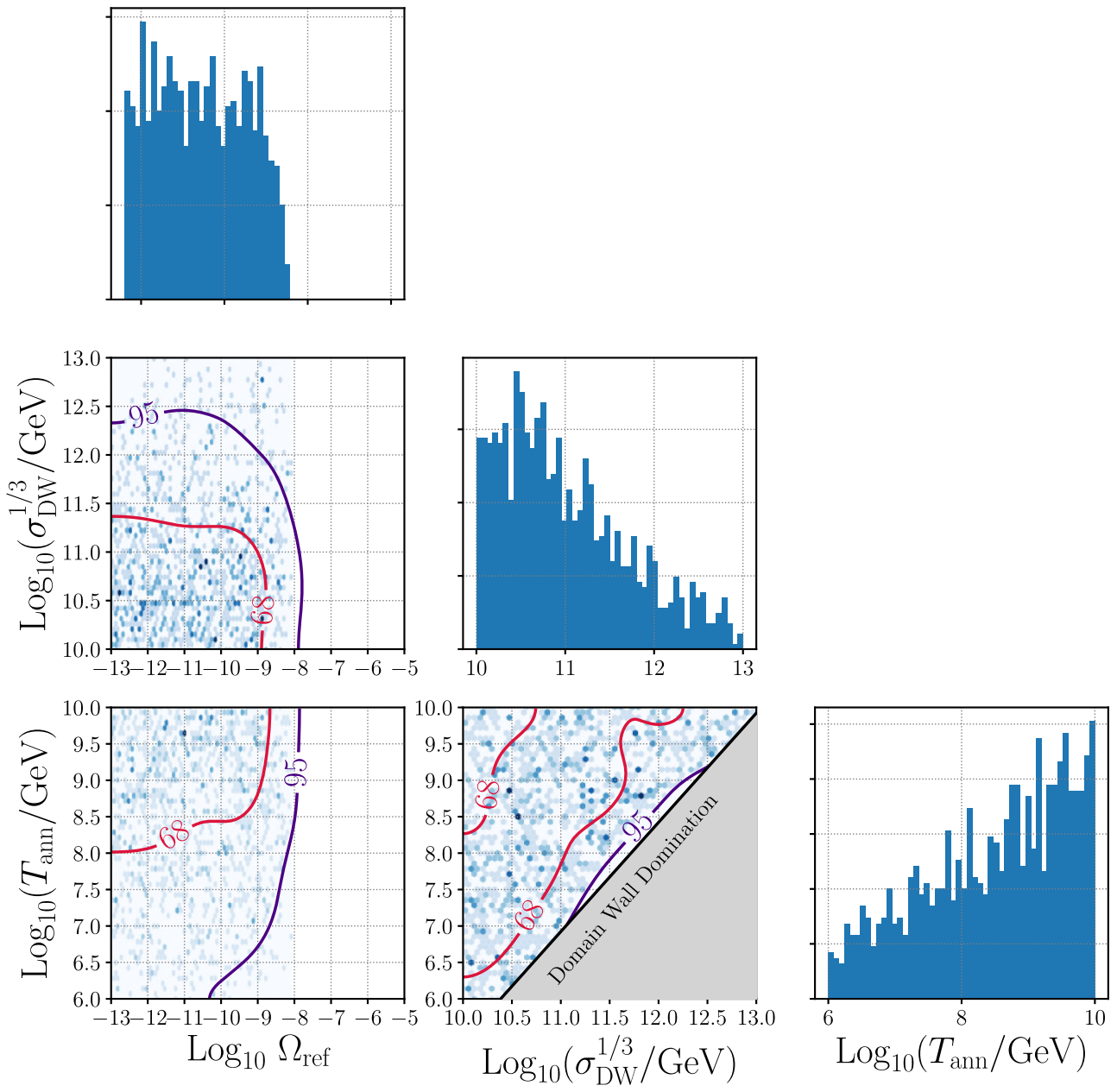

FIG. 7. Posteriors for the strength of the CBC background amplitude $\Omega_{\mathrm{ref}}$, the tension $\sigma_{\mathrm{DW}}$ of the DW network, and the temperature $T_{\mathrm{ann}}$ at which the network annihilates using LIGO-Virgo data from O1, O2, O3, and O4a. The gray region in the bottom corresponds to a region where the DW network dominates the Universe, i.e., $T_{\mathrm{ann}} \leq T_{\mathrm{dom}}$.

In conclusion, we find no evidence for a gravitational-wave signal from a DW network. The constraints on $\sigma_{\mathrm{DW}}$ and $T_{\mathrm{ann}}$ derived from gravitational-wave data exclude specific regions of parameter space, which can be used in the context of particle physics models.

## V. STIFF EQUATION OF STATE

### A. Motivation

Standard inflationary models in the slow-roll regime typically give rise to a gravitational-wave background that is too weak to be detected by the current and future gravitational-wave experiments. Indeed, the current constraints on the scale of inflation and on the tensor-to-scalar ratio implies that the typical primordial gravitational-wave flat spectrum lies below the sensitivity of current and future gravitational-wave experiments in the $\Lambda$CDM model.

However, the detailed form of the primordial spectrum and, hence, its detectability depends on the assumptions about the cosmological history in the period between reheating and the onset of BBN. In particular, adding an exotic era dominated by stiff energy [called the *stiff*-dominated (SD) epoch; see, e.g., Refs. [43–51] ] leads to an inflationary gravitational-wave background growing at higher frequencies (i.e., blue tilted), making it accessible to the LIGO-Virgo-KAGRA network [50–53], future third-generation detectors, and space-based laser interferometer experiments [47,215–225]. There are several concrete models in physics beyond the Standard Model that lead to cosmological periods with a stiff equation of state. We list here, for instance, quintessence models [43,44,226], axion scenarios [227–230], models with string theory moduli [231,232], Peccei-Quinn inflation models [233], etc. In the following, we introduce a model-independent parametrization for a stiff epoch and explore the constraints imposed on this scenario by the O1–O4a runs of LIGO-Virgo-KAGRA. The same model was previously constrained using the O1–O3 data in Ref. [53].

### B. Model

The Universe can be described as a cosmological fluid with two key parameters: density $\rho$ and pressure $P$. The equation of state parameter $w = P/\rho$ characterizes the Universe's behavior across different epochs. In the $\Lambda$CDM model, inflation is immediately followed by a period of radiation domination (RD), with $w = 1/3$, then a period of matter domination (MD) with $w = 0$, and, currently, a period of dark energy domination with $w = -1$. Modifications to this sequence can be made by inserting other eras between the end of inflation and the onset of BBN, provided the Universe is RD during BBN [46].

Our model introduces an unconventional sequence of epochs preceding the standard eras: an exotic RD era (denoted by RD1), an exotic MD era (denoted by MD1), and an exotic era driven by *stiff energy*, described by an equation of state parameter $1/3 \leq w_s \leq 1$ (denoted by SD). The most extreme case of such a SD era is called kination, in which $w_s$ is fixed to 1.

This cosmological model is motivated by high-energy physics and can have a variety of observational consequences. Indeed, this specific sequence of epochs (in the case of kination) naturally arises in axion models [227–230] (see also Refs. [234–236]), providing a theoretical motivation for this cosmological scenario. In addition, a stiff era leads to a blue tilt in the inflationary gravitational-wave spectrum, making it observationally interesting. Finally, including a MD era suppresses the gravitational-wave spectrum at higher frequencies, allowing the model to evade indirect constraints from CMB and BBN observations.

The unconventional cosmological history enhances the inflationary gravitational-wave background, which results in a spectral shape with the following asymptotic behavior [53]:

$$\Omega_{\rm SD}(f) = \Omega_{\rm SD}|^{(0)}_{\rm plateau} \begin{cases} \mathcal{A}_1 & \text{if } f \ll f_{\rm RD}, \\ \mathcal{A}_{\alpha_s}\left(\frac{f}{f_{\rm RD}}\right)^{2(1-\alpha_s)} & \text{if } f_{\rm RD} \ll f \ll f_{\rm SD}, \\ \mathcal{A}_2\left(\frac{f_{\rm SD}}{f_{\rm RD}}\right)^{2(1-\alpha_s)}\left(\frac{f_{\rm SD}}{f}\right)^2 & \text{if } f_{\rm SD} \ll f \ll f_{\rm MD}, \\ \mathcal{A}_1\left(\frac{f_{\rm SD}}{f_{\rm RD}}\right)^{2(1-\alpha_s)}\left(\frac{f_{\rm SD}}{f_{\rm MD}}\right)^2 & \text{if } f_{\rm MD} \ll f \end{cases} \tag{24}$$

with $\mathcal{A}_{\alpha_{\rm era}}$ a coefficient that depends on the equation of state at the moment when the gravitational-wave mode reenters the Hubble radius, given by [46]

$$\mathcal{A}_{\alpha_{\rm era}} \equiv \frac{\Gamma^2(\alpha_{\rm era}+1/2)}{\pi}\left(\frac{2}{\alpha_{\rm era}}\right)^{2\alpha_{\rm era}}, \quad \alpha_{\rm era} \equiv \frac{2}{1+3w_{\rm era}}, \tag{25}$$

and where [50]

$$\Omega^{(0)}_{\rm SD}|_{\rm plateau} \equiv G_{\rm k}\frac{\Omega^{(0)}_{\rm rad}}{12\pi^2}\left(\frac{H_{\rm inf}}{M_{\rm Pl}}\right)^2, \tag{26}$$

with $\Omega^{(0)}_{\rm rad} \approx 9\times 10^{-5}$ and with $M_{\rm Pl} = 1/\sqrt{8\pi G} \approx 2.44 \times 10^{18}$ GeV the reduced Planck mass. Moreover, the factor $G_{\rm k} = [g_{*,{\rm k}}/g_{*,0}][g_{s,0}/g_{s,{\rm k}}]^{4/3}$ encodes the change in relativistic degrees of freedom between today and the time when the mode $k$ enters the Hubble radius at $k = aH$. The detailed gravitational-wave background assuming instantaneous transitions between subsequent epochs can be found in Ref. [53], and it is the one used in the following. Note that, in more realistic models, transitions between different cosmological epochs are typically smooth rather than instantaneous (see, for instance, Ref. [227] for the case of a kination epoch). This smoothness slightly modifies the gravitational-wave spectrum around the corresponding transition frequencies. However, the precise form of this modification is model dependent and, as observed in previous analyses based on LIGO-Virgo data [53], does not significantly affect the resulting constraints. Within this approximation, there are five different parameters that influence the spectrum, as can be seen in Fig. 8.

The Hubble scale of inflation $H_{\rm inf}$ influences the size of the spectrum. CMB polarization experiments Planck 2018, BICEP2, Keck Array, and BICEP3 [237] constrain the tensor-to-scalar-ratio $r$ and, therefore, also the inflationary power spectrum as $H_{\rm inf} < H_{\rm inf,max} = 5.12 \times 10^{13}$ GeV.

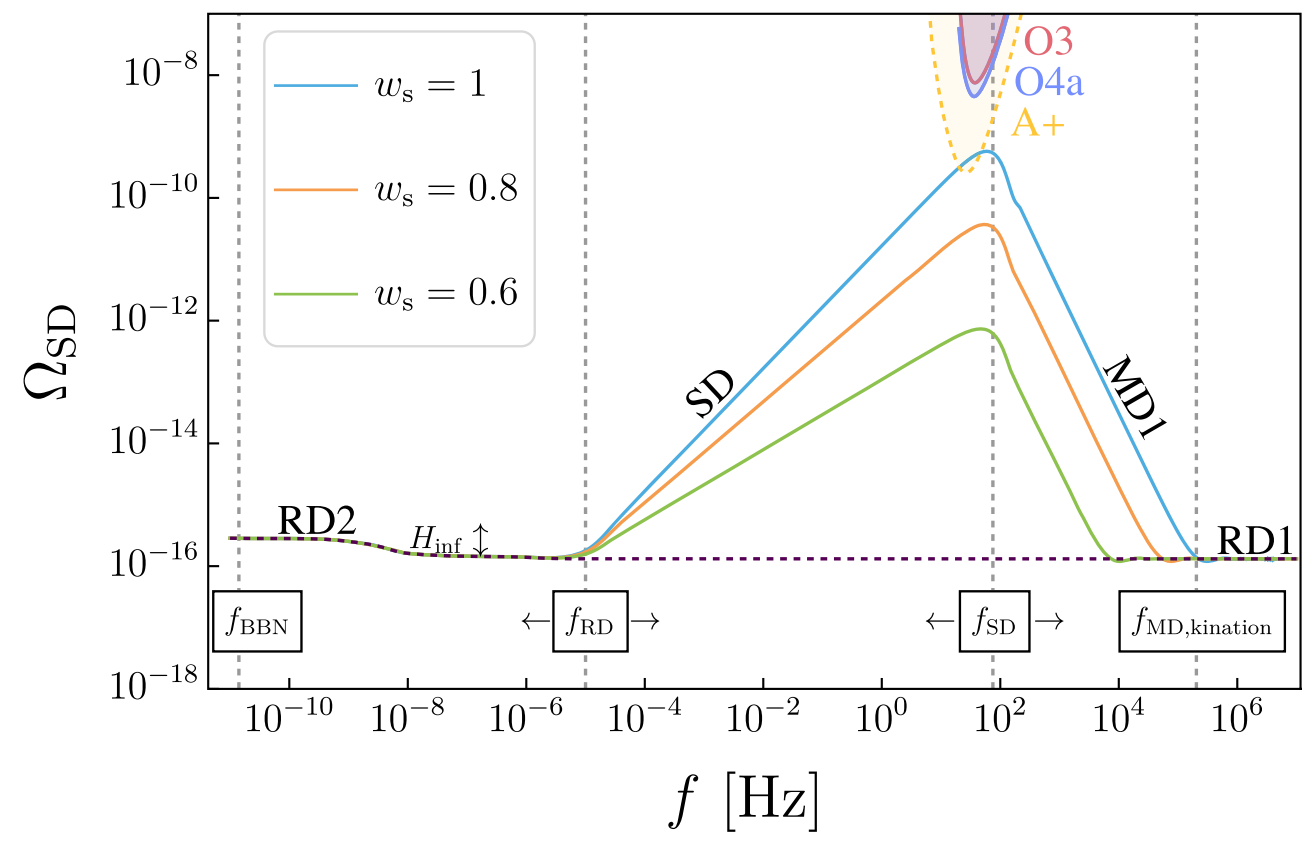

FIG. 8. The gravitational-wave background spectrum resulting from the exotic cosmology, for different choices of $w_s$. The power-law integrated sensitivity curves [213] for LIGO-Virgo O3 run [68], as well as the ones of the O4a run [19] and the LIGO A+ detector [214] are shown. The parameters that control the gravitational-wave spectrum are chosen as $H_{inf} = H_{inf,max}$, $f_{RD} = 10^{-5}$ Hz, $f_{SD} = 75$ Hz, and $f_{MD}$ is fixed, so that the right plateau and the left plateau have the same amplitude. The standard inflationary gravitational-wave background is denoted with a purple dashed line. The parameter $H_{inf}$ affects the overall gravitational-wave background amplitude, as indicated by the double arrow. The low-frequency part of the spectrum is shaped by $f_{RD}$: lower values of $f_{RD}$ shift the low-frequency plateau to the left, resulting in a stronger gravitational-wave background, while higher values shift it to the right, resulting in a weaker gravitational-wave background. The peak frequency of the spectrum depends on $f_{SD}$: lower values shift the peak to the left (weaker gravitational-wave background), and higher values shift it to the right (stronger gravitational-wave background).

The next-generation CMB experiment LiteBIRD [238] is expected to lead to an improvement in this bound as $H_{inf} < 1.21 \times 10^{13}$ GeV.

Then, $f_{RD}$ is the frequency corresponding to the moment of transition between SD and RD2 and influences the position of the *elbow* between plateau and increase in spectrum (here, RD2 denotes the standard radiation era occurring after the SD phase). Note that BBN should occur during the RD2 era, implying that $f_{RD} \geq f_{BBN} \simeq 1.41 \times 10^{-11}$ Hz.

Moreover, the equation of state parameter during the SD era, $w_s$, determines the slope of the increasing part of the spectrum. The range for the related parameter $\alpha_s \equiv 2/(1 + 3w_s)$ is $0.5 \leq \alpha_s \leq 1$, where the lower bound corresponds to kination and the upper bound to radiation.

Then, $f_{SD}$, which corresponds to the transition moment between MD1 and SD, influences the peak amplitude and the location of the peak.

Lastly, $f_{MD}$, which is the transition moment between RD1 and MD1, influences the elbow between the decreasing spectrum and the plateau on the right.

The gravitational-wave energy density can be constrained because of its contribution to the relativistic degrees of freedom in the Universe [23]. For the model considered here, we can estimate this bound as

$$\left(\frac{h^2\rho_{GW}}{\rho_c}\right)\bigg|_{\tau=\tau_0} \approx \frac{1}{2(1-\alpha_s)} h^2\Omega_{SD}(f_{peak}) < 1.3 \times 10^{-6}, \quad (27)$$

where the right-hand side is evaluated at the peak frequency $f_{peak} \sim f_{SD}$ and where we used the $2\sigma$ limit on $\Delta N_{eff}$ from the CMB plus BBN analysis [239]. Further constraints on a stiff epoch could possibly arise from the enhancement of scalar modes [236], depending on the inflationary model.

### C. Constraints using O1–O4a LIGO-Virgo data

We use the Bayesian analysis as described in Sec. I and employ the PYGWB package [122]. The model for the gravitational-wave energy density spectrum $\Omega_{GW}(f|\theta)$ combines the inflationary gravitational-wave background enhanced by a stiff era $\Omega_{SD}(f)$ and the astrophysical gravitational-wave background from unresolved CBCs $\Omega_{CBC}(f)$: $\Omega_{GW}(f|\theta) = \Omega_{SD}(f) + \Omega_{CBC}(f)$. Within the relevant frequency range, the CBC background takes the form of Eq. (2), where $f_{ref} = 25$ Hz is the reference frequency and $\Omega_{ref}$ is the amplitude of the CBC background at this frequency, and we fix the power law to $2/3$.

The parameter space is defined by $\theta = (\Omega_{ref}, H_{inf}, f_{MD}, f_{SD}, f_{RD}, \alpha_s)$. For the Bayesian analysis, however, $H_{inf}$ and $f_{MD}$ are fixed by using delta function priors centered around a constant value. First, we fix $H_{inf}$ to $H_{inf,max} = 5.12 \times 10^{13}$ GeV, which maximizes the amplitude of the gravitational-wave signal. Note that, for the signal in the LIGO-Virgo-KAGRA observational frequency band, there is a degeneracy between $H_{inf}$ and $f_{RD}$, which could be used to translate the results presented here to another value of $H_{inf}$ (for more details, see Ref. [53]). Second, we assume that $f_{MD}$ is higher than the maximal frequency detectable with LIGO-Virgo-KAGRA. Specifically, notice that, as soon as $f_{MD} \gtrsim 100$ Hz, the high-frequency portion of the gravitational-wave spectrum, which is set by $f_{MD}$, lies beyond the LIGO-Virgo-KAGRA observational band and, therefore, does not affect our analysis (see Ref. [53] for more details). For definiteness, we set $f_{MD} = f_i = 1.8 \times 10^8$ Hz, where $f_i$ represents the frequency associated with the end of inflation (assuming a constant Hubble scale during inflation and no entropy injection between RD1 and RD2). The priors for the other parameters are reported in Table IV.

In our analysis, we concretely consider two scenarios.

(i) *Kination model*—We fix $\alpha_s = 0.5$, which corresponds to the maximum value for $w_s$. This model is denoted as "kination + CBC," and its results are given in Fig. 9.

TABLE IV. Priors assumed for the Bayesian analysis. The prior for $\Omega_{\rm ref}$ comes from estimates of the CBC background [17]. The prior for $\alpha_s$ is determined by the allowed range of $w_s$: $1/3 \leq w_s \leq 1$. The prior ranges for $f_{\rm RD}$ and $f_{\rm SD}$ are set to satisfy $f_{\rm BBN} \leq f_{\rm RD} < f_{\rm SD}$. They are selected over a sufficiently wide range to ensure that they include regions of parameter space leading to gravitational-wave signals within the LIGO-Virgo-KAGRA frequency band. We have verified that the posteriors are not significantly affected if we make the priors larger.

| Parameters $\theta$ | Prior |
|---|---|
| $\Omega_{\rm ref}$ | LogUniform$[10^{-13}, 10^{-6}]$ |
| $f_{\rm RD}$/Hz | LogUniform$[10^{-10}, 10^{-5}]$ |
| $f_{\rm SD}$/Hz | LogUniform$[10^{-3}, 10^{6}]$ |
| $\alpha_s$ | Uniform$[0.5, 1]$ |

(ii) *General SD epoch model*—In this case, $w_s$ is allowed to vary, referred to as “SD + CBC.” Its results are given in Fig. 10.

Our study finds no evidence for either of these backgrounds, with log Bayes factors as follows: $\ln(\mathcal{B}^{\rm kination+CBC}_{\rm Noise}) = -1.16$ for the kination model and $\ln(\mathcal{B}^{\rm SD+CBC}_{\rm Noise}) = -0.62$ for the model with varying $\alpha_s$. We also compare a CBC-only background with a combined SD + CBC signal. For the kination model, we obtain $\ln(\mathcal{B}^{\rm kination+CBC}_{\rm CBC}) = -0.62$, suggesting a preference for a CBC-only background. For the general SD model, we find $\ln(\mathcal{B}^{\rm SD+CBC}_{\rm CBC}) = -0.09$, that indicates a small preference for a CBC background only.

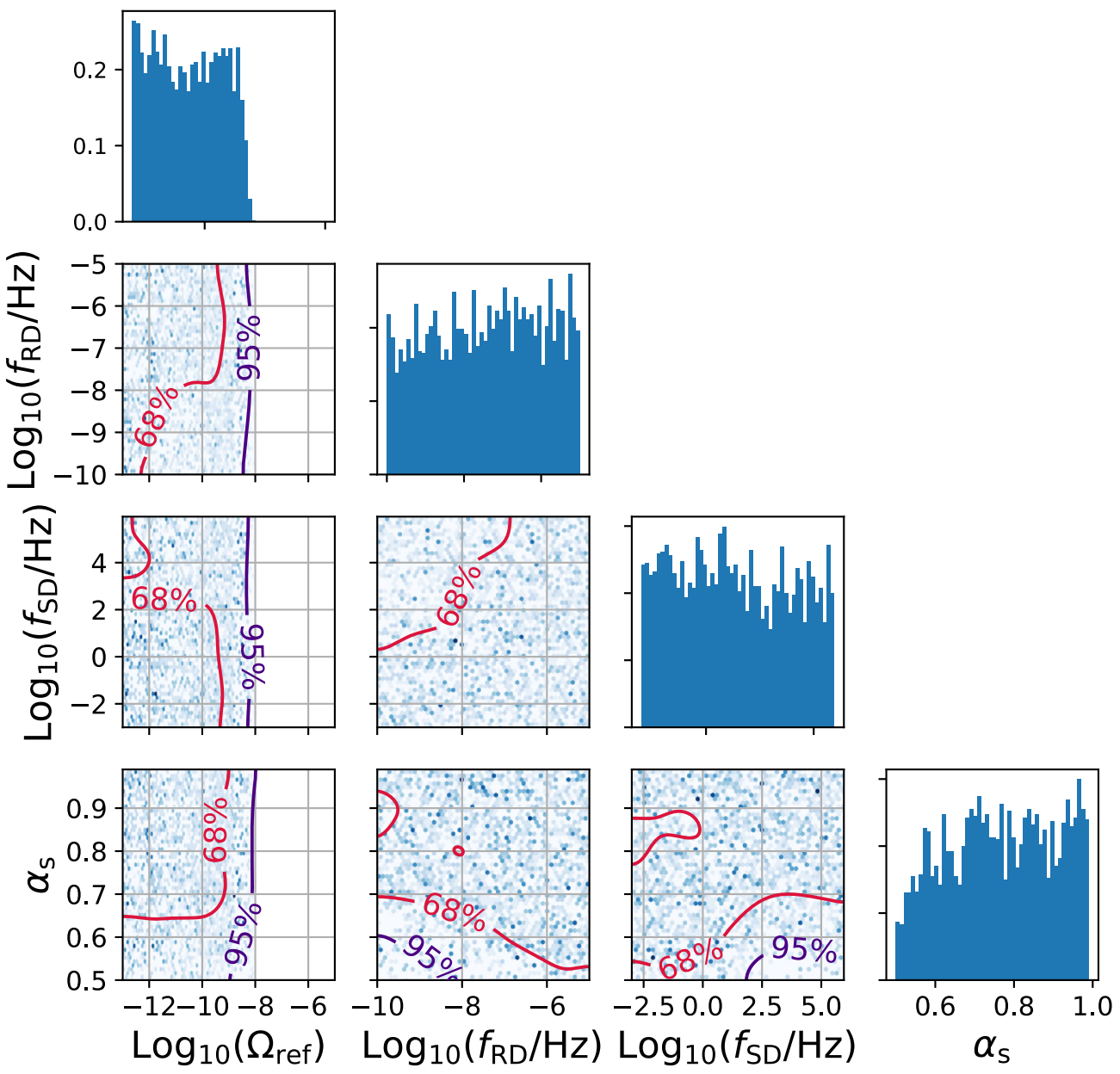

FIG. 10. Posteriors of the Bayesian analysis for an SD + CBC model. Contour regions in purple correspond to 95% CL and those in red to 68% CL.

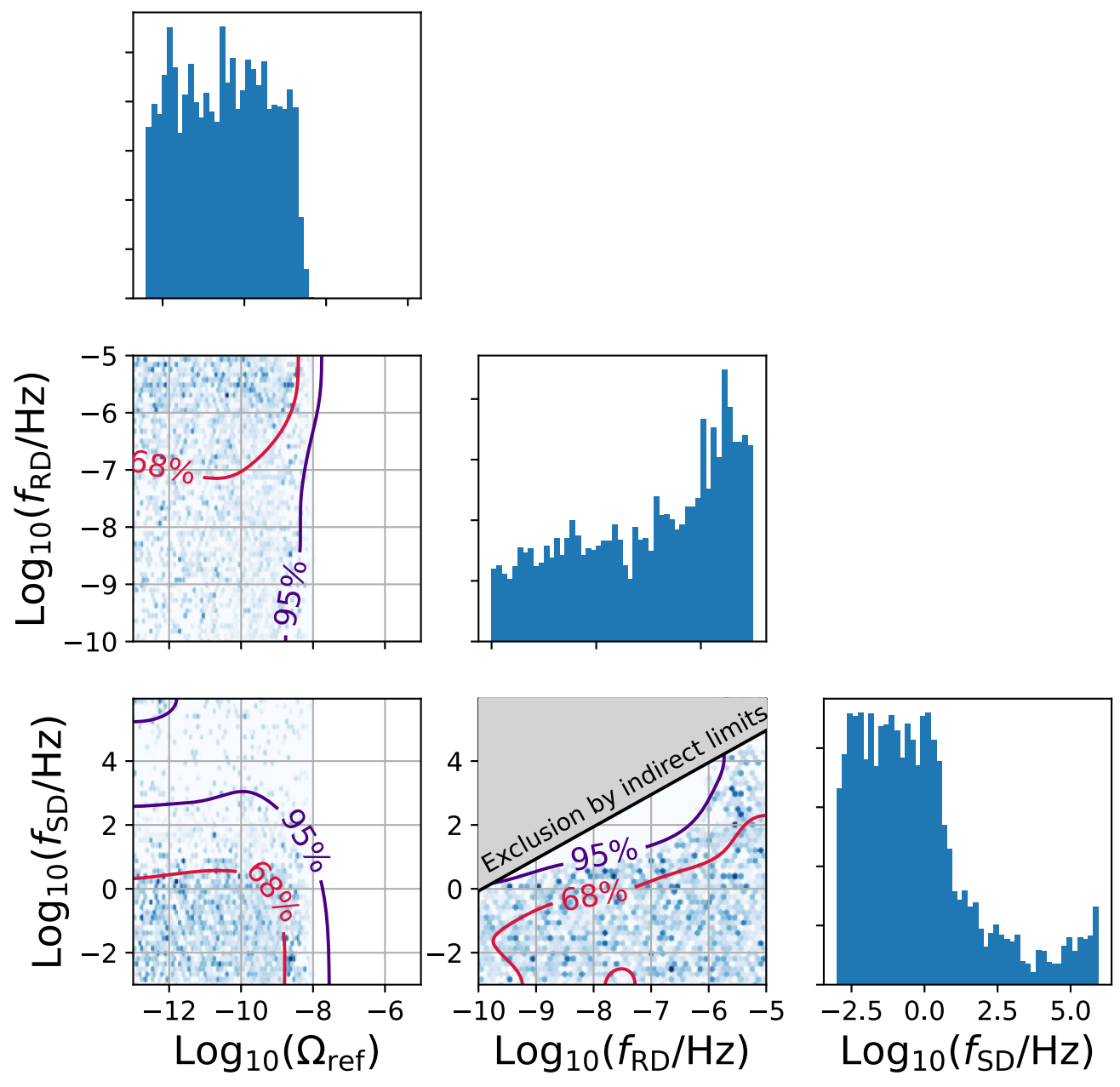

FIG. 9. Posteriors of the Bayesian analysis for the kination + CBC model. Here, $\alpha_s$ is fixed to $\alpha_s = 0.5$. For the contour regions, the same colors as in Fig. 10 are used. The gray region in the bottom panel is excluded by indirect limits from BBN and CMB as in Eq. (27).

In summary, we do not find evidence in the O1–O4a LIGO-Virgo data for either a CBC background or a gravitational-wave background coming from a stiff era. Consequently, we derive 95% CL upper limits on some of the parameters characterizing this unconventional cosmology. These are identified by the white regions in Figs. 9 and 10 (and slightly improve on the previous analysis performed only with O1–O3 data [53]). In Fig. 9, in the case of kination, we show that the data can exclude a portion of parameter space in the $f_{\rm RD}$ vs $f_{\rm SD}$ plane which would otherwise still be allowed by indirect limits. Our analysis, independently on the stiff epoch, also sets an upper limits on the amplitude of the astrophysical background, with the value $\Omega_{\rm ref} \leq 2.9 \times 10^{-9}$.

## VI. AXION INFLATION

### A. Motivation

Gravitational waves offer a novel tool to test inflationary models and constrain their parameters. One inflationary model motivated by high-energy physics is axion inflation, where a pseudoscalar axion, coupled to a gauge field, leads to the early Universe accelerated expansion [54–57]. This model offers rich opportunities for cosmological observations [240,241], including distinctive CMB features [242,243], the formation of primordial black holes [244] arising from the effective multifield dynamics induced by the gauge field background, and a chiral gravitational-wave background [55,245–248], which may be detectable by the LIGO-Virgo-KAGRA detectors.

Although U(1) gauge fields have been studied extensively, non-Abelian gauge fields, such as SU(2), present a compelling alternative. Their key difference is that the SU(2) gauge field can have an isotropic background value, whereas the U(1) cannot. As a result, gravitational-wave production can be computed using linear analysis in the case of SU(2), while it becomes a nonlinear process for U(1), generally requiring a more involved analysis to predict gravitational-wave amplitude [245,249]. In our analysis, we focus on the SU(2) gauge field.

Gravitational waves can be efficiently produced when background gauge fields induce linear couplings between metric and tensor perturbations [250,251]. While the original cosine inflaton potential is excluded by CMB observations [252,253], other more complex models [247,254–262] can evade CMB constraints and generate signals at currently detectable interferometric scales.

### B. Model

We consider chromonatural inflation [263,264], for which the action reads [56,263]

$$S = \int d^4x\sqrt{-\bar{g}}\left[\frac{M_{\rm Pl}^2}{2}R - \frac{1}{2}(\partial\phi)^2 - V(\phi) - \frac{1}{4}F^a_{\mu\nu}F^{a\mu\nu} + \frac{\alpha_f}{4}\phi F^a_{\mu\nu}\tilde{F}^{a\mu\nu}\right], \qquad (28)$$

where $M_{\rm Pl}$ is the reduced Planck mass, $\bar{g} = \det(g_{\mu\nu})$, $R$ denotes the Ricci scalar, and $\phi$ is the inflaton axion field with a scalar potential $V(\phi)$. The SU(2) gauge field $A^a_\mu$ has field strength $F^a_{\mu\nu} = \partial_\mu A^a_\nu - \partial_\nu A^a_\mu - g\epsilon^{abc}A^b_\mu A^c_\nu$, and its dual is $\tilde{F}^{a\mu\nu} = \epsilon^{\mu\nu\rho\sigma}F^a_{\rho\sigma}/(2\sqrt{-\bar{g}})$. The axion-gauge coupling is denoted by $\alpha_f$, and the gauge coupling by $g$.

We consider an isotropic ansatz for the homogeneous component of the gauge field:

$$A^a_0 = 0, \qquad A^a_i = \delta^a_i a(t)Q(t), \qquad (29)$$

where $Q \simeq (-\partial_\phi V/3\alpha_f gH)^{1/3}$ with $H \equiv \dot{a}/a$ the Hubble parameter [265–267].

The coupling between the inflaton and the SU(2) gauge field induces a tachyonic instability in one helicity mode of the gauge field. This leads to exponential amplification of that mode, which, in turn, sources a chiral (parity-violating) gravitational-wave background [56,253]. When the gauge coupling is small or the gauge field is weak, a non-Abelian SU($N$) gauge theory behaves approximately like $N^2 - 1$ independent copies of an Abelian U(1) gauge theory. In the non-Abelian regime, the background gauge field acquires a nonzero vacuum expectation value, which enables a linear coupling between gauge field tensor perturbations and the metric tensor perturbations. This linear coupling allows the enhanced helicity $+2$ mode of the gauge field to efficiently source gravitational waves during inflation. In contrast, in the Abelian case, such couplings arise only at the nonlinear level, making the gravitational-wave production less efficient. Therefore, we focus on the non-Abelian regime.

The non-Abelian regime is conservatively defined by the condition [261]

$$0.008e^{2.8\xi} \gtrsim 1/g. \qquad (30)$$

In this regime, the gravitational-wave background sourced can be analytically approximated as [261]

$$\Omega_{\rm SU(2)}(k) \simeq \frac{\sqrt{2}\Omega^{(0)}_{\rm rad}}{3}\left(\frac{\xi^3 H}{\pi M_{\rm Pl}}\right)^2_{\xi=\xi_{\rm cr}}\left(\frac{He^{(2-\sqrt{2})\pi\xi}}{g\sqrt{\xi}}\right)^2_{\xi=\xi_{\rm ref}}, \qquad (31)$$

where $\Omega^{(0)}_{\rm rad} = 9\times 10^{-5}$ and $\xi = \alpha_f\dot{\phi}/2H$. In Eq. (31), the first term is evaluated at $\xi{\rm cr} = \xi(x=1)$, while the second term is evaluated at $\xi_{\rm ref} = \xi[x = (2+\sqrt{2})\xi_{\rm cr}]$, with $x = -k\tau$ for conformal time $\tau$.

A straightforward way to evade the CMB constraints is to consider the piecewise linear potential originally proposed by Starobinsky [268,269]:

$$V(\phi) = \begin{cases} V_0 + A_+(\phi - \phi_0), & \text{for } \phi > \phi_0, \\ V_0 + A_-(\phi - \phi_0), & \text{for } \phi < \phi_0, \end{cases} \qquad (32)$$

where $V_0$ sets the energy scale of the potential and $A_+$ and $A_-$ determine the slopes on either side of $\phi_0$. Within the slow-roll approximation, the inflaton velocity remains approximately constant, leading to a constant velocity parameter $\xi$. For $\phi < \phi_0$, we denote the velocity by $\xi_0$, which in the slow-roll approximation is $\xi_0 = A_-\alpha_f/2V_0$. This property greatly simplifies analytical calculations and facilitates the computation of the resulting gravitational-wave spectrum.

In Fig. 11, we show the spectrum $\Omega_{\rm SU(2)}(k)$ for different parameter sets. Because of the constant velocity parameter $\xi$, the gravitational-wave amplitude remains constant over the relevant scales. We assume that the transition from the Abelian to the non-Abelian regime occurs near $\phi_0$, positioned between CMB and interferometer scales, which determines where the enhanced gravitational-wave production begins. For further discussion of the allowed parameter space and analyses of alternative models, see Ref. [270].

Although the SU(2) gauge field can enhance gravitational waves during inflation, observational constraints exist, which we summarize below.

#### *1. Cosmic microwave background*

The Hubble expansion rate during inflation at the CMB scale, $H_{\rm CMB}$, sets the amplitude of the vacuum contribution to the gravitational-wave background, $\Omega_{\rm vac}(k) = \Omega_{\rm rad}^{(0)} H^2/(12\pi^2 M_{\rm Pl}^2)$. The observable tensor-to-scalar ratio $r$ is related to $H_{\rm CMB}$ through

$$H_{\rm CMB} = 2.7 \times 10^{14} r^{1/2}\ {\rm GeV}. \tag{33}$$

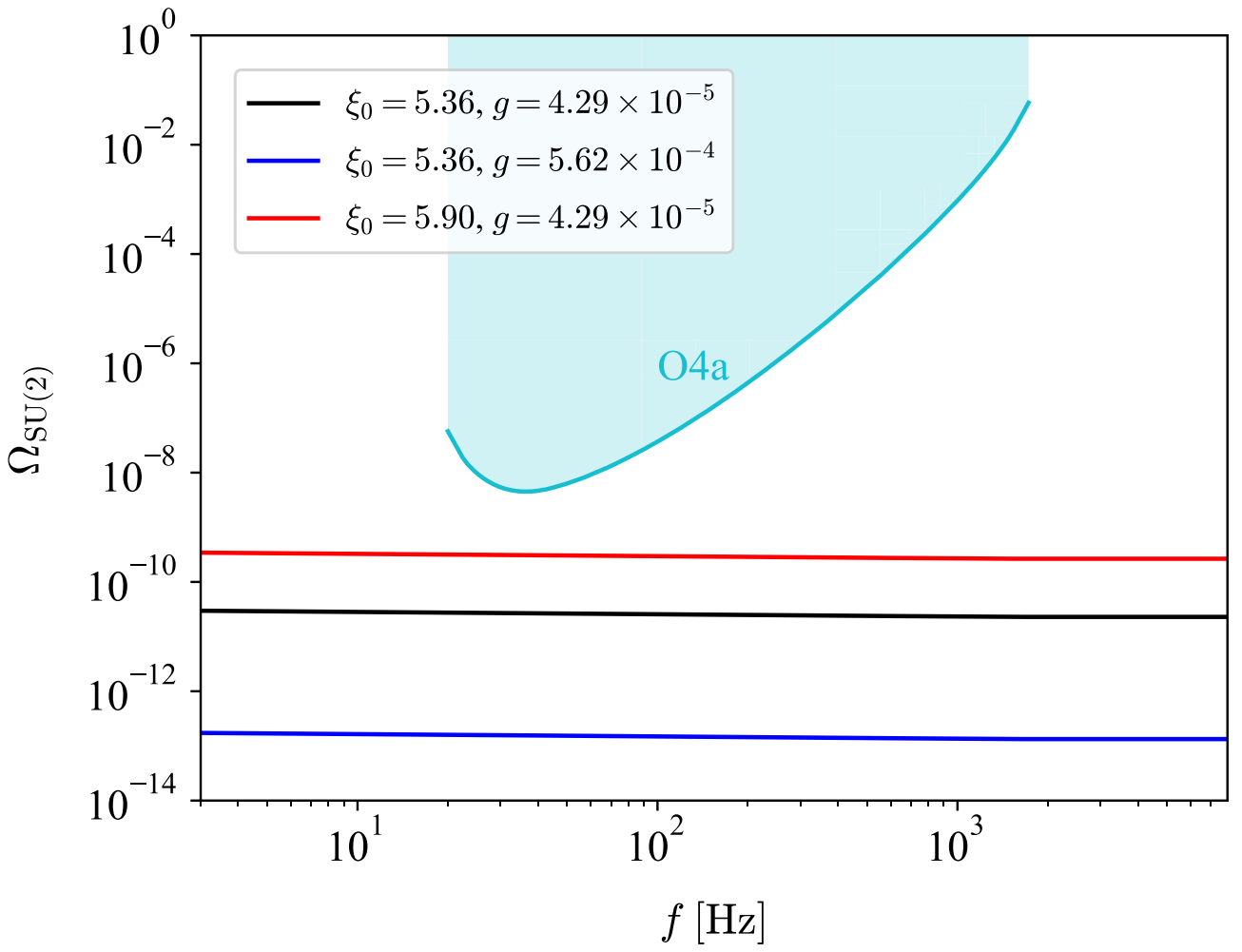

FIG. 11. Examples of Non-Abelian gravitational-wave background spectra for varying $\xi_0$, $g$, plotted with the O4a power-law integrated sensitivity curve [213].

The latest observational constraint, $r < 0.036$ at 95% CL [237], translates into $H_{\rm CMB} < 2.1 \times 10^{-5} M_{\rm Pl}$, thereby ruling out certain classes of inflationary models.

#### 2. *Primordial black hole overproduction*

Gauge-field-induced tensor modes can amplify primordial curvature fluctuations through second-order effects in cosmological perturbation theory. Once these fluctuations reenter the Hubble radius, they may collapse into primordial black holes, whose abundance is tightly constrained by cosmological and astrophysical observations [271,272]. If the curvature perturbations obey $\chi^2$ statistics, primordial black hole formation is more efficient than in the Gaussian case, excluding a substantial region of parameter space [245]. This bound may, however, be alleviated in certain cases. Lattice simulations of axion inflation with a U(1) gauge field suggest that, in the strong backreaction regime, curvature perturbations approach a Gaussian distribution [273], reducing the expected primordial black hole abundance. While a similar analysis has not yet been carried out for SU(2) gauge field models, it is plausible that backreaction effects could likewise relax primordial black hole constraints.

The strength of the backreaction is controlled by the parameter $\kappa$ [270,274]:

$$\kappa \simeq g\left(\frac{24\pi^2}{2.3 e^{3.9 m_Q}} \frac{m_Q^2}{1+m_Q^2}\right)^{-1/2}, \tag{34}$$

where $m_Q \equiv gQ/H$ plays the role of an effective mass. When backreaction is an efficient mechanism, $\kappa \simeq 1$, the curvature perturbations would approach a Gaussian distribution and the primordial black hole constraint may be relaxed. However, if the backreaction becomes too strong ($\kappa > 1$), the analytical expressions for the gravitational-wave spectrum are no longer reliable, and a dedicated numerical analysis would be required. We, thus, do not consider this latter case.

TABLE V. Prior distributions assumed for the parameters of the model and the CBC background.

| Parameter | Prior |
| --- | --- |
| $\Omega_{\rm ref}$ | LogUniform$[10^{-12}, 10^{-7}]$ |
| $N_{\rm CMB}$ | Uniform$[50, 60]$ |
| $f_0$/Hz | LogUniform$[10^{-6}, 10]$ |
| $\phi_{\rm end}/M_{\rm Pl}$ | Uniform$[0, 25]$ |
| $A_+/M_{\rm Pl}^3$ | LogUniform$[10^{-20}, 10^{-6}]$ |
| $A_-/M_{\rm Pl}^3$ | LogUniform$[10^{-20}, 10^{-6}]$ |
| $V_0/M_{\rm Pl}^4$ | LogUniform$[10^{-20}, 10^{-6}]$ |
| $\alpha_f/M_{\rm Pl}^{-1}$ | Uniform$[0, 250]$ |
| $g$ | LogUniform$[10^{-5}, 1]$ |

### C. Constraints using O1–O4a LIGO-Virgo data

We perform a Bayesian parameter estimation search for a combined non-Abelian and CBC background using PYGWB [122]. The model has eight free parameters, and the prior ranges are summarized in Table V.

Figure 12 shows the 95% constraints obtained from the joint SU(2) + CBC analysis. We show both cases where all eight parameters are searched and where $H_{\rm CMB}$ is fixed at $H_{\rm CMB} = 10^{-5} M_{\rm Pl}, 10^{-6} M_{\rm Pl}, 10^{-7} M_{\rm Pl}$. The constraint is shown in the $\log_{10} g$-$\xi_0$ plane, and other parameters are marginalized over. As discussed in the previous section, the viable region for $\xi$ and $g$ is limited to a certain area around the diagonal line in Fig. 12. The gravitational-wave amplitude is proportional to $g^{-2}$ and an exponentially increasing function of $\xi_0$ [see Eq. (31)], and it becomes larger toward the upper-left region in Fig. 12. Consequently, the gravitational-wave observations constrain the parameter space above the green lines.

95% upper bounds on the SU(2)+CBC model and CBC background s parameters obtained under different Hubble constant $H_{\rm CMB}$ prior assumptions are shown in Table VI. Parameter estimation log Bayes evidence factor for the searched combined model and CBC background under different Hubble constant prior assumptions are shown in Table VII.

The constraint is most sensitive to the value of $H_{\rm CMB}$, as it affects the overall amplitude. As we can see from the figure, we obtain tighter constraints on $g$ and $\xi_0$ when $H_{\rm CMB}$ is large, and vice versa. We can also observe that, when $H_{\rm CMB}$ is left free and marginalized over, the constraint is relatively weak. This is because, due to our prior allowing very small $H_{\rm CMB}$ values, small $H_{\rm CMB}$ likely dominate the results when $H_{\rm CMB}$ is marginalized over. In other words, the constraint is dependent on the prior of $H_{\rm CMB}$, and fixing $H_{\rm CMB}$ eliminates this ambiguity.

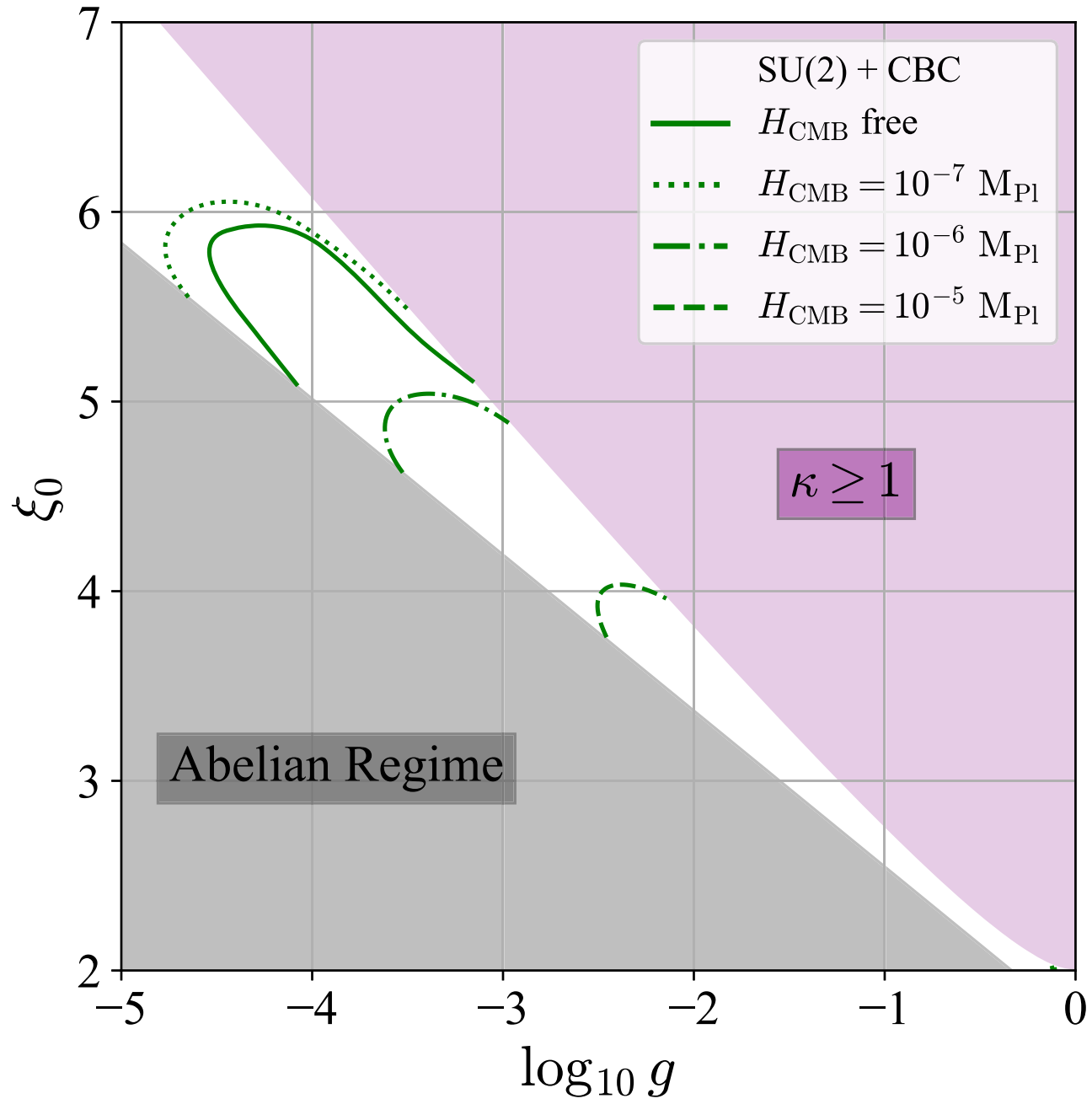

FIG. 12. Marginalized posterior in the $\xi_0 - \log_{10} g$ plane obtained from a search for early Universe axion inflation with an overlaid astrophysical background, assuming fixed values of $H_{\rm CMB} = 10^{-7} M_{\rm Pl}, 10^{-6} M_{\rm Pl}, 10^{-5} M_{\rm Pl}$ (dotted, dot-dashed, and dashed green lines, respectively) and also treating it as a free parameter (solid green line). The gray shaded region denotes the Abelian regime, where efficient gravitational-wave production is not expected. The purple shaded region corresponds to the parameter space where strong backreaction is anticipated, namely, $\kappa \geq 1$, and the gravitational-wave amplitude estimates may no longer be reliable.

This result can be understood physically as follows. In Fig. 12, if we fix $g$ and gradually increase $\xi_0$ from a small value, the energy transfer from the inflaton to the gauge field is initially too weak, keeping the system in the Abelian regime, where gravitational-wave production remains inefficient. However, beyond a certain threshold, the non-Abelian nature of the gauge field becomes significant, leading to enhanced gravitational-wave generation. However, if $\xi_0$ becomes too large, backreaction effects become dominant, violating the assumptions of our analysis, or the resulting gravitational-wave signal would contradict LIGO-Virgo observations, leading to exclusion.

TABLE VII. Parameter estimation log Bayes evidence factor for the searched combined model and CBC background under different Hubble constant $H_{\rm CMB}$ prior assumptions.

| $H_{\rm CMB}$ | $\log \mathcal{B}^{\rm GWB}_{\rm noise}$ |
|---|---|
| Free | $-0.515 \pm 0.028$ |
| $10^{-7} M_{\rm Pl}$ | $-0.541 \pm 0.029$ |
| $10^{-6} M_{\rm Pl}$ | $-0.555 \pm 0.029$ |
| $10^{-5} M_{\rm Pl}$ | $-0.511 \pm 0.028$ |

The result presented here applies specifically to the form of the potential given in Eq. (32). Different inflationary potentials yield different gravitational-wave spectra, as the latter are determined by the evolution of the scalar field (namely, $\xi$ evolves differently). We chose the double linear potential model because a linear potential leads to a constant solution for the velocity parameter $\xi$, and the two-stage inflation allows us to avoid concerns about CMB constraints. This provides a relatively simple picture in which the gravitational-wave spectrum is determined by the velocity parameter at the second stage, $\xi_0$. For discussions on cosine-type potentials and the $R^2$ potential, we refer the reader to Refs. [261,270].

## VII. SECOND-ORDER SCALAR PERTURBATIONS

### A. Motivation

The scalar-induced gravitational-wave background, arising from large-amplitude primordial curvature perturbations, provides an observational test for probing directly the epoch of inflation [58]. This topic has recently gained significant attention due to its connection with primordial black holes [275,276]. In scenarios where primordial curvature fluctuations are amplified during inflation, primordial black holes form through the collapse of extremely dense regions shortly after the corresponding modes enter the Hubble radius [277,278]. Associated with this process, gravitational waves are sourced by the second-order terms of scalar perturbations, in the context of cosmological perturbation theory [59–63]. Thus, an upper bound on the gravitational-wave background can provide constraints on primordial curvature perturbations [279–282].

TABLE VI. 95% upper bounds on the $\mathrm{SU}(2)$ + CBC model and CBC background's parameters obtained under different Hubble constant $H_{\rm CMB}$ prior assumptions.

| Parameters | $H_{\rm CMB}$ free | $H_{\rm CMB} = 10^{-5} M_{\rm Pl}$ | $H_{\rm CMB} = 10^{-6} M_{\rm Pl}$ | $H_{\rm CMB} = 10^{-7} M_{\rm Pl}$ |
|---|---|---|---|---|
| $\Omega_{\rm ref}$ | $2.48 \times 10^{-9}$ | $3.18 \times 10^{-9}$ | $3.37 \times 10^{-9}$ | $2.72 \times 10^{-9}$ |
| $g$ | 0.411 | 0.421 | 0.322 | 0.325 |
| $V_0$ | $4.41 \times 10^{-9}$ | $2.36 \times 10^{-10}$ | $2.37 \times 10^{-12}$ | $2.50 \times 10^{-14}$ |
| $A_+$ | $2.77 \times 10^{-12}$ | $3.73 \times 10^{-11}$ | $3.77 \times 10^{-13}$ | $3.69 \times 10^{-15}$ |
| $A_-$ | $2.01 \times 10^{-10}$ | $1.60 \times 10^{-11}$ | $2.23 \times 10^{-13}$ | $2.36 \times 10^{-15}$ |
| $\xi_0$ | 5.936 | 4.104 | 4.976 | 5.884 |
| $H_{\rm CMB}/M_{\rm Pl}$ | $4.18 \times 10^{-5}$ | ... | ... | ... |

The amplification of the primordial curvature spectrum can be achieved through various mechanisms. Within the framework of single-field inflation, an enhanced power spectrum on small scales can occur, for instance, via the hilltop-type or running mass models [283,284], the former being phenomenologically equivalent to the latter. These models predict a gravitational-wave spectrum, probably detectable by DECIGO [285] and LISA, but certainly not within the LIGO-Virgo-KAGRA band detectability. Primordial black holes might be overproduced if the gravitational waves are induced by the curvature perturbations that follow the Gaussian distribution. In particular, excess of primordial black holes rules out a region of the hilltop-type and running mass parameter space [283,284]. Hence, one may prefer models associated with the non-Gaussianity of the curvature perturbations that suppress the primordial black hole production rate. One such model is the axionlike curvaton model leading to a blue-tilted power spectrum of curvature perturbations [286,287] and inducing a gravitational-wave spectrum within the pulsar timing array detectability [288]. One should, however, keep in mind that the position of the peak in the curvature power spectrum is model dependent, and, hence, an enhancement can result for gravitational-wave signals within the LIGO-Virgo-KAGRA frequency band. For instance, an inflationary model with an ultraslow-roll phase, achieved through a plateau region in the inflationary potential [289–291], provides a flexibility to adjust the scale of enhancement. Another possibility is to consider multifield inflation, which can predict enhanced curvature perturbations through mechanisms such as hybrid inflation with a tachyonic instability [292,293] or turns in field space, corresponding to a bending of the inflationary trajectory [294,295].

### B. Model

Although extensive phenomenological studies have been conducted on the scalar-induced gravitational-wave background, we provide constraints based on the simplest assumptions: a log-normal spectral shape and a Gaussian distribution of the primordial curvature perturbations. The peak in the primordial curvature power spectrum is assumed to take the form [296]

$$\mathcal{P}_\zeta(k) = \frac{A}{\sqrt{2\pi}\Delta}\exp\left[-\frac{\ln^2(k/k_*)}{2\Delta^2}\right]. \tag{35}$$

It is defined by its position $k_*$ with its width controlled by the parameter $\Delta$ and its amplitude characterized by $A$. In the $\Delta \to 0$ limit, Eq. (35) reduces to a Dirac delta function $\mathcal{P}_\zeta(k) = A\delta[\ln(k/k_*)]$. The assumption for the primordial curvature power, Eq. (35), provides constraints in a model-independent manner and serves as a good approximation for many inflationary models. Furthermore, given the relatively narrow frequency band, our constraints are not sensitive to the detailed shape of the spectrum. Inflationary models typically predict an enhanced curvature perturbation spectrum over a wider range of scales, which can be well approximated by a log-normal peak with a large width within the LIGO-Virgo frequency coverage.

In our analysis, we focus on a Gaussian distribution for the primordial curvature perturbations. Non-Gaussianity can significantly modify the spectrum of the scalar-induced gravitational-wave background [297–303]; however, the precise form of non-Gaussianity depends on the inflation model and a general parametrization is lacking.

Hence, assuming a Gaussian distribution for the curvature perturbations, the energy-density spectrum of the scalar-induced gravitational-wave background can be computed using the approximate analytical expression [304,305]

$$\begin{aligned}\Omega_{\rm Scalar}(k)h^2 &\simeq 1.62\times10^{-5}\left(\frac{\Omega_{\rm rad}^{(0)}h^2}{4.18\times10^{-5}}\right)\left(\frac{g_*}{106.75}\right)\left(\frac{g_{*,s}}{106.75}\right)^{-4/3}\\ &\times\frac{1}{12}\int_{-1}^{1}dx\int_{1}^{\infty}dy\,\mathcal{P}_\zeta\left(k\frac{y-x}{2}\right)\mathcal{P}_\zeta\left(k\frac{x+y}{2}\right)F(x,y),\end{aligned} \tag{36}$$

where $\Omega_{\rm rad}^{(0)}$ is the present value of the energy density fraction of radiation and $g_*$ and $g_{*,s}$ are the effective number of degrees of freedom for energy density and entropy density, respectively. The function $F(x,y)$ is given by

$$\begin{aligned}F(x,y) &= \frac{288(x^2+y^2-6)^2(x^2-1)^2(y^2-1)^2}{(x-y)^8(x+y)^8}\\ &\times\left[\left(x^2-y^2+\frac{x^2+y^2-6}{2}\ln\left|\frac{y^2-3}{x^2-3}\right|\right)^2\right.\\ &\left.+\frac{\pi^2}{4}(x^2+y^2-6)^2\theta(y-\sqrt{3})\right],\end{aligned} \tag{37}$$

where $\theta$ denotes the Heaviside step function. The frequency range of our search corresponds to wave numbers between approximately $10^{16}$ and $10^{19}$ $\mathrm{Mpc}^{-1}$. These scales reentered the Hubble horizon when the temperature exceeded $10^8$ GeV, allowing us to set $g_* = g_{*,s} = 106.75$ within the Standard Model framework.

Figure 13 shows the spectrum assuming a log-normal curvature power spectrum for different values of the width parameter, $\Delta = 0, 0.1, 1$, while keeping $A$ and $k_*$ fixed. As $\Delta$ increases, the peak becomes broader and the spectral amplitude decreases. The integrated amplitude $A$ sets the overall normalization, with the spectrum scaling as $\Omega_{\rm Scalar}(f) \propto A^2$, while the peak scale $k_*$ determines the frequency at which the spectrum reaches its maximum.

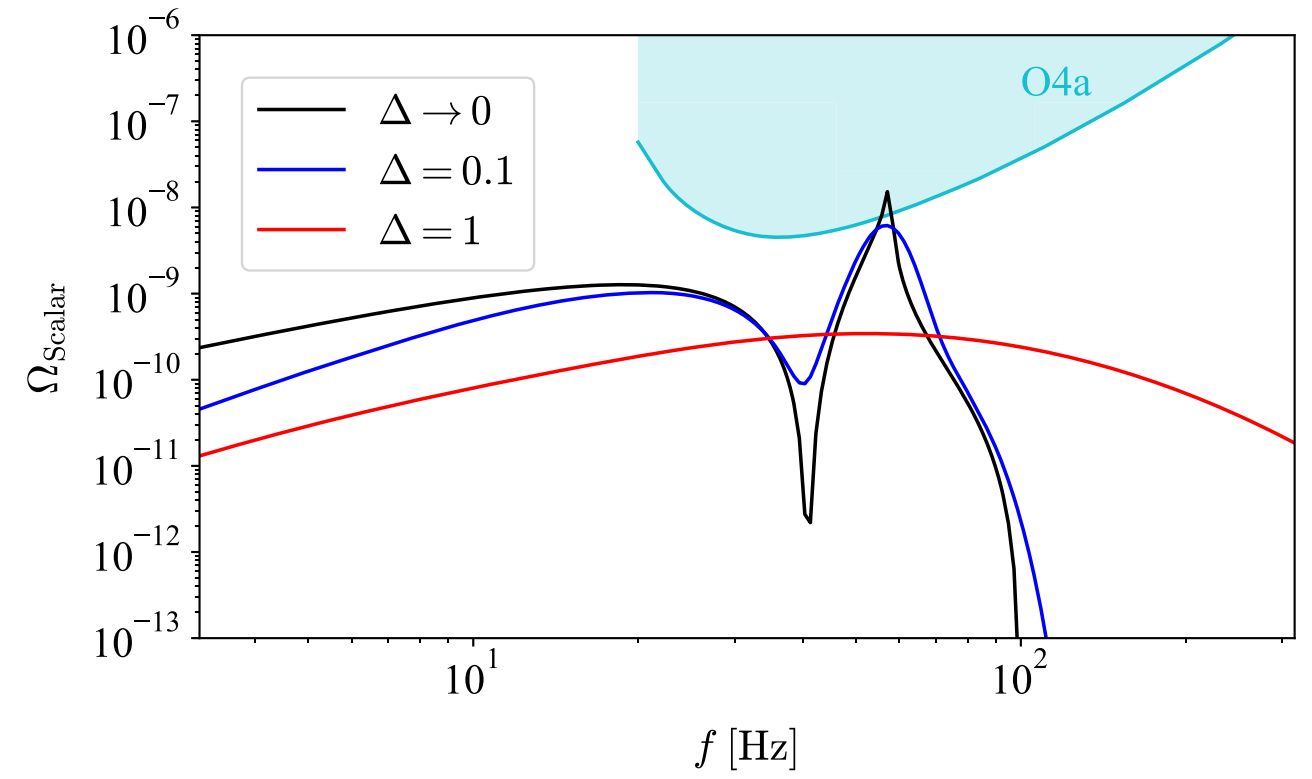

FIG. 13. Spectrum for different values of width $\Delta = 0$, 0.1, 1 plotted with the O4a power-law integrated sensitivity curve [213]. We assume $A = 0.01$ and $f_* = 50$ Hz.

The peak scale is set by the specific mechanism that enhances scalar perturbations during inflation, and the gravitational-wave spectrum peaks at approximately the same wave number as the curvature power spectrum. If primordial black holes form after the corresponding mode reenters the Hubble radius, their mass can be related to the peak frequency as

$$f_* \equiv \frac{ck_*}{2\pi} = 25\left(\frac{k_*}{1.6\times10^{16}\ \mathrm{Mpc}^{-1}}\right)\ \mathrm{Hz}$$
$$\simeq 25\gamma_H^{1/2}\left(\frac{M_{\rm PBH}}{5.3\times10^{-20}M_\odot}\right)^{-1/2}\ \mathrm{Hz}, \qquad (38)$$

where $M_\odot \simeq 2\times10^{30}$ kg is the solar mass and $\gamma_H \coloneqq M_{\rm PBH}/M_H$, typically of the order of unity, accounts for the difference between the primordial black hole mass $M_{\rm PBH}$ and the horizon mass $M_H$.

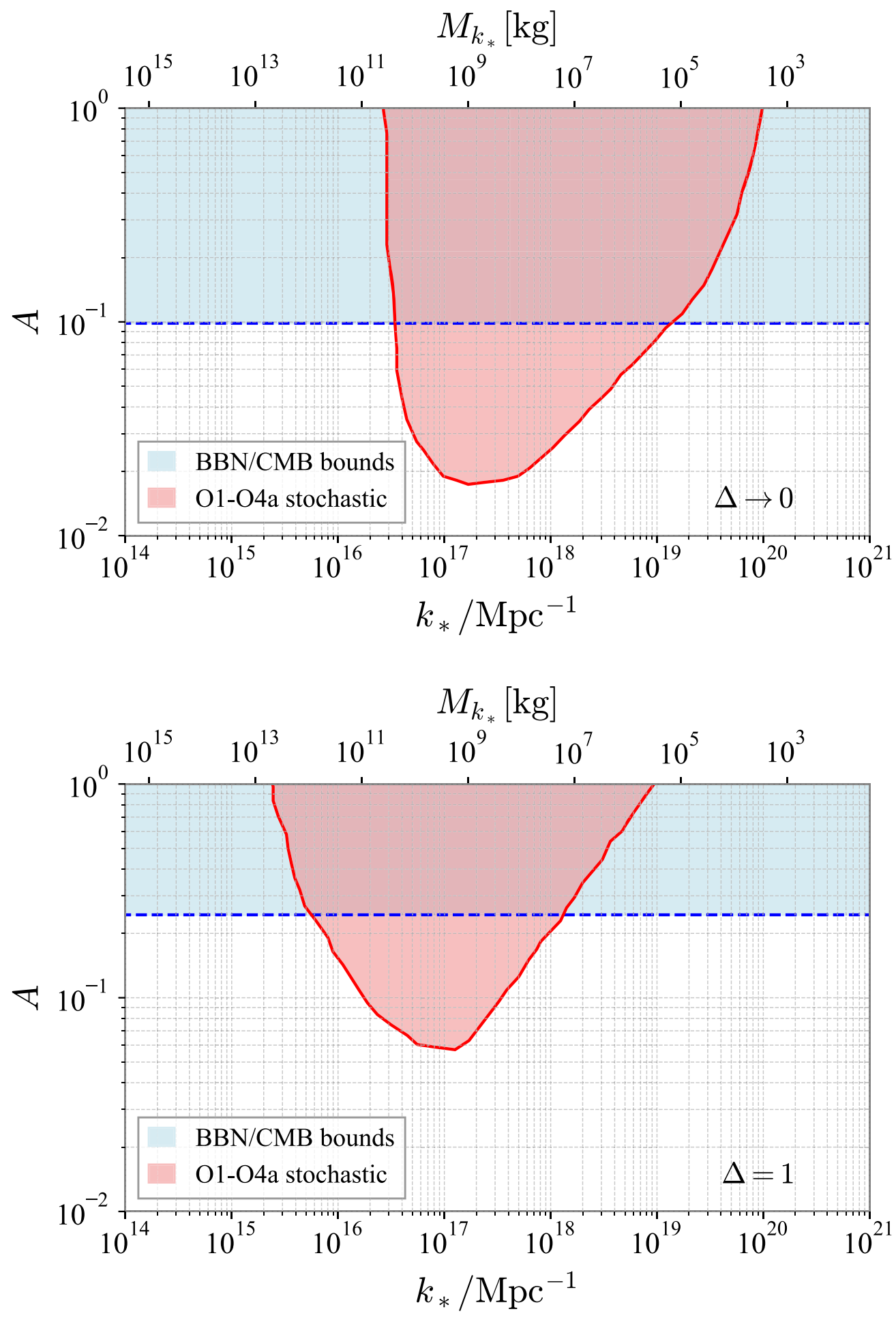

FIG. 14. Constraints on the curvature perturbation amplitude for $\Delta = 0$ (delta function case) and $\Delta = 1$.

TABLE VIII. Prior distributions assumed for the parameters of the scalar-induced gravitational-wave background and the CBC background.

| Parameter | Prior |
|---|---|
| $\Omega_{\rm ref}$ | LogUniform$[10^{-13}, 10^{-5}]$ |
| $A$ | LogUniform$[10^{-6}, 10^{0.5}]$ |
| $f_*$/Hz | LogUniform$[10^{-2}, 10^{6}]$ |
| $\Delta$ | Fixed at 0 and 1 |

### C. Constraints using O1–O4a LIGO-Virgo data

The result of the Bayesian parameter estimation, obtained using PYGWB [122], are shown in Fig. 14. The prior distributions are given in Table VIII. The bounds set on $A$ from the O4a data (shaded red region) are compared with BBN and CMB constraints for $\Delta \to 0$ and for $\Delta = 1$. The bottom and top horizontal axis represent the peak scale of the curvature perturbation $k_*$ and the primordial black hole masses related with the scale calculated using Eq. (38), respectively. The shaded blue region represents indirect bounds from BBN and CMB on the abundance of the stochastic gravitational-wave background. We calculate the bound by using the recent joint CMB + BBN analysis, which indicates that $\int d\ln f h^2\Omega_{\rm GW}(f) < 1.3\times10^{-6}$ at $2\sigma$ for $f > 2\times10^{-11}$ Hz [239]. Since marginalized constraints tend to be prior dependent, here we run the Bayesian search by fixing the value of $\Delta$ and taking $k_*$ and $A$ as free parameters. The upper bounds on $A$ for different combinations of $\Delta$ and $k_*$ are summarized in Table IX.

TABLE IX. 95% CL upper bounds on the power $A$ of the curvature spectrum for fixed values of the peak position $k_*$ and width $\Delta$.

| | $k_* = 10^{15}\ \mathrm{Mpc}^{-1}$ | $k_* = 10^{17}\ \mathrm{Mpc}^{-1}$ | $k_* = 10^{19}\ \mathrm{Mpc}^{-1}$ |
|---|---|---|---|
| $\Delta \to 0$ | 1.44 | 0.01 | 0.15 |
| $\Delta = 1$ | 0.96 | 0.05 | 2.12 |

## VIII. PRIMORDIAL BLACK HOLES

### A. Motivation

Several mechanisms have been proposed for the formation of primordial black holes. One of the most extensively studied scenarios involves the amplification of small-scale perturbations during inflation. Other proposed mechanisms [64] include formation during phase transitions, an early matter-dominated era, scalar field instabilities, and the collapse of topological defects, among others. Their masses can span a wide range, from asteroid-like scales to supermassive sizes, depending on the formation scenario. Primordial black holes are compelling candidates for dark matter [64,306,307] and may contribute to the observed population of binary black holes [308–316].

Just like astrophysical black holes, primordial black holes can form binaries and emit gravitational waves. The resulting gravitational-wave background, produced by a collection of unresolved events, provides a powerful probe of source populations in the early Universe [317–326]. Moreover, it offers a unique opportunity to test the existence of primordial black hole binaries, since the abundance of astrophysical black holes is expected to be low at high redshifts during the cosmic dark ages, before star formation. In contrast, primordial black holes could form binaries during this epoch, generating gravitational waves from the early Universe.

Here, two main formation channels are considered: one operating in the early Universe and another in the late Universe, as discussed in the following subsection. A key factor determining the merger rate is the number density of primordial black holes, typically quantified by $f_{\rm PBH}$, the fraction of dark matter composed of primordial black holes today. While current observations rule out $f_{\rm PBH} = 1$, primordial black holes could still make up a significant fraction of dark matter.

### B. Model

The mass function of primordial black holes is commonly parametrized using monochromatic or log-normal distributions. This provides a good approximation for primordial black holes produced by a peak in the primordial power spectrum [327]. We define the log-normal mass function $p(m)$ as

$$p(m) = \frac{1}{\rho_{\rm PBH}} \frac{d\rho_{\rm PBH}}{d\ln m}, \tag{39}$$

where $\rho_{\rm PBH}$ stands for the primordial black hole energy density. For a log-normal distribution, the mass function takes the form

$$p(m) = \frac{1}{\sqrt{2\pi}\sigma} \exp\left[-\frac{(\ln m - \ln \mu)^2}{2\sigma^2}\right], \tag{40}$$

where $\mu$ is the median mass and $\sigma$ controls the width of the distribution in logarithmic space. The mass function is normalized such that $\int p(m) d\ln m = 1$.

We calculate the gravitational-wave background generated by an ensemble of binary events by summing the energy spectra of individual binaries, considering the redshift of gravitational waves since emission and using the merger rate distribution [11]:

$$\Omega_{\rm PBH}(f) = \frac{f}{\rho_{c,0}} \int_0^{z_{\rm max}} dz \int d\ln m_1\, d\ln m_2 \frac{p(m_1)p(m_2)}{(1+z)H(z)} \times \frac{d^2 R_{\rm EB/LB}}{d\ln m_1 d\ln m_2} \frac{dE_{\rm GW}}{df_{\rm r}}. \tag{41}$$

In Eq. (41), the integration is over the masses $m_1$ and $m_2$ of the binaries, and $f_{\rm r} = (1+z)f$ is the gravitational-wave frequency in the source frame. The quantity $d^2 R_{\rm EB/LB}/d\ln m_1/d\ln m_2$ represents the differential merger rate per unit time, comoving volume, and mass interval. Different binary formation mechanisms are denoted by EB (early binary) and LB (late binary). Early binaries form during the radiation-dominated era shortly after primordial black hole formation, while late binaries form later via dynamical capture in primordial black hole clusters during the matter-dominated era. We model the single-source energy spectrum $dE_{\rm GW}/df_r$ by the phenomenological fitting function of Ref. [328], which captures the inspiral, merger, and ringdown phases. Although binaries at high redshift emit gravitational waves early, the nearest binaries typically dominate the background power unless the merger rate rises sharply with redshift. Since the merger rate in our model does not increase steeply, we take $z_{\rm max} = 100$, which is sufficient to include all relevant contributions to the gravitational-wave background.

Several uncertainties may influence the shape of the gravitational-wave background, such as the binary formation mechanism and the potential disruption of binaries by a third body. The formation scenario can also affect properties like eccentricity, spin, and precession, which, in turn, modify the waveform. In what follows, we consider only nonspinning binaries.

The gravitational-wave background is composed of primordial black hole merger events occurring across a range of redshifts. If nearby binaries are the dominant ones, the spectrum exhibits a peak at the characteristic merger frequency, determined mainly by the primordial black hole masses. For equal-mass binaries, the energy spectrum has a peak at

$$f \simeq 8.3 \times 10^3 \left(\frac{M_{\rm PBH}}{M_\odot}\right)^{-1} \text{Hz}. \tag{42}$$

Hence, LIGO-Virgo-KAGRA detectors are sensitive to binaries with masses ranging from subsolar scales up to $\mathcal{O}(10^2)M_\odot$. The possibility of probing such a background with the LIGO-Virgo-KAGRA detectors has been investigated in the literature [312,319–321,326,329–331].

There are two major binary formation channels. In the early binary formation scenario, a binary originates from a pair of closely spaced primordial black holes, where the tidal influence of a third nearby object imparts the angular momentum necessary for binary formation. The merger rate for early binaries at cosmic time $t$ is given by [309,332–336]

$$\frac{d^2R_{\rm EB}}{d\ln m_1 d\ln m_2} = \frac{1.6\times 10^6}{\rm Gpc^3\ yr} f_{\rm PBH}^{53/37}\left[\frac{t}{t_0}\right]^{-34/37} \times \left(\frac{m_1+m_2}{M_\odot}\right)^{-32/37}\left[\frac{m_1 m_2}{(m_1+m_2)^2}\right]^{-34/37} \times S(m_1, m_2, f_{\rm PBH}), \quad (43)$$

where $t_0$ is the age of the Universe. A rate suppression factor $S$ has been introduced to take into account two main mechanisms that can suppress binary formation, one resulting from local matter inhomogeneities and nearby primordial black holes $S_1(m_1, m_2, f_{\rm PBH})$ and the other from clustering due to their initial Poissonian fluctuations $S_2(f_{\rm PBH})$ [311,312,336]. We employ suppression factors modeled with analytical methods [312,336].

In the late binary formation scenario, binaries can form within dense environments if clusters of primordial black holes develop during the matter-dominated era [308]. Analytical expressions for the merger rate can be derived by considering the two-body capture process within a cluster, under the assumption that the merger timescale of the resulting binary is much shorter than the age of the Universe [337,338]:

$$\frac{d^2R_{\rm LB}}{d\ln m_1 d\ln m_2} = \frac{R_{\rm clust}}{\rm Gpc^3\ yr} f_{\rm PBH}^2 \frac{(m_1+m_2)^{10/7}}{(m_1 m_2)^{5/7}}. \quad (44)$$

The parameter $R_{\rm clust}$ captures the enhancement of the primordial black hole merger rate due to local clustering [318,339], which depends on their velocity dispersion and density contrast. We consider three representative values: $R_{\rm clust} = [1, 4\times 10^2, 10^3]$, corresponding to (i) modest clustering consistent with ΛCDM [308], (ii) the level needed to match observed binary merger rates, and (iii) an optimistic scenario with highly efficient cluster formation [339].

With O4a sensitivity, the merger rate of late binaries is typically subdominant compared to early binaries, except for $m_{\rm PBH} \gtrsim 100 M_\odot$. This behavior is also illustrated in Fig. 15, which shows example spectra for both formation channels.

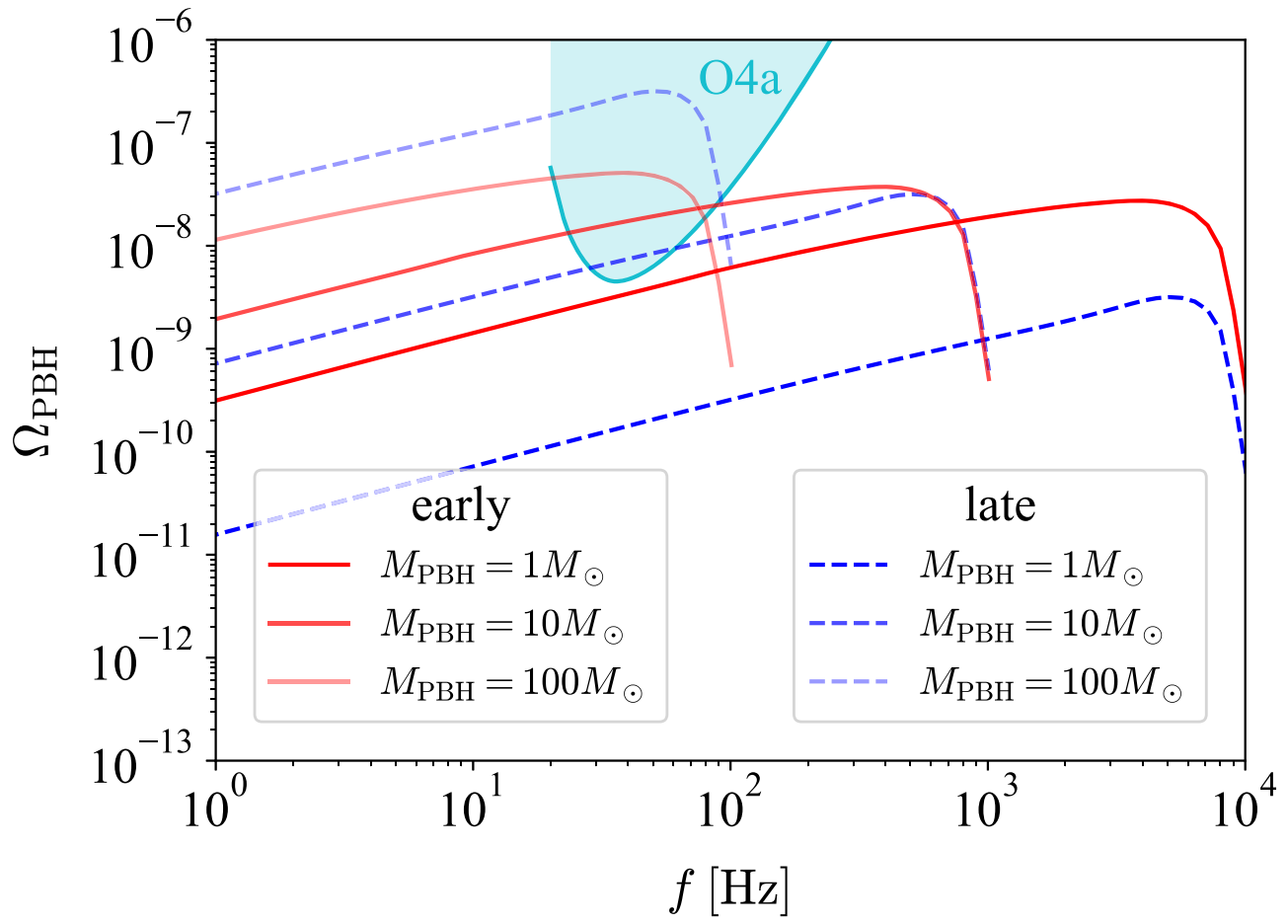

FIG. 15. Gravitational-wave background spectrum for monochromatic mass function, for both early (red solid line) and late (blue dashed line) binary formation channels. Different curves show different primordial black hole masses. We also plot the O4a power-law integrated sensitivity curve [213]. Here, we assume $f_{\rm PBH} = 1$, fixed suppression factor $S = 0.002$, and $R_{\rm clust} = 400$.

### C. Constraints using O1–O4a LIGO-Virgo data

The prior ranges for the parameters are summarized in Table X. Note that we adopt a relatively narrow prior for the width of the mass function $\sigma$, since the merger rate is known to be reliable only when the mass function is sharply peaked, particularly in the early binary formation scenario. The Bayesian analysis implies that no substantial gravitational-wave background sourced by primordial black holes or CBCs has been detected. We establish upper limits on the CBC energy density parameter $\Omega_{\rm ref} \sim 3\times 10^{-9}$ at the 95% CL level, which are consistent with the constraints from the isotropic background search [19].

Even in the absence of a detection, we can still constrain $f_{\rm PBH}$ as a function of $\mu$. We show our constraint in Fig. 16, along with those from other analyses. Our constraint is derived by marginalizing over the mass function width $\sigma$ and $\Omega_{\rm ref}$. For $M_{\rm PBH} \gtrsim 2\times 10^2 M_\odot$, the spectral amplitude of the gravitational-wave background is predominantly attributed to the late binary formation channel. However, the O4a sensitivity loses its constraining power sharply in this mass range due to the limited frequency range. These findings underscore the ongoing and future importance of gravitational-wave background searches as a tool for probing phenomena within the mass range of $[1, 3\times 10^2]M_\odot$. The resulting bounds are complementary to existing ones from individual binary events, microlensing surveys, and cosmic microwave background observations. The posterior distributions of each parameter are shown in Fig. 17, and the 95% upper bounds on $f_{\rm PBH}$ for various combinations of $\mu$ and $R_{\rm clust}$ are presented in Table XI.

TABLE X. Prior distributions for the model parameters in Bayesian analysis.

| Parameter | Prior |
|---|---|
| $\Omega_{\rm ref}$ | LogUniform$[10^{-12}, 10^{-7}]$ |
| $\mu/M_\odot$ | LogUniform$[10^{-1}, 10^{3}]$ |
| $\sigma$ | LogUniform$[10^{-2}, 1]$ |
| $f_{\rm PBH}$ | LogUniform$[10^{-5}, 1]$ |

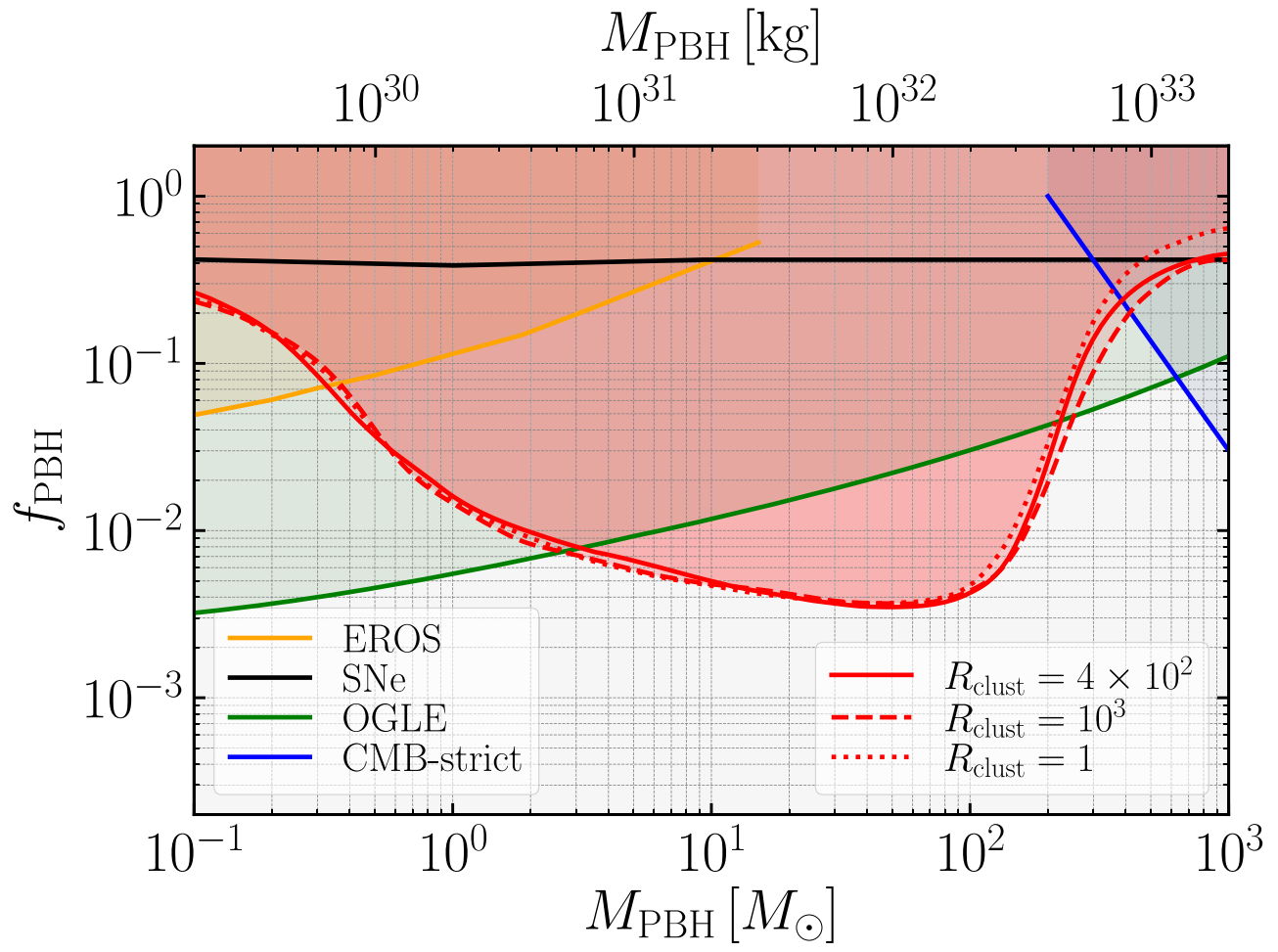

FIG. 16. The 95% CL constraints on $f_{\rm PBH}$ as a function of $M_{\rm PBH}$ $(=\mu)$ from the Bayesian analysis, shown for $R_{\rm clust} = 1, 4\times10^2, 10^3$ (dotted, solid, and dashed red curves, respectively). We also show other observational constraints such as supernova lensing constraints (SNe, black) [340], massive compact halo object constraints (EROS, yellow) [341], the recent optical gravitational lensing experiment constraints (OGLE, green) [342], and cosmic microwave background constraints (CMB, blue) [343].

We note that the parameters exhibit a degeneracy with the CBC contribution $\Omega_{\rm ref}$, because both primordial and astrophysical sources produce a gravitational-wave spectrum with the same frequency dependence, $\propto f^{2/3}$, during the inspiral phase. This similarity is particularly relevant when the average black hole mass is $\lesssim 10 M_\odot$. However, our results indicate the highest sensitivity to primordial black hole masses around approximately $100 M_\odot$, where the dominant contribution comes from the merger phase. In this regime, the spectrum has a characteristic frequency dependence, and its shape is sensitive to the assumed merger rate and mass distribution, allowing us to break the degeneracy between parameters.

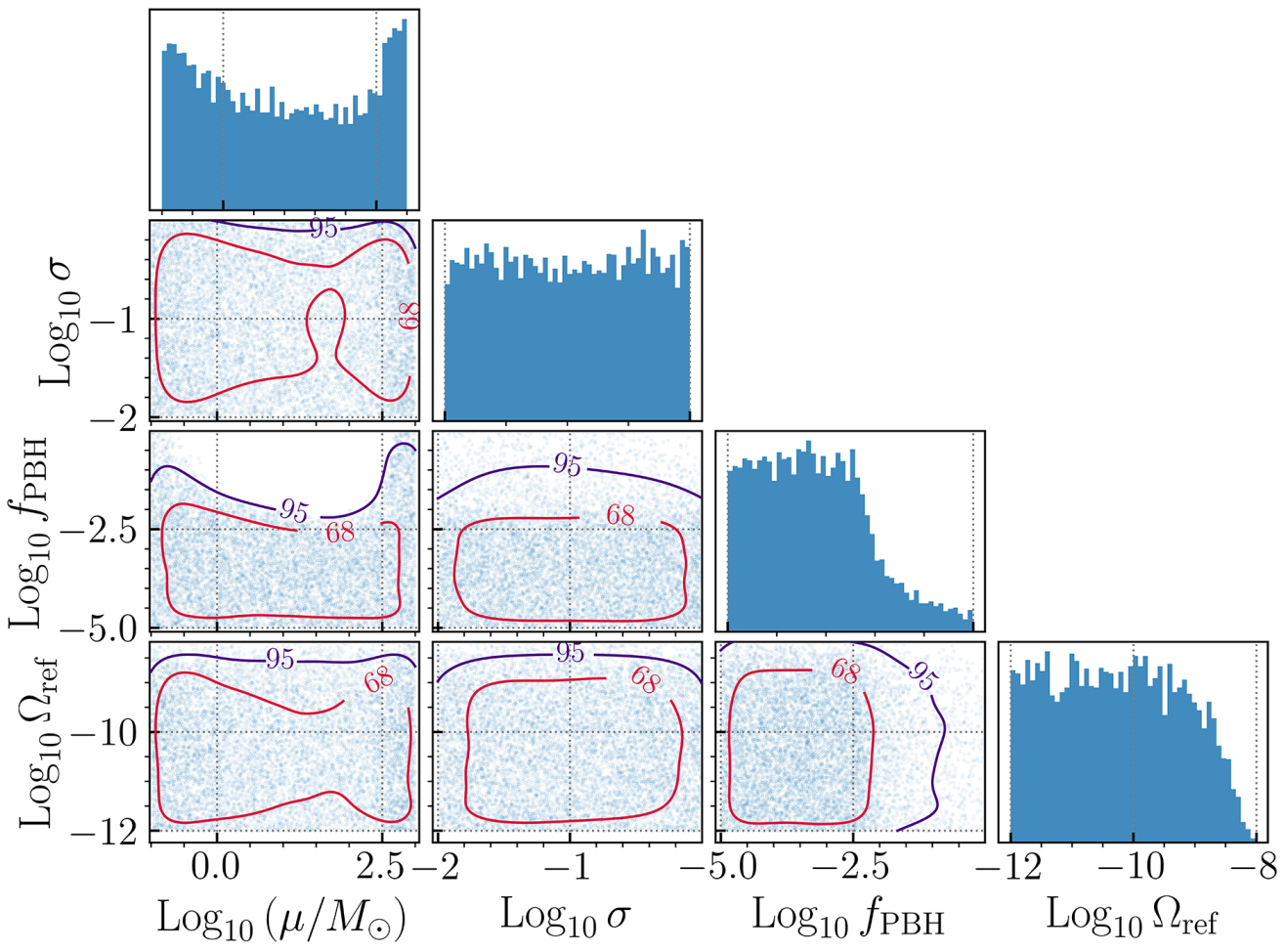

FIG. 17. Corner plots of the posterior distributions for the gravitational-wave background from primordial black hole binaries, assuming $R_{\rm clust} = 4\times10^2$. The results for other values of $R_{\rm clust}$ are very similar.

TABLE XI. 95% CL on $f_{\rm PBH}$ for various masses.

| $R_{\rm clust}$ | $\mu = 1[M_\odot]$ | $\mu = 30[M_\odot]$ | $\mu = 10^3[M_\odot]$ |
|---|---|---|---|
| 1 | $1.5\times10^{-2}$ | $3.7\times10^{-3}$ | $6.4\times10^{-1}$ |
| $4\times10^2$ | $1.6\times10^{-2}$ | $3.6\times10^{-3}$ | $4.5\times10^{-1}$ |
| $10^3$ | $1.4\times10^{-2}$ | $3.8\times10^{-3}$ | $4.1\times10^{-1}$ |

Note that if primordial black hole binaries are produced from curvature perturbations during inflation, they are accompanied by gravitational waves. Such gravitational waves from primordial black holes can be detectable by LIGO-Virgo detectors, provided their mass is $M_{\rm PBH} \leq 10^{16}$ g [344]. Various studies have addressed the possibility that some of the detected CBCs events originate from primordial black hole mergers (see, e.g., Refs. [345–349]). More recently, pulsar timing array data have been examined in the framework of large curvature perturbations generating subsolar primordial black holes and a gravitational-wave background background at nanohertz frequencies [350,351].

## IX. PARITY VIOLATION

In the previous sections, we searched for signatures in the gravitational-wave background spectrum originating from beyond the standard particle physics and early Universe models. In the following, we investigate parity violation, either conducting a model-independent search or focusing within the context of one of the exotic particle physics models that we have previously considered.

### A. Motivation

Several string-theory models and scalar-tensor models of gravity can result in circularly polarized gravitational waves, most notably models inspired by Chern-Simons gravity [352–355] and by inflationary scenarios coupled to Abelian gauge fields [356–361]. Chirality is also expected in various models of early Universe phase transitions [362–368] and axion inflation sourced by non-Abelian gauge fields, commonly referred to as chromonatural inflation [56,56,251,261,263,369].

We describe a generic parity-violation search based on a power-law energy density gravitational-wave spectrum [65] and then detail more theoretically motivated polarized gravitational-wave background models of early Universe turbulence and chromonatural inflation.

## B. Model

In searching for parity-violating models, we adopt the formalism [65] that uses modified cross-correlation estimator

$$\langle \hat{C}_{d_1 d_2} \rangle = \int_{-\infty}^{\infty} df \int_{-\infty}^{\infty} df' \delta_T(f-f') \langle s^*_{d_1}(f) s_{d_2}(f') \rangle \tilde{Q}(f') = \frac{3H_0^2 T}{10\pi^2} \int_0^{\infty} df \frac{\Omega'_{\rm GW}(f) \gamma_I^{d_1 d_2}(f) \tilde{Q}(f)}{f^3}, \quad (45)$$

where

$$\Omega'_{\rm GW} = \Omega_{\rm GW} \left[ 1 + \Pi(f) \frac{\gamma_V^{d_1 d_2}(f)}{\gamma_I^{d_1 d_2}(f)} \right],$$

with

$$\gamma_I^{d_1 d_2}(f) = \frac{5}{8\pi} \int d\hat{\Omega} (F^+_{d_1} F^{+*}_{d_2} + F^\times_{d_1} F^{\times *}_{d_2}) e^{2\pi i f \hat{\Omega}\cdot\Delta\vec{x}},$$
$$\gamma_V^{d_1 d_2}(f) = -\frac{5}{8\pi} \int d\hat{\Omega} (F^+_{d_1} F^{\times *}_{d_2} - F^\times_{d_1} F^{+*}_{d_2}) e^{2\pi i f \hat{\Omega}\cdot\Delta\vec{x}}. \quad (46)$$

We denote $T$ the measurement time, $\delta_T(f) = \sin(\pi f T)/(\pi f)$, $s_d(f)$ the strain time series of the two gravitational-wave detectors (denoted by $d_1$ and $d_2$). $\tilde{Q}(f)$ is a filter, and $F^A_n$ stands for the contraction of the tensor modes of polarization $A = +, \times$ to the $n$th detector's geometry. We denote by $\gamma_I^{d_1 d_2}$ the usual (unpolarized isotropic gravitational-wave background) overlap reduction function of detectors $d_1$ and $d_2$ [76] and $\gamma_V^{d_1 d_2}$ as the overlap function associated with the parity-violation term [370]. The polarization degree

$$\Pi(f) = V(f)/I(f) = \frac{P_R(f) - P_L(f)}{P_R(f) + P_L(f)} \quad (47)$$

ranges from $-1$ (fully left polarization) to 1 (fully right polarization), with $\Pi = 0$ corresponding to an unpolarized isotropic gravitational-wave background. We indicate by $I$ and $V$ the Stokes parameters, and $P_{R/L}$ denote the right- and left-hand gravitational-wave power spectra. Note that allowing $\Pi = 0$ in Eqs. (45) and (47) returns the formalism to one utilized in standard, isotropic searches [19].

### 1. Model independent

We conduct a generic search for a parity-violating gravitational-wave background exhibiting power-law behavior, Eq. (2), with $f_{\rm ref} = 25$ Hz. We use a log-uniform amplitude prior from $10^{-13}$ and $10^{-5}$, while the model spectral index prior is a Gaussian distribution centered at 0 with a standard deviation of 3.5. We search for this model using the O1–O4a gravitational-wave data and place upper limits on its parameters.

TABLE XII. Prior distribution for the model-independent parity-violation searches.

| Parameter | Prior |
|---|---|
| $\Omega^{\rm PV}_{\rm ref}$ | LogUniform$[10^{-13}, 10^{-5}]$ |
| $\alpha$ | Gaussian[0, 3.5] |
| $\Pi$ | Uniform$[-1, 1]$ |
| $\beta$ | Uniform$[-2, 0]$ |

We investigate both a simplified model with constant $\Pi$ and a model in which the polarization varies with frequency. In the constant polarization case, we search uniformly for $\Pi$ between $-1$ and 1. Additionally, two searches that fix $\Pi = -1$ and 1 are conducted to compare constraints under maximal chiral assumptions. For the frequency-dependent model, we use $\Pi(f) = \pm(f/1\ {\rm Hz})^\beta$ with a uniform prior $\beta$ between $-2$ and 0. This is motivated from theoretical models where $\Pi$ decays with increasing frequency [371,372]. We refer the reader to Table XII. In our analysis, we consider only frequencies larger than 1 Hz—at lower frequencies, terrestrial detectors are limited by seismic noise—and, hence, the form of $\Pi(f)$ guarantees that the physically allowed bound $|\Pi| \leq 1$ is valid.

### 2. Early Universe turbulence

A parity-violating turbulent source during a phase transition will produce circularly polarized gravitational waves. Depending on the helicity strength, there are two types of turbulent gravitational-wave spectra [373,374]. When energy dissipation at small scales dominates, it leads to a helical Kolmogorov spectrum, and we consider this type of polarization.

Parity violation at the electroweak scale can be realized in extensions of the Standard Model of particle physics, manifesting as helical (or chiral) turbulent motion [375,376]. Circularly polarized gravitational waves are generated by parity-violating turbulent sources [377]. Their spectrum has a broken-power-law spectrum with a peak at the characteristic frequency of the source. We search gravitational-wave data for models [378–380]

$$\Omega_{\rm Turbulence}(f) = \begin{cases} \Omega_{\rm peak}(f/f_{\rm peak}), & f \leq f_{\rm peak}, \\ \Omega_{\rm peak}(f/f_{\rm peak})^{-8/3}, & f > f_{\rm peak}. \end{cases} \quad (48)$$

The peak frequency $f_{\rm peak}$ is related to the temperature $T_*$ at which the first-order phase transition takes place. At an energy scale of $T_* \sim 10^8$ GeV, the predicted chiral turbulence spectrum would exhibit a peak within the current LIGO-Virgo-KAGRA observational band. We, therefore, search for $f_{\rm peak}$ over a broad range (10–2000) Hz. The prior distributions are given in Table XIII.

TABLE XIII. Prior distribution for the turbulence parity-violation searches.

| Parameter | Prior |
|---|---|
| $\Omega_{\mathrm{peak}}$ | LogUniform$[10^{-13}, 10^{-5}]$] |
| $f_{\mathrm{peak}}/\mathrm{Hz}$ | Uniform[5, 2000] |
| $\beta$ | Uniform[−2, 0] |

Previous numerical studies calculated the net circular polarization of gravitational waves under various initial turbulent conditions, determining the degree of polarization as a function of the wave number $k$. They identified models where $\Pi$ depends on the frequency [371,377]. We model the polarization as the power-law functional form described previously.

### 3. Non-Abelian axion inflation

The Chern-Simons interaction term sources exponential production of gravitational waves through the induced linear couplings between metric and gauge field tensor perturbations and is given in Eq. (31).

The total gravitational-wave spectrum includes the vacuum contribution

$$\Omega_{\mathrm{Vacuum}}(k) = \frac{\Omega_{R,0}}{12\pi^2}\frac{H^2}{M_{\mathrm{Pl}}^2}. \tag{49}$$

We consider a piecewise linear model potential, given previously in Eq. (32). The inflaton velocity, in the slow-roll approximation, is approximately constant:

$$\xi = \begin{cases} \xi_{\mathrm{CMB}} = A_+\alpha_f/2V_0, & \text{for } \phi > \phi_0, \\ \xi_0 = A_-\alpha_f/2V_0, & \text{for } \phi < \phi_0, \end{cases} \tag{50}$$

where CMB constraints set un upper bound of $\xi_{\mathrm{CMB}} < 2.5$ at 95% CL [381]. More studies for this model can be found in Ref. [269].

TABLE XIV. Prior distribution for the SU(2) axion inflation parity-violation searches. Note: the same as the previously listed prior table in Sec. VI, just with added $\Pi$ prior.

| Parameter | Prior |
|---|---|
| $\Omega_{\mathrm{ref}}$ | LogUniform$[10^{-13}, 10^{-5}]$ |
| $N_{\mathrm{CMB}}$ [e-folds] | Uniform[50, 60] |
| $f_0/\mathrm{Hz}$ | LogUniform$[10^{-6}, 10]$ |
| $\phi_{\mathrm{end}}/M_{\mathrm{Pl}}$ | Uniform[0, 25] |
| $A_+/M_{\mathrm{Pl}}^3$ | LogUniform$[10^{-20}, 10^{-6}]$ |
| $A_-/M_{\mathrm{Pl}}^3$ | LogUniform$[10^{-20}, 10^{-6}]$ |
| $V_0/M_{\mathrm{Pl}}^4$ | LogUniform$[10^{-20}, 10^{-6}]$ |
| $\alpha_f/M_{\mathrm{Pl}}^{-1}$ | Uniform[0, 250] |
| $g$ | LogUniform$[10^{-5}, 1]$] |
| $\Pi$ | Uniform[0, 1] |

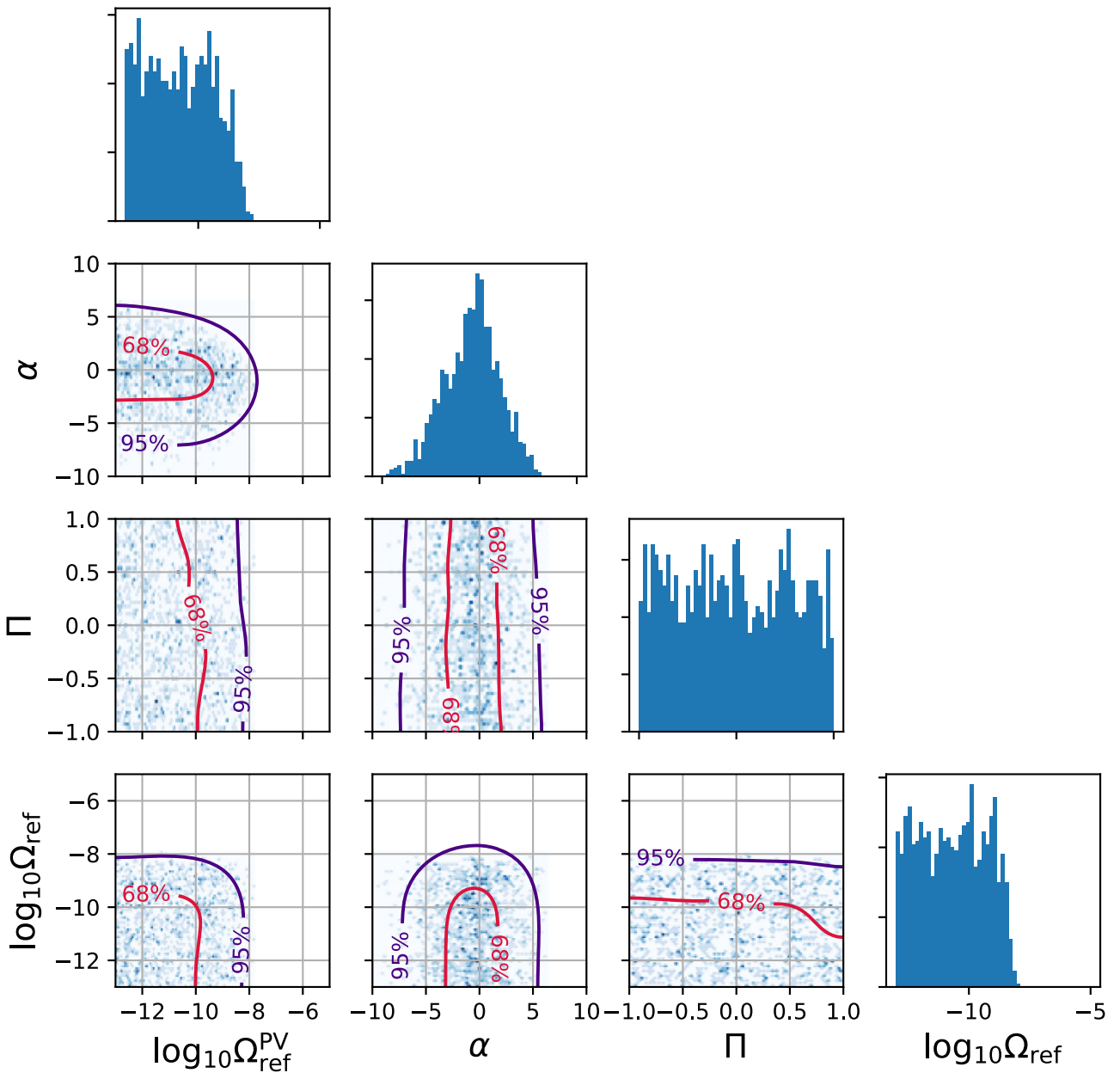

FIG. 18. Posterior distributions for a power-law gravitational-wave background model with $\Pi(f) = \mathrm{const}$, assuming an overlaying CBC background.

The Chern-Simons term not only sources significant production of gravitational waves, the spin-2 fluctuation of the gauge field leads to an asymmetry between its left- and right-handed polarization states. Thus, the gauge field can produce a chiral gravitational-wave signal within the ground detectors' frequency band.

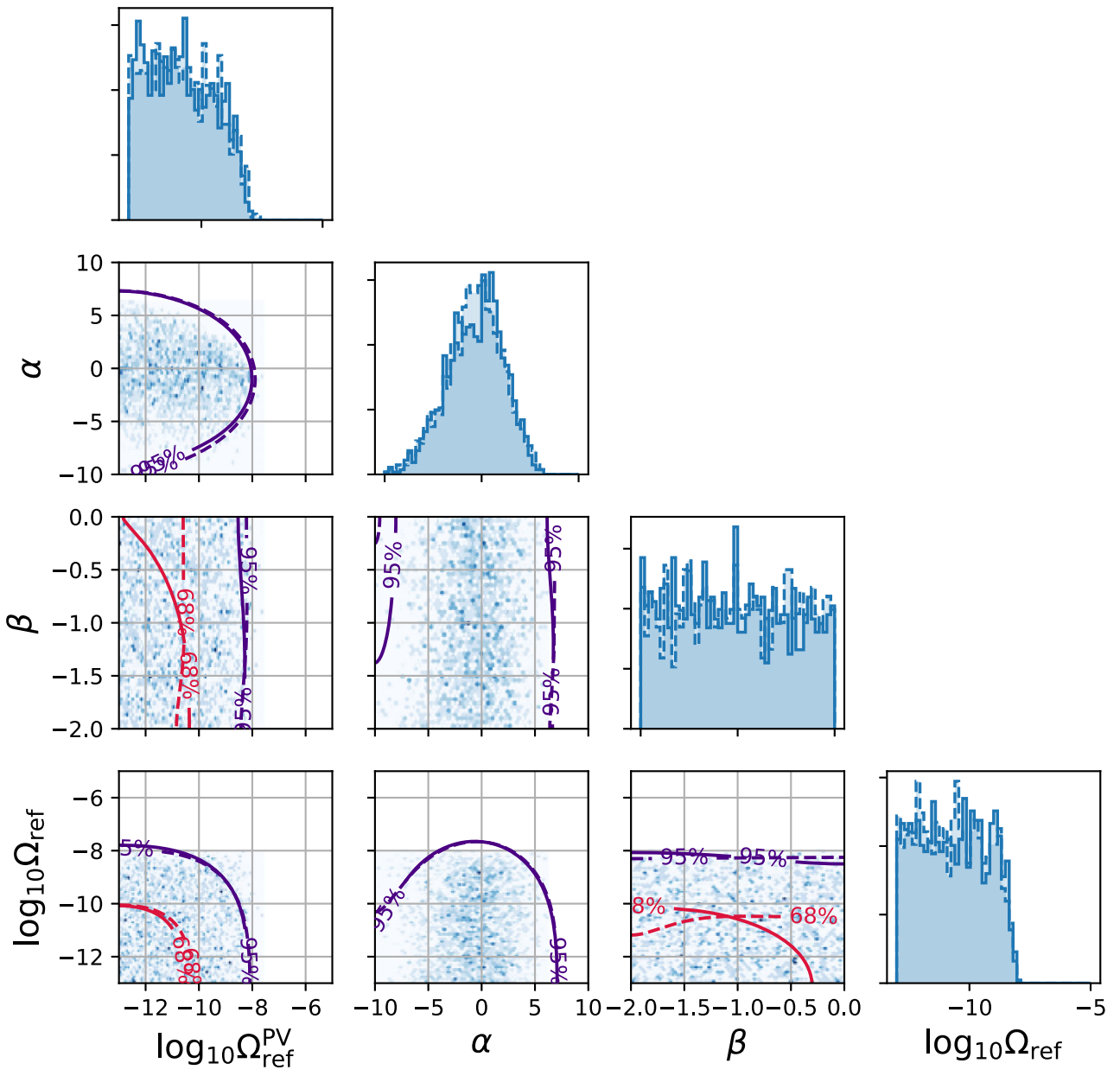

FIG. 19. Posterior distributions for a power-law gravitational-wave background model with $\Pi(f) = \pm(f/\mathrm{Hz})^{\beta}$ (positive power law in red, negative in purple) assuming an overlaying CBC background.

TABLE XV. General parity-violating model search results with an assumed overlaying CBC background.

| PV model | 95% upper bound of $\Omega_{\rm ref}^{\rm PV}$ | 95% upper bound of $\Omega_{\rm ref}$ | $\ln \mathcal{B}_{\rm noise}^{\rm PV\,Model+CBC}$ |
|---|---|---|---|
| $\Pi = {\rm const}$ | $2.54 \times 10^{-9}$ | $2.61 \times 10^{-9}$ | $-1.175 \pm 0.047$ |
| $\Pi = +(f/{\rm Hz})^{\beta}$ | $2.41 \times 10^{-9}$ | $2.95 \times 10^{-9}$ | $-1.222 \pm 0.043$ |
| $\Pi = -(f/{\rm Hz})^{\beta}$ | $2.76 \times 10^{-9}$ | $2.46 \times 10^{-9}$ | $-1.203 \pm 0.044$ |

In Ref. [250], it was shown that the enhanced helicity is model dependent and relies on the inflaton vacuum expectation value sign; right-handed tensor modes corresponding to positive vacuum expectation value, left-handed modes for negative vacuum expectation value. For the toy model studied, the polarization can be well approximated as

$$\Pi \simeq \frac{[\bar{\rho}_{\rm YM}/\bar{\rho}]\mathcal{G}_+^2(m_Q)}{[\bar{\rho}_{\rm YM}/\bar{\rho}]\mathcal{G}_+^2(m_Q)+1} = {\rm const} > 0, \qquad (51)$$

where $\bar{\rho}_{\rm YM}/\bar{\rho} \lesssim \epsilon^2$ for slow-roll parameter $\epsilon$, effective mass $m_Q$ is approximated as $\xi \simeq m_Q + m_Q^{-1}$, and the explicit functional form of $\mathcal{G}_s(m_Q)$ is detailed in Ref. [250]. It is easy to show that $\Pi(f) \simeq 1$ for $\xi_0 \gtrsim 4$ and $\bar{\rho}_{\rm YM}/\bar{\rho} \gtrsim 5 \times 10^{-5}$ over the ground detectors' frequency band.

We perform a search for parity-violating axion inflation, a model investigated in Sec. VI by introducing an additional parameter for the polarization amplitude $\Pi$ with a uniform prior range of [0, 1]. For the other parameters, we use the same prior range as in Table V. Note that we impose a non-negative prior on the polarization amplitude, as we expect $\Pi \geq 0$ for the studied toy model. We refer the reader to Table XIV.

## C. Constraints using O1–O4a LIGO-Virgo data

We present the results for the models where a search was conducted. We find no evidence for parity violation; constraints on such models are set.

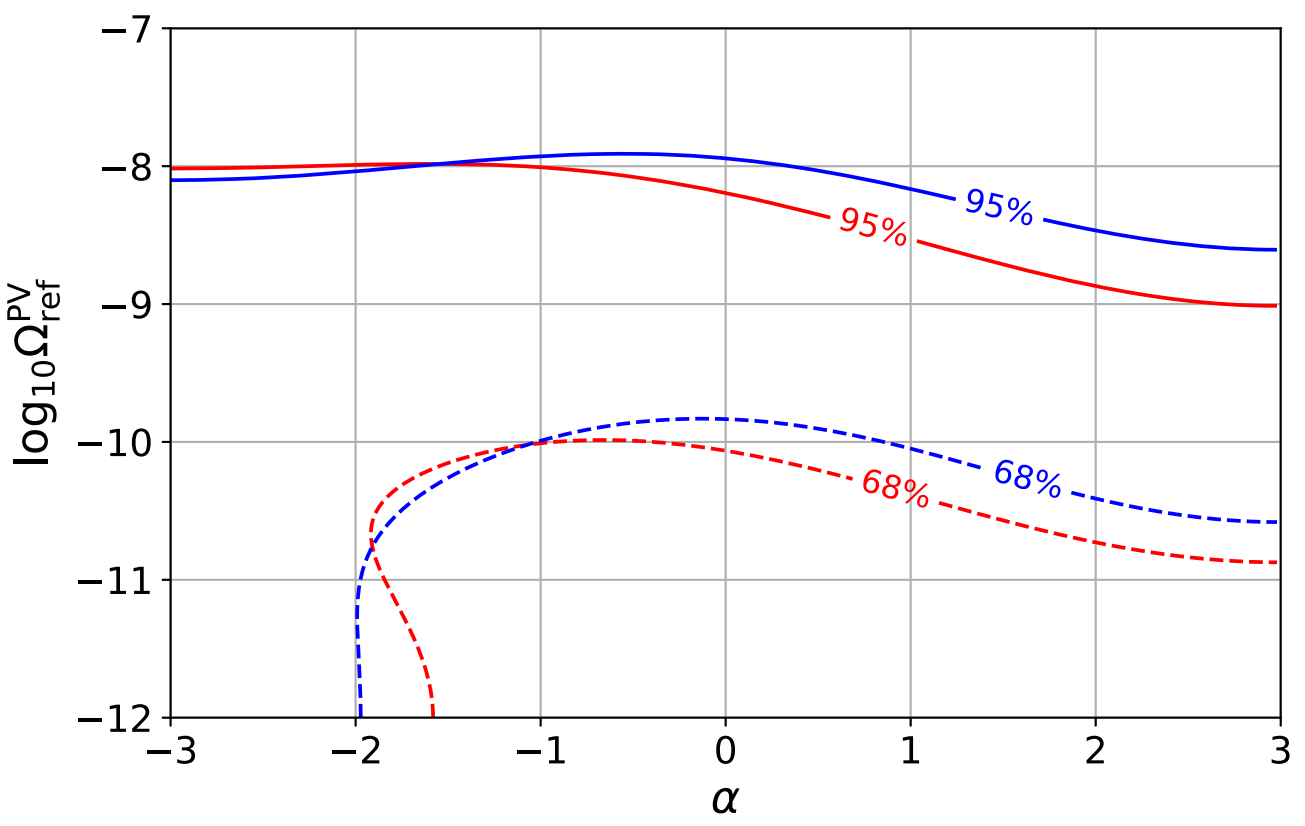

FIG. 20. $\Omega_{\rm ref}^{\rm PV} - \alpha$ confidence curve at 95% (solid line) and 68% (dashed line) level for assumed $\Pi = 1$ (red) and $\Pi = -1$ (blue) polarization.

### 1. Model independent

We plot the results and the 65% and 95% confidence contours for the searched general models with an assumed CBC background in Figs. 18 and 19. Although no constraints can be placed on the parity-violating (PV) associated parameter, we set upper bounds on the background's strength parameter $\Omega_{\rm ref}^{\rm PV}$. We list the 95% upper bound on $\Omega_{\rm ref}^{\rm PV}$ and $\Omega_{\rm ref}$ plus the logarithmic Bayes factor for each general search in Table XV.

Figure 20 displays the 68% and 95% confidence contours of the resulting $\Omega_{\rm ref}^{\rm PV} - \alpha$ posteriors from an assumed $\Pi = \pm 1$ polarization. One can see that more stringent constraints can be made for an assumed entirely right-handed polarization ($2.82 \times 10^{-9} < \Omega_{\rm ref}^{\rm PV,95\%}$, red) than for a left-handed polarization ($2.94 \times 10^{-9} < \Omega_{\rm ref}^{\rm PV,95\%}$, blue); this was also found using the O3 data [382]. This preference can be explained by the ratio between the standard and parity-violating associated overlap reduction functions $\varsigma^{d_1 d_2} \equiv \gamma_V^{d_1 d_2}/\gamma_I^{d_1 d_2}$. While $\varsigma^{\rm HV}$ and $\varsigma^{\rm LV}$ are roughly periodic in the considered frequency range, $\varsigma^{\rm HL}$ is preferentially positive. Preferentially positive $\varsigma^{\rm HL}$ combined with $\Pi > 0$ results in enhanced modified $\Omega_{\rm GW}$ [Eq. (47)], hence leading to stricter constraints on right-hand polarized signals.

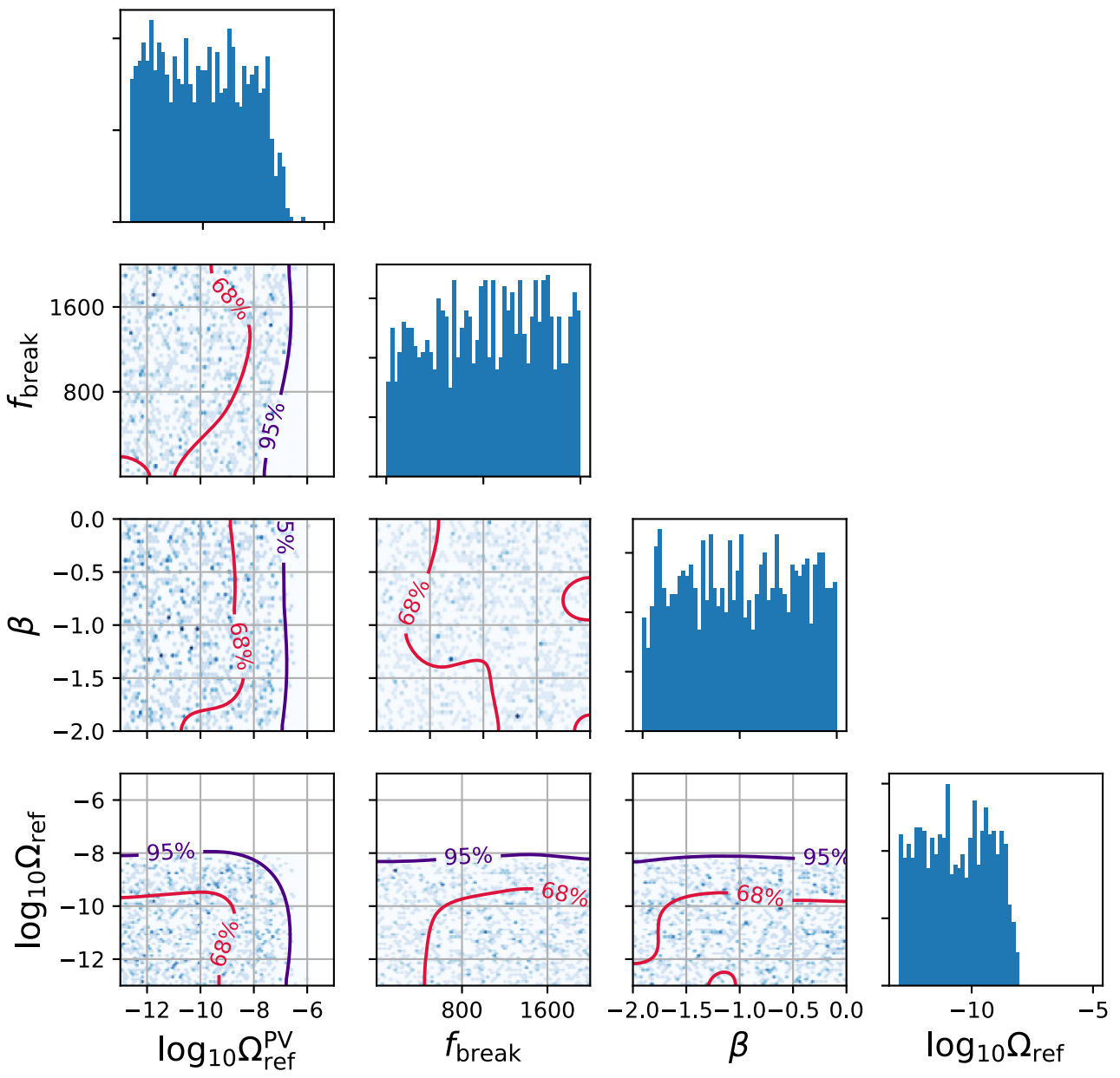

FIG. 21. Posterior distributions for early Universe turbulence with an overlaying CBC background model with $\Pi(f) = (f/{\rm Hz})^{\beta}$.

TABLE XVI. Parameter estimation 95% confidence upper bound for the searched chiral and nonchiral models.

| Parameter | $\Pi = 0$ | $\Pi \neq 0$ |
|---|---|---|
| $g$ | 0.411 | 0.385 |
| $V_0/M_{\rm Pl}^4$ | $4.41 \times 10^{-9}$ | $3.47 \times 10^{-9}$ |
| $A_+/M_{\rm Pl}^3$ | $2.77 \times 10^{-12}$ | $1.62 \times 10^{-12}$ |
| $A_-/M_{\rm Pl}^3$ | $2.01 \times 10^{-10}$ | $2.55 \times 10^{-10}$ |
| $\xi_0$ | 5.936 | 5.843 |
| $H_{\rm CMB}/M_{\rm Pl}$ | $4.18 \times 10^{-5}$ | $3.94 \times 10^{-5}$ |
| $\Omega_{\rm ref}$ | $2.48 \times 10^{-9}$ | $2.80 \times 10^{-9}$ |

A general power-law search assuming no parity violation ($\Pi = 0$) yields a logarithmic Bayes factor of $\ln \mathcal{B}^{\Pi=0+\rm CBC}_{\rm noise} = -1.194 \pm 0.042$. In combination with Table XV, we find no statistical preference between polarized ($\Pi \neq 0$) and nonpolarized ($\Pi = 0$) power-law models.

#### 2. Early Universe turbulence

We plot the resulting constraints from a parity-violating turbulence with overlaying CBC model search in Fig. 21. We calculate a log Bayes factor of $\log \mathcal{B}^{\rm Turb+CBC}_{\rm noise} = -0.830 \pm 0.035$, indicating no evidence for a chiral turbulent background. No constraints on parity-violating parameter $\beta$ could be made. We find the 95% upper bound of the gravitational-wave background strength to be $\Omega_{\rm peak} < 5.39 \times 10^{-8}$—larger compared to power-law background model constraints due to the allowed broken-power-law spectra being able to peak at frequencies with poor sensitivity.

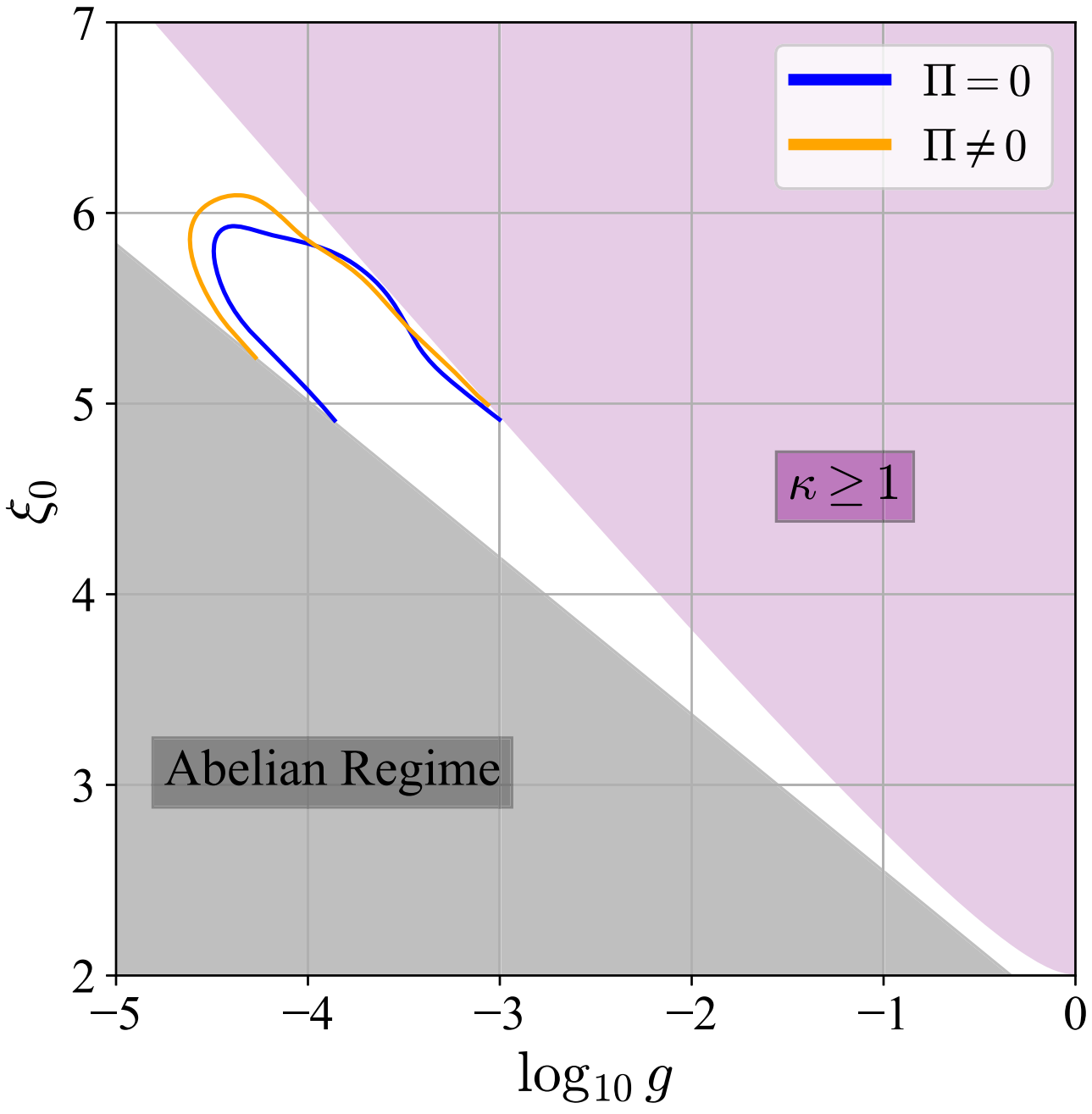

FIG. 22. Parameter estimation 95% confidence limit contours for the SU(2) gauge field model. The blue and orange curves show the 95% confidence limit contours for chiral ($\Pi \neq 0$) and nonchiral ($\Pi = 0$) searches, respectively.

#### 3. Non-Abelian axion inflation

We find a log Bayes factor of $\log \mathcal{B}^{\rm PV\,SU(2)+CBC}_{\rm noise} = -0.545 \pm 0.029$ and, thus, no evidence of such a model nor a preference for a polarized model over a nonchiral model. We list the 95% confidence limits on both searched models in Table XVI, and there do not appear to be large discrepancies in the parameter estimation between the searched models. In Fig. 22, we show the 2D posterior $\xi_0 - H_{\rm CMB}$ results. Similarly, there are small differences between the searched models, highlighting the lack of statistical preference between chiral and nonchiral models. It is important to highlight that these results do not exclude other parity-violating models of axion inflation based on other scalar potential models.

## X. CONCLUSIONS

The LIGO-Virgo-KAGRA Collaboration uses the O4a data from LIGO Hanford and LIGO Livingston to search for a gravitational-wave background signal, in addition to the data from LIGO-Virgo from O1, O2, and O3. In this publication, we report the results of dedicated searches of various particle physics models and cosmological scenarios which could contribute to the gravitational-wave background. These are first-order phase transitions, cosmic strings, domain walls, stiff equation of state, axion inflation, second-order scalar perturbations, primordial black holes, and parity violation. They can all lead to a gravitational-wave background potentially detectable by the LIGO-Virgo-KAGRA network.

First-order phase transitions could have occurred within the first one-trillionth of a second after the Big Bang, and their gravitational-wave imprints would be an important key to determining the correct theory beyond the Standard Model. They could generate gravitational waves from processes such as bubble collisions, sound waves propagating in the early Universe plasma, and magnetohydrodynamic turbulence. We place new constraints on the strength, temperature, and duration of these transitions. We also present a search for a generic broken-power-law gravitational-wave spectrum, and we obtain constraints on the peak amplitude of $\mathcal{O}(10^{-7})$.

Cosmic strings are one-dimensional topological defects that can be generated after phase transitions followed by spontaneously symmetry breaking. Cosmic string loops oscillate because of their tension and shrink as a result of the emission of gravitational waves. We constrain the string tension, a parameter related to the temperature of the symmetry breaking. In our study, we consider cosmic strings following the Nambu-Goto dynamics. For three out of the four loop distribution models we have studied (dubbed model $B$, model $C1$, and model $C2$), we exclude cosmic strings with a tension $G\mu$ greater than $\mathcal{O}(10^{-15})$.

Domain walls are two-dimensional topological defects. Our analysis constrains the domain wall tension, and the temperature at which they collapse, resulting in their annihilation, to avoid domain wall dominance in the Universe. For sufficiently large domain wall tension, we rule out $10^7$ GeV $< T_{\mathrm{ann}} < 10^9$ GeV.

High-energy physics can motivate a cosmological model with a stiff equation of state ($1/3 \leq w_{\mathrm{s}} \leq 1$). We derive 95% CL upper limits on some of the parameters characterizing this unconventional cosmology.

Gravitational waves offer a novel tool to test inflationary models and constrain their parameters. While single-field slow-roll inflation within the ΛCDM cosmological model predicts a gravitational-wave background that is too weak to be observed with current detectors, other inflationary models may produce a detectable gravitational-wave background. We consider an axion inflation model, where a pseudoscalar axion is coupled to a gauge field and we impose constraints on the gauge coupling and the inflaton velocity.

The scalar-induced gravitational-wave background, arising from large-amplitude primordial curvature perturbations, provides an observational test for probing directly the epoch of inflation. In scenarios where primordial curvature fluctuations are amplified during inflation, primordial black holes form through the collapse of extremely dense regions shortly after the corresponding modes enter the Hubble radius. We impose an upper bound on a gravitational-wave background, leading to constraints on primordial curvature perturbations, stronger than the one imposed by Big Bang nucleosynthesis and cosmic microwave background at a scale of approximately $10^{17}$ Mpc.

Several mechanisms have been proposed for the formation of primordial black holes. Just like astrophysical black holes, primordial black holes can form binaries and emit gravitational waves. A key factor determining the merger rate is the number density of primordial black holes, typically quantified by $f_{\mathrm{PBH}}$, the fraction of dark matter composed of primordial black holes today. We find 95% UL constraints on $f_{\mathrm{PBH}}$ as a function of the primordial black hole mass. In particular, we set $f_{\mathrm{PBH}} < 10^{-2}$ for primordial black holes with masses in the range $(1\text{–}100)M_{\odot}$.

Several string-theory models and scalar-tensor models of gravity can result in circularly polarized gravitational waves. We study a generic parity-violation search based on a power-law energy density gravitational-wave spectrum and then detail more theoretically motivated polarized gravitational-wave background models of early Universe turbulence and axion inflation. We impose new constraints on parity-violation parameters.

In searching for a cosmologically produced gravitational-wave background, we also account for the presence of an astrophysical CBC background, composed of black holes and neutron stars. This is a background that LIGO-Virgo-KAGRA will likely detect before the cosmological background [15]. For the current generation of ground-based detectors, it will be challenging to separate astrophysical and cosmological contributions [383,384]. Various methods have been proposed for signal separation with future detectors [385–389], such as the Einstein Telescope [390] and the Cosmic Explorer [391].

No gravitational-wave background signal has been detected for any of the cosmological and high-energy physics models considered here, leading to constraints on their parameters. No CBC-produced gravitational-wave background has been detected either. Nevertheless, our analyses demonstrate that the LIGO-Virgo data can already be used to derive new constraints on a variety of beyond the Standard Model theories, thereby enabling the testing of early Universe scenarios and particle physics models at energy scales otherwise inaccessible. We expect the constraints presented in this publication to be useful for particle physics and cosmological model building.

The LIGO-Virgo-KAGRA Collaboration will continue to improve gravitational-wave background limits using data from the remainder of O4, with updated results to follow its completion. Although the sensitivity changes across O4a, O4b, and O4c are modest, the extended observing time will improve the sensitivity to the energy density of the gravitational-wave background. The subsequent O5 run will push these limits even further, deepening their impact on cosmology and high-energy physics.

## ACKNOWLEDGMENTS

This material is based upon work supported by NSF's LIGO Laboratory, which is a major facility fully funded by the National Science Foundation. The authors also gratefully acknowledge the support of the Science and Technology Facilities Council (STFC) of the United Kingdom, the Max-Planck-Society (MPS), and the State of Niedersachsen/Germany for support of the construction of Advanced LIGO and construction and operation of the GEO 600 detector. Additional support for Advanced LIGO was provided by the Australian Research Council. The authors gratefully acknowledge the Italian Istituto Nazionale di Fisica Nucleare (INFN), the French Centre National de la Recherche Scientifique (CNRS), and the Netherlands Organization for Scientific Research (NWO) for the construction and operation of the Virgo detector and the creation and support of the EGO consortium. The authors also gratefully acknowledge research support from these agencies as well as by the Council of Scientific and Industrial Research of India, the Department of Science and Technology, India, the Science and Engineering Research Board (SERB), India, the Ministry of Human Resource Development, India, the Spanish Agencia Estatal de Investigación (AEI), the Spanish Ministerio de Ciencia, Innovación y Universidades, the European Union NextGenerationEU/PRTR (PRTR-C17.I1), the ICSC—CentroNazionale di Ricerca in High Performance Computing, Big Data and Quantum Computing, funded by the European Union NextGenerationEU, the Comunitat Autonòma de les Illes Balears through the Conselleria d'Educació i Universitats, the Conselleria d'Innovació, Universitats, Ciència i Societat Digital de la Generalitat Valenciana and the CERCA Programme Generalitat de Catalunya, Spain, the Polish National Agency for Academic Exchange, the National Science Centre of Poland and the European Union–European Regional Development Fund; the Foundation for Polish Science (FNP), the Polish Ministry of Science and Higher Education, the Swiss National Science Foundation (SNSF), the Russian Science Foundation, the European Commission, the European Social Funds (ESF), the European Regional Development Funds (ERDF), the Royal Society, the Scottish Funding Council, the Scottish Universities Physics Alliance, the Hungarian Scientific Research Fund (OTKA), the French Lyon Institute of Origins (LIO), the Belgian Fonds de la Recherche Scientifique (FRS-FNRS), Actions de Recherche Concertées (ARC) and Fonds Wetenschappelijk Onderzoek—Vlaanderen (FWO), Belgium, the Paris Île-de-France Region, the National Research, Development and Innovation Office of Hungary (NKFIH), the National Research Foundation of Korea, the Natural Sciences and Engineering Research Council of Canada (NSERC), the Canadian Foundation for Innovation (CFI), the Brazilian Ministry of Science, Technology, and Innovations, the International Center for Theoretical Physics South American Institute for Fundamental Research (ICTP-SAIFR), the Research Grants Council of Hong Kong, the National Natural Science Foundation of China (NSFC), the Israel Science Foundation (ISF), the U.S.–Israel Binational Science Fund (BSF), the Leverhulme Trust, the Research Corporation, the National Science and Technology Council (NSTC), Taiwan, the U.S. Department of Energy, and the Kavli Foundation. The authors gratefully acknowledge the support of the NSF, STFC, INFN, and CNRS for provision of computational resources. This work was supported by MEXT, the JSPS Leading-edge Research Infrastructure Program, JSPS Grant-in-Aid for Specially Promoted Research 26000005, JSPS Grant-in-Aid for Scientific Research on Innovative Areas 2402: 24103006, 24103005, and 2905: JP17H06358, JP17H06361, and JP17H06364, JSPS Core-to-Core Program A. Advanced Research Networks, JSPS Grants-in-Aid for Scientific Research (S) 17H06133 and 20H05639, JSPS Grant-in-Aid for Transformative Research Areas (A) 20A203: JP20H05854, the joint research program of the Institute for Cosmic Ray Research, University of Tokyo, the National Research Foundation (NRF), the Computing Infrastructure Project of the Global Science experimental Data hub Center (GSDC) at KISTI, the Korea Astronomy and Space Science Institute (KASI), the Ministry of Science and ICT (MSIT) in Korea, Academia Sinica (AS), the AS Grid Center (ASGC) and the National Science and Technology Council (NSTC) in Taiwan under grants including the Science Vanguard Research Program, the Advanced Technology Center (ATC) of NAOJ, and the Mechanical Engineering Center of KEK.

## DATA AVAILABILITY

The data that support the findings of this article are openly available [4]; embargo periods may apply.

---

A. G. Abac,[1] I. Abouelfettouh,[2] F. Acernese,[3,4] K. Ackley,[5] C. Adamcewicz,[6] S. Adhicary,[7] D. Adhikari,[8,9] N. Adhikari,[10] R. X. Adhikari,[11] V. K. Adkins,[12] S. Afroz,[13] A. Agapito,[14] D. Agarwal,[15] M. Agathos,[16] N. Aggarwal,[17] S. Aggarwal,[18] O. D. Aguiar,[19] I.-L. Ahrend,[20] L. Aiello,[21,22] A. Ain,[23] P. Ajith,[24] T. Akutsu,[25,26] S. Albanesi,[27,28] W. Ali,[29,30] S. Al-Kershi,[8,9] C. Alléné,[31] A. Allocca,[32,4] S. Al-Shammari,[33] P. A. Altin,[34] S. Alvarez-Lopez,[35] W. Amar,[31] O. Amarasinghe,[33] A. Amato,[36,37] F. Amicucci,[38,39] C. Amra,[40] A. Ananyeva,[11] S. B. Anderson,[11] W. G. Anderson,[11] M. Andia,[41] M. Ando,[42] M. Andrés-Carcasona,[43] T. Andrić,[44,45,8,9] J. Anglin,[46] S. Ansoldi,[47,48] J. M. Antelis,[49] S. Antier,[41] M. Aoumi,[50] E. Z. Appavuravther,[51,52] S. Appert,[11] S. K. Apple,[53] K. Arai,[11] A. Araya,[42] M. C. Araya,[11] M. Arca Sedda,[44,45] J. S. Areeda,[54] N. Aritomi,[2] F. Armato,[29,30] S. Armstrong,[55] N. Arnaud,[56] M. Arogeti,[57] S. M. Aronson,[12] G. Ashton,[58] Y. Aso,[25,59] L. Asprea,[28] M. Assiduo,[60,61] S. Assis de Souza Melo,[62] S. M. Aston,[63] P. Astone,[38] F. Attadio,[39,38] F. Aubin,[64] K. AultONeal,[65] G. Avallone,[66] E. A. Avila,[49] S. Babak,[20] C. Badger,[67] S. Bae,[68] S. Bagnasco,[28] L. Baiotti,[69] R. Bajpai,[70] T. Baka,[71,37] A. M. Baker,[6] K. A. Baker,[72] T. Baker,[73] G. Baldi,[74,75] N. Baldicchi,[76,51] M. Ball,[77] G. Ballardin,[62] S. W. Ballmer,[78] S. Banagiri,[6] B. Banerjee,[44] D. Bankar,[79] T. M. Baptiste,[12] P. Baral,[10] M. Baratti,[80,81] J. C. Barayoga,[11] B. C. Barish,[11] D. Barker,[2] N. Barman,[79] P. Barneo,[82,83,84] F. Barone,[85,4] B. Barr,[86] L. Barsotti,[35] M. Barsuglia,[20] D. Barta,[87] A. M. Bartoletti,[88] M. A. Barton,[86] I. Bartos,[46] A. Basalaev,[8,9] R. Bassiri,[89] A. Basti,[81,80] M. Bawaj,[76,51] P. Baxi,[90] J. C. Bayley,[86] A. C. Baylor,[10] P. A. Baynard II,[57] M. Bazzan,[91,92] V. M. Bedakihale,[93] F. Beirnaert,[94] M. Bejger,[95] D. Belardinelli,[22] A. S. Bell,[86] D. S. Bellie,[96] L. Bellizzi,[80,81] W. Benoit,[18] I. Bentara,[56] J. D. Bentley,[97] M. Ben Yaala,[55] S. Bera,[98,99] F. Bergamin,[33] B. K. Berger,[89] S. Bernuzzi,[27] M. Beroiz,[11] D. Bersanetti,[29] T. Bertheas,[100] A. Bertolini,[37,36] J. Betzwieser,[63] D. Beveridge,[72] G. Bevilacqua,[101] N. Bevins,[102] R. Bhandare,[103] R. Bhatt,[11] D. Bhattacharjee,[104,105] S. Bhattacharyya,[106] S. Bhaumik,[46] V. Biancalana,[101] A. Bianchi,[37,107] I. A. Bilenko,[108] G. Billingsley,[11] A. Binetti,[109] S. Bini,[11,74,75] C. Binu,[110] S. Biot,[111] O. Birnholtz,[112] S. Biscoveanu,[96] A. Bisht,[9] M. Bitossi,[62,80] M.-A. Bizouard,[113] S. Blaber,[114] J. K. Blackburn,[11] L. A. Blagg,[77] C. D. Blair,[72,63] D. G. Blair,[72] S. Blasi,[115,323] N. Bode,[8,9] N. Boettner,[97] G. Boileau,[113] M. Boldrini,[38] G. N. Bolingbroke,[116] A. Bolliand,[117,40] L. D. Bonavena,[46] R. Bondarescu,[82] F. Bondu,[118] E. Bonilla,[89] M. S. Bonilla,[54] A. Bonino,[119] R. Bonnand,[31,117] A. Borchers,[8,9] S. Borhanian,[7] V. Boschi,[80] S. Bose,[120] V. Bossilkov,[63] Y. Bothra,[37,107] A. Boudon,[56] L. Bourg,[57] T. D. Boybeyi,[18]

M. Boyle,[121] A. Bozzi,[62] C. Bradaschia,[80] P. R. Brady,[10] A. Branch,[63] M. Branchesi,[44,45] I. Braun,[104] T. Briant,[122] A. Brillet,[113,†] M. Brinkmann,[8,9] P. Brockill,[10] E. Brockmueller,[8,9] A. F. Brooks,[11] B. C. Brown,[46] D. D. Brown,[116] M. L. Brozzetti,[76,51] S. Brunett,[11] G. Bruno,[15] R. Bruntz,[123] J. Bryant,[119] Y. Bu,[124] F. Bucci,[61] J. Buchanan,[123] O. Bulashenko,[82,83] T. Bulik,[125] H. J. Bulten,[37] A. Buonanno,[126,1] K. Burtnyk,[2] R. Buscicchio,[127,128] D. Buskulic,[31] C. Buy,[100] R. L. Byer,[89] G. S. Cabourn Davies,[73] R. Cabrita,[15] V. Cáceres-Barbosa,[7] L. Cadonati,[57] G. Cagnoli,[129] C. Cahillane,[78] A. Calafat,[98] T. A. Callister,[130] E. Calloni,[32,4] S. R. Callos,[77,29] G. Caneva Santoro,[43] K. C. Cannon,[42] H. Cao,[35] L. A. Capistran,[131] E. Capocasa,[20] E. Capote,[2,11] G. Capurri,[81,80] G. Carapella,[66,132] F. Carbognani,[62] M. Carlassara,[8,9] J. B. Carlin,[124] T. K. Carlson,[133] M. F. Carney,[104] M. Carpinelli,[127,62] G. Carrillo,[77] J. J. Carter,[8,9] G. Carullo,[119,134] A. Casallas-Lagos,[135] J. Casanueva Diaz,[62] C. Casentini,[136,22] S. Y. Castro-Lucas,[137] S. Caudill,[133] M. Cavaglià,[105] R. Cavalieri,[62] A. Ceja,[54] G. Cella,[80] P. Cerdá-Durán,[138,139] E. Cesarini,[22] N. Chabbra,[34] W. Chaibi,[113] A. Chakraborty,[13] P. Chakraborty,[8,9] S. Chakraborty,[103] S. Chalathadka Subrahmanya,[97] J. C. L. Chan,[140] M. Chan,[114] K. Chang,[141] S. Chao,[142,141] P. Charlton,[143] E. Chassande-Mottin,[20] C. Chatterjee,[144] Debarati Chatterjee,[79] Deep Chatterjee,[35] M. Chaturvedi,[103] S. Chaty,[20] A. Chen,[145] A. H.-Y. Chen,[146] D. Chen,[147] H. Chen,[142] H. Y. Chen,[148] S. Chen,[144] Yanbei Chen,[149] Yitian Chen,[121] H. P. Cheng,[150] P. Chessa,[76,51] H. T. Cheung,[90] S. Y. Cheung,[6] F. Chiadini,[151,132] G. Chiarini,[8,9,92] A. Chiba,[152] A. Chincarini,[29] M. L. Chiofalo,[81,80] A. Chiummo,[4,62] C. Chou,[146] S. Choudhary,[72] N. Christensen,[113,153] S. S. Y. Chua,[34] G. Ciani,[74,75] P. Ciecielag,[95] M. Cieślar,[125] M. Cifaldi,[22] B. Cirok,[154] F. Clara,[2] J. A. Clark,[11,57] T. A. Clarke,[6] P. Clearwater,[155] S. Clesse,[111] F. Cleva,[113,117] E. Coccia,[44,45,43] E. Codazzo,[156,157] P.-F. Cohadon,[122] S. Colace,[30] E. Colangeli,[73] M. Colleoni,[98] C. G. Collette,[158] J. Collins,[63] S. Colloms,[86] A. Colombo,[159,128] C. M. Compton,[2] G. Connolly,[77] L. Conti,[92] T. R. Corbitt,[12] I. Cordero-Carrión,[160] S. Corezzi,[76,51] N. J. Cornish,[161] I. Coronado,[162] A. Corsi,[163] R. Cottingham,[63] M. W. Coughlin,[18] A. Couineaux,[38] P. Couvares,[11,57] D. M. Coward,[72] R. Coyne,[164] A. Cozzumbo,[44] J. D. E. Creighton,[10] T. D. Creighton,[165] P. Cremonese,[98] S. Crook,[63] R. Crouch,[2] J. Csizmazia,[2] J. R. Cudell,[166] T. J. Cullen,[11] A. Cumming,[86] E. Cuoco,[167,168] M. Cusinato,[138] L. V. Da Conceição,[169] T. Dal Canton,[41] S. Dal Pra,[170] G. Dálya,[100] B. D'Angelo,[29] S. Danilishin,[36,37] S. D'Antonio,[38] K. Danzmann,[9,8,9] K. E. Darroch,[123] L. P. Dartez,[63] R. Das,[106] A. Dasgupta,[93] V. Dattilo,[62] A. Daumas,[20] N. Davari,[171,172] I. Dave,[103] A. Davenport,[137] M. Davier,[41] T. F. Davies,[72] D. Davis,[11] L. Davis,[72] M. C. Davis,[18] P. Davis,[173,174] E. J. Daw,[175] M. Dax,[1] J. De Bolle,[94] M. Deenadayalan,[79] J. Degallaix,[176] M. De Laurentis,[32,4] F. De Lillo,[23] S. Della Torre,[128] W. Del Pozzo,[81,80] A. Demagny,[31] F. De Marco,[39,38] G. Demasi,[177,61] F. De Matteis,[21,22] N. Demos,[35] T. Dent,[178] A. Depasse,[15] N. DePergola,[102] R. De Pietri,[179,180] R. De Rosa,[32,4] C. De Rossi,[62] M. Desai,[35] R. DeSalvo,[181] A. DeSimone,[182] R. De Simone,[151,132] A. Dhani,[1] R. Diab,[46] M. C. Díaz,[165] M. Di Cesare,[32,4] G. Dideron,[183] T. Dietrich,[1] L. Di Fiore,[4] C. Di Fronzo,[72] M. Di Giovanni,[39,38] T. Di Girolamo,[32,4] D. Diksha,[37,36] J. Ding,[20,184] S. Di Pace,[39,38] I. Di Palma,[39,38] D. Di Piero,[185,48] F. Di Renzo,[56] Divyajyoti,[33] A. Dmitriev,[119] J. P. Docherty,[86] Z. Doctor,[96] N. Doerksen,[169] E. Dohmen,[2] A. Doke,[133] A. Domiciano De Souza,[186] L. D'Onofrio,[38] F. Donovan,[35] K. L. Dooley,[33] T. Dooney,[71] S. Doravari,[79] O. Dorosh,[187] W. J. D. Doyle,[123] M. Drago,[39,38] J. C. Driggers,[2] L. Dunn,[124] U. Dupletsa,[44] P.-A. Duverne,[20] D. D'Urso,[171,156] P. Dutta Roy,[46] H. Duval,[115] S. E. Dwyer,[2] C. Eassa,[2] M. Ebersold,[188,31] T. Eckhardt,[97] G. Eddolls,[78] A. Effler,[63] J. Eichholz,[34] H. Einsle,[113] M. Eisenmann,[25] M. Emma,[58] K. Endo,[152] R. Enficiaud,[1] L. Errico,[32,4] R. Espinosa,[165] M. Esposito,[4,32] R. C. Essick,[189] H. Estellés,[1] T. Etzel,[11] M. Evans,[35] T. Evstafyeva,[183] B. E. Ewing,[7] J. M. Ezquiaga,[140] F. Fabrizi,[60,61] V. Fafone,[21,22] S. Fairhurst,[33] A. M. Farah,[130] B. Farr,[77] W. M. Farr,[190,191] G. Favaro,[91] M. Favata,[192] M. Fays,[166] M. Fazio,[55] J. Feicht,[11] M. M. Fejer,[89] R. Felicetti,[185,48] E. Fenyvesi,[87,193] J. Fernandes,[194] T. Fernandes,[195,138] D. Fernando,[110] S. Ferraiuolo,[196,39,38] T. A. Ferreira,[12] F. Fidecaro,[81,80] P. Figura,[95] A. Fiori,[80,81] I. Fiori,[62] M. Fishbach,[189] R. P. Fisher,[123] R. Fittipaldi,[197,132] V. Fiumara,[198,132] R. Flaminio,[31] S. M. Fleischer,[199] L. S. Fleming,[200] E. Floden,[18] H. Fong,[114] J. A. Font,[138,139] F. Fontinele-Nunes,[18] C. Foo,[1] B. Fornal,[201] K. Franceschetti,[179] F. Frappez,[31] S. Frasca,[39,38] F. Frasconi,[80] J. P. Freed,[65] Z. Frei,[202] A. Freise,[37,107] O. Freitas,[195,138] R. Frey,[77] W. Frischhertz,[63] P. Fritschel,[35] V. V. Frolov,[63] G. G. Fronzé,[28] M. Fuentes-Garcia,[11] S. Fujii,[203] T. Fujimori,[204] T. Fujita,[205] P. Fulda,[46] M. Fyffe,[63] B. Gadre,[71] J. R. Gair,[1] S. Galaudage,[186] V. Galdi,[206] R. Gamba,[7] A. Gamboa,[1] S. Gamoji,[181] D. Ganapathy,[207] A. Ganguly,[79] B. Garaventa,[29] J. García-Bellido,[208] C. García-Quirós,[188] J. W. Gardner,[34] K. A. Gardner,[114] S. Garg,[42] J. Gargiulo,[62] X. Garrido,[41] A. Garron,[98] F. Garufi,[32,4] P. A. Garver,[89] C. Gasbarra,[21,22] B. Gateley,[2] F. Gautier,[209] V. Gayathri,[10] T. Gayer,[78] G. Gemme,[29]

A. Gennai,[80] V. Gennari,[100] J. George,[103] R. George,[148] O. Gerberding,[97] L. Gergely,[154] Archisman Ghosh,[94] Sayantan Ghosh,[194] Shaon Ghosh,[192] Shrobana Ghosh,[8,9] Suprovo Ghosh,[210] Tathagata Ghosh,[79] J. A. Giaime,[12,63] K. D. Giardina,[63] D. R. Gibson,[200] C. Gier,[55] S. Gkaitatzis,[81,80] J. Glanzer,[11] F. Glotin,[41] J. Godfrey,[77] R. V. Godley,[8,9] O. Godwin,[11] A. S. Goettel,[33] E. Goetz,[114] J. Golomb,[11] S. Gomez Lopez,[39,38] B. Goncharov,[44] G. González,[12] P. Goodarzi,[211] S. Goode,[6] A. W. Goodwin-Jones,[15] M. Gosselin,[62] R. Gouaty,[31] D. W. Gould,[34] K. Govorkova,[35] A. Grado,[76,51] V. Graham,[86] A. E. Granados,[18] M. Granata,[176] V. Granata,[212,132] S. Gras,[35] P. Grassia,[11] J. Graves,[57] C. Gray,[2] R. Gray,[86] G. Greco,[51] A. C. Green,[37,107] L. Green,[213] S. M. Green,[73] S. R. Green,[214] C. Greenberg,[133] A. M. Gretarsson,[65] H. K. Griffin,[18] D. Griffith,[11] H. L. Griggs,[57] G. Grignani,[76,51] C. Grimaud,[31] H. Grote,[33] S. Grunewald,[1] D. Guerra,[138] D. Guetta,[215] G. M. Guidi,[60,61] A. R. Guimaraes,[12] H. K. Gulati,[93] F. Gulminelli,[173,174] H. Guo,[145] W. Guo,[72] Y. Guo,[37,36] Anuradha Gupta,[216] I. Gupta,[7] N. C. Gupta,[93] S. K. Gupta,[46] V. Gupta,[18] N. Gupte,[1] J. Gurs,[97] N. Gutierrez,[176] N. Guttman,[6] F. Guzman,[131] D. Haba,[217] M. Haberland,[1] S. Haino,[218] E. D. Hall,[35] E. Z. Hamilton,[98] G. Hammond,[86] M. Haney,[37] J. Hanks,[2] C. Hanna,[7] M. D. Hannam,[33] O. A. Hannuksela,[219] A. G. Hanselman,[130] H. Hansen,[2] J. Hanson,[63] S. Hanumasagar,[57] R. Harada,[42] A. R. Hardison,[182] S. Harikumar,[187] K. Haris,[37,71] I. Harley-Trochimczyk,[131] T. Harmark,[134] J. Harms,[44,45] G. M. Harry,[220] I. W. Harry,[73] J. Hart,[104] B. Haskell,[95,221,222] C. J. Haster,[213] K. Haughian,[86] H. Hayakawa,[50] K. Hayama,[223] M. C. Heintze,[63] J. Heinze,[119] J. Heinzel,[35] H. Heitmann,[113] F. Hellman,[207] A. F. Helmling-Cornell,[77] G. Hemming,[62] O. Henderson-Sapir,[116] M. Hendry,[86] I. S. Heng,[86] M. H. Hennig,[86] C. Henshaw,[57] M. Heurs,[8,9] A. L. Hewitt,[224,225] J. Heynen,[15] J. Heyns,[35] S. Higginbotham,[33] S. Hild,[36,37] S. Hill,[86] Y. Himemoto,[226] N. Hirata,[227] C. Hirose,[228] D. Hofman,[176] B. E. Hogan,[65] N. A. Holland,[37,107] K. Holley-Bockelmann,[144] I. J. Hollows,[175] D. E. Holz,[130] L. Honet,[111] D. J. Horton-Bailey,[207] J. Hough,[86] S. Hourihane,[11] N. T. Howard,[144] E. J. Howell,[72] C. G. Hoy,[73] C. A. Hrishikesh,[21] P. Hsi,[35] H.-F. Hsieh,[142] H.-Y. Hsieh,[142] C. Hsiung,[229] S.-H. Hsu,[146] W.-F. Hsu,[109] Q. Hu,[86] H. Y. Huang,[141] Y. Huang,[7] Y. T. Huang,[78] A. D. Huddart,[230] B. Hughey,[65] V. Hui,[31] S. Husa,[98] R. Huxford,[7] L. Iampieri,[39,38] G. A. Iandolo,[36] M. Ianni,[22,21] G. Iannone,[132] J. Iascau,[77] K. Ide,[231] R. Iden,[217] A. Ierardi,[44,45] S. Ikeda,[147] H. Imafuku,[42] Y. Inoue,[141] G. Iorio,[91] P. Iosif,[185,48] M. H. Iqbal,[34] J. Irwin,[86] R. Ishikawa,[231] M. Isi,[190,191] K. S. Isleif,[232] Y. Itoh,[204,233] M. Iwaya,[203] B. R. Iyer,[24] C. Jacquet,[100] P.-E. Jacquet,[122] T. Jacquot,[41] S. J. Jadhav,[234] S. P. Jadhav,[155] M. Jain,[133] T. Jain,[224] A. L. James,[11] K. Jani,[144] J. Janquart,[15] N. N. Janthalur,[234] S. Jaraba,[235] P. Jaranowski,[236] R. Jaume,[98] W. Javed,[33] A. Jennings,[2] M. Jensen,[2] W. Jia,[35] J. Jiang,[150] H.-B. Jin,[237,238] G. R. Johns,[123] N. A. Johnson,[46] M. C. Johnston,[213] R. Johnston,[86] N. Johny,[8,9] D. H. Jones,[34] D. I. Jones,[210] R. Jones,[86] H. E. Jose,[77] P. Joshi,[7] S. K. Joshi,[79] G. Joubert,[56] J. Ju,[239] L. Ju,[72] K. Jung,[240] J. Junker,[34] V. Juste,[111] H. B. Kabagoz,[63,35] T. Kajita,[241] I. Kaku,[204] V. Kalogera,[96] M. Kalomenopoulos,[213] M. Kamiizumi,[50] N. Kanda,[233,204] S. Kandhasamy,[79] G. Kang,[242] N. C. Kannachel,[6] J. B. Kanner,[11] S. A. KantiMahanty,[18] S. J. Kapadia,[79] D. P. Kapasi,[54] M. Karthikeyan,[133] M. Kasprzack,[11] H. Kato,[152] T. Kato,[203] E. Katsavounidis,[35] W. Katzman,[63] R. Kaushik,[103] K. Kawabe,[2] R. Kawamoto,[204] D. Keitel,[98] L. J. Kemperman,[116] J. Kennington,[7] F. A. Kerkow,[18] R. Kesharwani,[79] J. S. Key,[243] R. Khadela,[8,9] S. Khadka,[89] S. S. Khadkikar,[7] F. Y. Khalili,[108] F. Khan,[8,9] T. Khanam,[163] M. Khursheed,[103] N. M. Khusid,[190,191] W. Kiendrebeogo,[113,244] N. Kijbunchoo,[116] C. Kim,[245] J. C. Kim,[246] K. Kim,[247] M. H. Kim,[239] S. Kim,[248] Y.-M. Kim,[247] C. Kimball,[96] K. Kimes,[54] M. Kinnear,[33] J. S. Kissel,[2] S. Klimenko,[46] A. M. Knee,[114] E. J. Knox,[77] N. Knust,[8,9] K. Kobayashi,[203] S. M. Koehlenbeck,[89] G. Koekoek,[37,36] K. Kohri,[249,250] K. Kokeyama,[33,251] S. Koley,[44,166] P. Kolitsidou,[119] A. E. Koloniari,[252] K. Komori,[42] A. K. H. Kong,[142] A. Kontos,[253] L. M. Koponen,[119] M. Korobko,[97] X. Kou,[18] A. Koushik,[23] N. Kouvatsos,[67] M. Kovalam,[72] T. Koyama,[152] D. B. Kozak,[11] S. L. Kranzhoff,[36,37] V. Kringel,[8,9] N. V. Krishnendu,[119] S. Kroker,[254] A. Królak,[255,187] K. Kruska,[8,9] J. Kubisz,[256] G. Kuehn,[8,9] S. Kulkarni,[216] A. Kulur Ramamohan,[34] Achal Kumar,[46] Anil Kumar,[234] Praveen Kumar,[178] Prayush Kumar,[24] Rahul Kumar,[2] Rakesh Kumar,[93] J. Kume,[257,258,42] K. Kuns,[35] N. Kuntimaddi,[33] S. Kuroyanagi,[208,259] S. Kuwahara,[42] K. Kwak,[240] K. Kwan,[34] S. Kwon,[42] G. Lacaille,[86] D. Laghi,[188,100] A. H. Laity,[164] E. Lalande,[260] M. Lalleman,[23] P. C. Lalremruati,[261] M. Landry,[2] B. B. Lane,[35] R. N. Lang,[35] J. Lange,[148] R. Langgin,[213] B. Lantz,[89] I. La Rosa,[98] J. Larsen,[199] A. Lartaux-Vollard,[41] P. D. Lasky,[6] J. Lawrence,[165] M. Laxen,[63] C. Lazarte,[138] A. Lazzarini,[11] C. Lazzaro,[157,156] P. Leaci,[39,38] L. Leali,[18] Y. K. Lecoeuche,[114] H. M. Lee,[262] H. W. Lee,[263] J. Lee,[78] K. Lee,[239] R.-K. Lee,[142] R. Lee,[35] Sungho Lee,[247] Sunjae Lee,[239] Y. Lee,[141] I. N. Legred,[11] J. Lehmann,[8,9] L. Lehner,[183] M. Le Jean,[176,117] A. Lemaître,[264] M. Lenti,[61,177] M. Leonardi,[74,75,265] M. Lequime,[40] N. Leroy,[41] M. Lesovsky,[11]

N. Letendre,[31] M. Lethuillier,[56] Y. Levin,[6] K. Leyde,[73] A. K. Y. Li,[11] K. L. Li,[266] T. G. F. Li,[109] X. Li,[149] Y. Li,[96] Z. Li,[86] A. Lihos,[123] E. T. Lin,[142] F. Lin,[141] L. C.-C. Lin,[266] Y.-C. Lin,[142] C. Lindsay,[200] S. D. Linker,[181] A. Liu,[219] G. C. Liu,[229] Jian Liu,[72] F. Llamas Villarreal,[165] J. Llobera-Querol,[98] R. K. L. Lo,[140] J.-P. Locquet,[109] S. C. G. Loggins,[267] M. R. Loizou,[133] L. T. London,[67] A. Longo,[60,61] D. Lopez,[166] M. Lopez Portilla,[71] A. Lorenzo-Medina,[178] V. Loriette,[41] M. Lormand,[63] G. Losurdo,[268,80] E. Lotti,[133] T. P. Lott IV,[57] J. D. Lough,[8,9] H. A. Loughlin,[35] C. O. Lousto,[110] N. Low,[124] N. Lu,[34] L. Lucchesi,[80] H. Lück,[9,8,9] D. Lumaca,[22] A. P. Lundgren,[269,270] A. W. Lussier,[260] R. Macas,[73] M. MacInnis,[35] D. M. Macleod,[33] I. A. O. MacMillan,[11] A. Macquet,[41] K. Maeda,[152] S. Maenaut,[109] S. S. Magare,[79] R. M. Magee,[11] E. Maggio,[1] R. Maggiore,[37,107] M. Magnozzi,[29,30] M. Mahesh,[97] M. Maini,[164] S. Majhi,[79] E. Majorana,[39,38] C. N. Makarem,[11] D. Malakar,[105] J. A. Malaquias-Reis,[19] U. Mali,[189] S. Maliakal,[11] A. Malik,[103] L. Mallick,[169,189] A.-K. Malz,[58] N. Man,[113] M. Mancarella,[99] V. Mandic,[18] V. Mangano,[171,156] B. Mannix,[77] G. L. Mansell,[78] M. Manske,[10] M. Mantovani,[62] M. Mapelli,[91,92,271] C. Marinelli,[101] F. Marion,[31] A. Mariotti,[115] A. S. Markosyan,[89] A. Markowitz,[11] E. Maros,[11] S. Marsat,[100] F. Martelli,[60,61] I. W. Martin,[86] R. M. Martin,[192] B. B. Martinez,[131] D. A. Martinez,[54] M. Martinez,[43,272] V. Martinez,[129] A. Martini,[74,75] J. C. Martins,[19] D. V. Martynov,[119] E. J. Marx,[35] L. Massaro,[36,37] A. Masserot,[31] M. Masso-Reid,[86] S. Mastrogiovanni,[38] T. Matcovich,[51] M. Matiushechkina,[8,9] L. Maurin,[209] N. Mavalvala,[35] N. Maxwell,[2] G. McCarrol,[63] R. McCarthy,[2] D. E. McClelland,[34] S. McCormick,[63] L. McCuller,[11] S. McEachin,[123] C. McElhenny,[123] G. I. McGhee,[86] J. McGinn,[86] K. B. M. McGowan,[144] J. McIver,[114] A. McLeod,[72] I. McMahon,[188] T. McRae,[34] R. McTeague,[86] D. Meacher,[10] B. N. Meagher,[78] R. Mechum,[110] Q. Meijer,[71] A. Melatos,[124] C. S. Menoni,[137] F. Mera,[2] R. A. Mercer,[10] L. Mereni,[176] K. Merfeld,[163] E. L. Merilh,[63] J. R. Mérou,[98] J. D. Merritt,[77] M. Merzougui,[113] C. Messick,[10] B. Mestichelli,[44] M. Meyer-Conde,[273] F. Meylahn,[8,9] A. Mhaske,[79] A. Miani,[74,75] H. Miao,[274] C. Michel,[176] Y. Michimura,[42] H. Middleton,[119] D. P. Mihaylov,[104] S. J. Miller,[11] M. Millhouse,[57] E. Milotti,[185,48] V. Milotti,[91] Y. Minenkov,[22] E. M. Minihan,[65] Ll. M. Mir,[43] L. Mirasola,[156,157] M. Miravet-Tenés,[138] C.-A. Miritescu,[43] A. Mishra,[24] C. Mishra,[106] T. Mishra,[46] A. L. Mitchell,[37,107] J. G. Mitchell,[65] S. Mitra,[79] V. P. Mitrofanov,[108] K. Mitsuhashi,[25] R. Mittleman,[35] O. Miyakawa,[50] S. Miyoki,[50] A. Miyoko,[65] G. Mo,[35] L. Mobilia,[60,61] S. R. P. Mohapatra,[11] S. R. Mohite,[7] M. Molina-Ruiz,[207] M. Mondin,[181] M. Montani,[60,61] C. J. Moore,[224] D. Moraru,[2] A. More,[79] S. More,[79] C. Moreno,[135] E. A. Moreno,[35] G. Moreno,[2] A. Moreso Serra,[82] S. Morisaki,[42,203] Y. Moriwaki,[152] G. Morras,[208] A. Moscatello,[91] M. Mould,[35] B. Mours,[64] C. M. Mow-Lowry,[37,107] L. Muccillo,[177,61] F. Muciaccia,[39,38] D. Mukherjee,[119] Samanwaya Mukherjee,[24] Soma Mukherjee,[165] Subroto Mukherjee,[93] Suvodip Mukherjee,[13] N. Mukund,[35] A. Mullavey,[63] H. Mullock,[114] J. Mundi,[220] C. L. Mungioli,[72] M. Murakoshi,[231] P. G. Murray,[86] D. Nabari,[74,75] S. L. Nadji,[8,9] A. Nagar,[28,275] N. Nagarajan,[86] K. Nakagaki,[50] K. Nakamura,[25] H. Nakano,[276] M. Nakano,[11] D. Nanadoumgar-Lacroze,[43] D. Nandi,[12] V. Napolano,[62] P. Narayan,[216] I. Nardecchia,[22] T. Narikawa,[203] H. Narola,[71] L. Naticchioni,[38] R. K. Nayak,[261] L. Negri,[71] A. Nela,[86] C. Nelle,[77] A. Nelson,[131] T. J. N. Nelson,[63] M. Nery,[8,9] A. Neunzert,[2] S. Ng,[54] L. Nguyen Quynh,[277] S. A. Nichols,[12] A. B. Nielsen,[278] Y. Nishino,[25,42] A. Nishizawa,[279] S. Nissanke,[280,37] W. Niu,[7] F. Nocera,[62] J. Noller,[281] M. Norman,[33] C. North,[33] J. Novak,[117,235,282] R. Nowicki,[144] J. F. Nuño Siles,[208] L. K. Nuttall,[73] K. Obayashi,[231] J. Oberling,[2] J. O'Dell,[230] E. Oelker,[35] M. Oertel,[235,117,283,282] G. Oganesyan,[44,45] T. O'Hanlon,[63] M. Ohashi,[50] F. Ohme,[8,9] R. Oliveri,[117,283,282] R. Omer,[18] B. O'Neal,[123] M. Onishi,[152] K. Oohara,[284] B. O'Reilly,[63] M. Orselli,[51,76] R. O'Shaughnessy,[110] S. O'Shea,[86] S. Oshino,[50] C. Osthelder,[11] I. Ota,[12] D. J. Ottaway,[116] A. Ouzriat,[56] H. Overmier,[63] B. J. Owen,[285] R. Ozaki,[231] A. E. Pace,[7] R. Pagano,[12] M. A. Page,[25] A. Pai,[194] L. Paiella,[44] A. Pal,[286] S. Pal,[261] M. A. Palaia,[80,81] M. Pálfi,[202] P. P. Palma,[39,21,22] C. Palomba,[38] P. Palud,[20] H. Pan,[142] J. Pan,[72] K. C. Pan,[142] P. K. Panda,[234] Shiksha Pandey,[7] Swadha Pandey,[35] P. T. H. Pang,[37,71] F. Pannarale,[39,38] K. A. Pannone,[54] B. C. Pant,[103] F. H. Panther,[72] M. Panzeri,[60,61] F. Paoletti,[80] A. Paolone,[38,287] A. Papadopoulos,[86] E. E. Papalexakis,[211] L. Papalini,[80,81] G. Papigkiotis,[252] A. Paquis,[41] A. Parisi,[76,51] B.-J. Park,[247] J. Park,[288] W. Parker,[63] G. Pascale,[8,9] D. Pascucci,[94] A. Pasqualetti,[62] R. Passaquieti,[81,80] L. Passenger,[6] D. Passuello,[80] O. Patane,[2] A. V. Patel,[141] D. Pathak,[79] A. Patra,[33] B. Patricelli,[81,80] B. G. Patterson,[33] K. Paul,[106] S. Paul,[77] E. Payne,[11] T. Pearce,[33] M. Pedraza,[11] A. Pele,[11] F. E. Peña Arellano,[289] X. Peng,[119] Y. Peng,[57] S. Penn,[290] M. D. Penuliar,[54] A. Perego,[74,75] Z. Pereira,[133] C. Périgois,[291,92,91] G. Perna,[91] A. Perreca,[74,75,44] J. Perret,[20] S. Perriès,[56] J. W. Perry,[37,107] D. Pesios,[252] S. Peters,[166] S. Petracca,[206] C. Petrillo,[76] H. P. Pfeiffer,[1] H. Pham,[63] K. A. Pham,[18] K. S. Phukon,[119] H. Phurailatpam,[219] M. Piarulli,[100] L. Piccari,[39,38] O. J. Piccinni,[34] M. Pichot,[113]

M. Piendibene,[81,80] F. Piergiovanni,[60,61] L. Pierini,[38] G. Pierra,[38] V. Pierro,[292,132] M. Pietrzak,[95] M. Pillas,[166] F. Pilo,[80] L. Pinard,[176] I. M. Pinto,[292,132,293,32] M. Pinto,[62] B. J. Piotrzkowski,[10] M. Pirello,[2] M. D. Pitkin,[224,86] A. Placidi,[51] E. Placidi,[39,38] M. L. Planas,[98] W. Plastino,[212,22] C. Plunkett,[35] R. Poggiani,[81,80] E. Polini,[35] J. Pomper,[80,81] L. Pompili,[1] J. Poon,[219] E. Porcelli,[37] E. K. Porter,[20] C. Posnansky,[7] R. Poulton,[62] J. Powell,[155] G. S. Prabhu,[79] M. Pracchia,[166] B. K. Pradhan,[79] T. Pradier,[64] A. K. Prajapati,[93] K. Prasai,[294] R. Prasanna,[234] P. Prasia,[79] G. Pratten,[119] G. Principe,[185,48] G. A. Prodi,[74,75] P. Prosperi,[80] P. Prosposito,[21,22] A. C. Providence,[65] A. Puecher,[1] J. Pullin,[12] P. Puppo,[38] M. Pürrer,[164] H. Qi,[16] J. Qin,[34] G. Quéméner,[174,117] V. Quetschke,[165] P. J. Quinonez,[65] N. Qutob,[57] R. Rading,[232] I. Rainho,[138] S. Raja,[103] C. Rajan,[103] B. Rajbhandari,[110] K. E. Ramirez,[63] F. A. Ramis Vidal,[98] M. Ramos Arevalo,[165] A. Ramos-Buades,[98,37] S. Ranjan,[57] K. Ransom,[63] P. Rapagnani,[39,38] A. Rase,[115] B. Ratto,[65] A. Ravichandran,[133] A. Ray,[96] V. Raymond,[33] M. Razzano,[81,80] J. Read,[54] T. Regimbau,[31] S. Reid,[55] C. Reissel,[35] D. H. Reitze,[11] A. I. Renzini,[127,11] B. Revenu,[295,41] A. Revilla Peña,[82] R. Reyes,[181] L. Ricca,[15] F. Ricci,[39,38] M. Ricci,[38,39] A. Ricciardone,[81,80] J. Rice,[78] J. W. Richardson,[211] M. L. Richardson,[116] A. Rijal,[65] K. Riles,[90] H. K. Riley,[33] S. Rinaldi,[271] J. Rittmeyer,[97] C. Robertson,[230] F. Robinet,[41] M. Robinson,[2] A. Rocchi,[22] L. Rolland,[31] J. G. Rollins,[11] A. E. Romano,[296] R. Romano,[3,4] A. Romero,[31] I. M. Romero-Shaw,[224] J. H. Romie,[63] S. Ronchini,[7] T. J. Roocke,[116] L. Rosa,[4,32] T. J. Rosauer,[211] C. A. Rose,[57] D. Rosińska,[125] M. P. Ross,[53] M. Rossello-Sastre,[98] S. Rowan,[86] S. K. Roy,[190,191] S. Roy,[15] D. Rozza,[127,128] P. Ruggi,[62] N. Ruhama,[240] E. Ruiz Morales,[297,208] K. Ruiz-Rocha,[144] S. Sachdev,[57] T. Sadecki,[2] P. Saffarieh,[37,107] S. Safi-Harb,[169] M. R. Sah,[13] S. Saha,[142] T. Sainrat,[64] S. Sajith Menon,[215,39,38] K. Sakai,[298] Y. Sakai,[273] M. Sakellariadou,[67] S. Sakon,[7] O. S. Salafia,[159,128,127] F. Salces-Carcoba,[11] L. Salconi,[62] M. Saleem,[148] F. Salemi,[39,38] M. Sallé,[37] S. U. Salunkhe,[79] S. Salvador,[174,173] A. Salvarese,[148] A. Samajdar,[71,37] A. Sanchez,[2] E. J. Sanchez,[11] L. E. Sanchez,[11] N. Sanchis-Gual,[138] J. R. Sanders,[182] E. M. Sänger,[1] F. Santoliquido,[44,45] F. Sarandrea,[28] T. R. Saravanan,[79] N. Sarin,[6] P. Sarkar,[8,9] A. Sasli,[252] P. Sassi,[51,76] B. Sassolas,[176] B. S. Sathyaprakash,[7,33] R. Sato,[228] S. Sato,[152] Yukino Sato,[152] Yu Sato,[152] O. Sauter,[46] R. L. Savage,[2] T. Sawada,[50] H. L. Sawant,[79] S. Sayah,[176] V. Scacco,[21,22] D. Schaetzl,[11] M. Scheel,[149] A. Schiebelbein,[189] M. G. Schiworski,[78] P. Schmidt,[119] S. Schmidt,[71] R. Schnabel,[97] M. Schneewind,[8,9] R. M. S. Schofield,[77] K. Schouteden,[109] B. W. Schulte,[8,9] B. F. Schutz,[33,8,9] E. Schwartz,[299] M. Scialpi,[300] J. Scott,[86] S. M. Scott,[34] R. M. Sedas,[63] T. C. Seetharamu,[86] M. Seglar-Arroyo,[43] Y. Sekiguchi,[301] D. Sellers,[63] N. Sembo,[204] A. S. Sengupta,[302] E. G. Seo,[86] J. W. Seo,[109] V. Sequino,[32,4] M. Serra,[38] A. Sevrin,[115] T. Shaffer,[2] U. S. Shah,[57] M. A. Shaikh,[262] L. Shao,[303] A. K. Sharma,[98] Preeti Sharma,[12] Prianka Sharma,[103] Ritwik Sharma,[18] S. Sharma Chaudhary,[105] P. Shawhan,[126] N. S. Shcheblanov,[304,264] E. Sheridan,[144] Z.-H. Shi,[142] M. Shikauchi,[42] R. Shimomura,[305] H. Shinkai,[305] S. Shirke,[79] D. H. Shoemaker,[35] D. M. Shoemaker,[148] R. W. Short,[2] S. ShyamSundar,[103] A. Sider,[158] H. Siegel,[190,191] D. Sigg,[2] L. Silenzi,[36,37] L. Silvestri,[39,170] M. Simmonds,[116] L. P. Singer,[306] Amitesh Singh,[216] Anika Singh,[11] D. Singh,[207] N. Singh,[98] S. Singh,[217,59] A. M. Sintes,[98] V. Sipala,[171,156] V. Skliris,[33] B. J. J. Slagmolen,[34] D. A. Slater,[199] T. J. Slaven-Blair,[72] J. Smetana,[119] J. R. Smith,[54] L. Smith,[86,185,48] R. J. E. Smith,[6] W. J. Smith,[144] S. Soares de Albuquerque Filho,[60] M. Soares-Santos,[188] K. Somiya,[217] I. Song,[142] S. Soni,[35] V. Sordini,[56] F. Sorrentino,[29] H. Sotani,[307] F. Spada,[80] V. Spagnuolo,[37] A. P. Spencer,[86] P. Spinicelli,[62] A. K. Srivastava,[93] F. Stachurski,[86] C. J. Stark,[123] D. A. Steer,[308] N. Steinle,[169] J. Steinlechner,[36,37] S. Steinlechner,[36,37] N. Stergioulas,[252] P. Stevens,[41] M. StPierre,[164] M. D. Strong,[12] A. Strunk,[2] A. L. Stuver,[102,†] M. Suchenek,[95] S. Sudhagar,[95] Y. Sudo,[231] N. Sueltmann,[97] L. Suleiman,[54] K. D. Sullivan,[12] J. Sun,[242] L. Sun,[34] S. Sunil,[93] J. Suresh,[113] B. J. Sutton,[67] P. J. Sutton,[33] K. Suzuki,[217] M. Suzuki,[203] B. L. Swinkels,[37] A. Syx,[117] M. J. Szczepańczyk,[309] P. Szewczyk,[125] M. Tacca,[37] H. Tagoshi,[310] S. C. Tait,[11] K. Takada,[310] H. Takahashi,[311] R. Takahashi,[25] A. Takamori,[42] S. Takano,[312] H. Takeda,[313,314] K. Takeshita,[217] I. Takimoto Schmiegelow,[44,45] M. Takou-Ayaoh,[78] C. Talbot,[130] M. Tamaki,[203] N. Tamanini,[100] D. Tanabe,[141] K. Tanaka,[50] S. J. Tanaka,[231] S. Tanioka,[33] D. B. Tanner,[46] W. Tanner,[8,9] L. Tao,[211] R. D. Tapia,[7] E. N. Tapia San Martín,[37] C. Taranto,[21,22] A. Taruya,[315] J. D. Tasson,[153] J. G. Tau,[110] D. Tellez,[54] R. Tenorio,[98] H. Themann,[181] A. Theodoropoulos,[138] M. P. Thirugnanasambandam,[79] L. M. Thomas,[11] M. Thomas,[63] P. Thomas,[2] J. E. Thompson,[210] S. R. Thondapu,[103] K. A. Thorne,[63] E. Thrane,[6] J. Tissino,[44,45] A. Tiwari,[79] Pawan Tiwari,[44] Praveer Tiwari,[194] S. Tiwari,[188] V. Tiwari,[119] M. R. Todd,[78] M. Toffano,[91] A. M. Toivonen,[18] K. Toland,[86] A. E. Tolley,[73] T. Tomaru,[25] V. Tommasini,[11] T. Tomura,[50] H. Tong,[6] C. Tong-Yu,[141] A. Torres-Forné,[138,139] C. I. Torrie,[11] I. Tosta e Melo,[316] E. Tournefier,[31] M. Trad Nery,[113] K. Tran,[123] A. Trapananti,[52,51] R. Travaglini,[168]

F. Travasso,[52,51] G. Traylor,[63] M. Trevor,[126] M. C. Tringali,[62] A. Tripathee,[90] G. Troian,[185,48] A. Trovato,[185,48] L. Trozzo,[4] R. J. Trudeau,[11] T. Tsang,[33] S. Tsuchida,[317] L. Tsukada,[213] K. Turbang,[115,23] M. Turconi,[113] C. Turski,[94] H. Ubach,[82,83] T. Uchiyama,[50] R. P. Udall,[11] T. Uehara,[318] K. Ueno,[42] V. Undheim,[278] L. E. Uronen,[219] T. Ushiba,[50] M. Vacatello,[80,81] H. Vahlbruch,[8,9] N. Vaidya,[11] G. Vajente,[11] A. Vajpeyi,[6] J. Valencia,[98] M. Valentini,[107,37] S. A. Vallejo-Peña,[296] S. Vallero,[28] V. Valsan,[10] M. van Dael,[37,319] E. Van den Bossche,[115] J. F. J. van den Brand,[36,107,37] C. Van Den Broeck,[71,37] M. van der Sluys,[37,71] A. Van de Walle,[41] J. van Dongen,[37,107] K. Vandra,[102] M. VanDyke,[120] H. van Haevermaet,[23] J. V. van Heijningen,[37,107] P. Van Hove,[64] J. Vanier,[260] M. VanKeuren,[104] J. Vanosky,[2] N. van Remortel,[23] M. Vardaro,[36,37] A. F. Vargas,[124] V. Varma,[133] A. N. Vazquez,[89] A. Vecchio,[119] G. Vedovato,[92] J. Veitch,[86] P. J. Veitch,[116] S. Venikoudis,[15] R. C. Venterea,[18] P. Verdier,[56] M. Vereecken,[15] D. Verkindt,[31] B. Verma,[133] Y. Verma,[103] S. M. Vermeulen,[11] F. Vetrano,[60] A. Veutro,[38,39] A. Viceré,[60,61] S. Vidyant,[78] A. D. Viets,[88] A. Vijaykumar,[189] A. Vilkha,[110] N. Villanueva Espinosa,[138] V. Villa-Ortega,[178] E. T. Vincent,[57] J.-Y. Vinet,[113] S. Viret,[56] S. Vitale,[35] H. Vocca,[76,51] D. Voigt,[97] E. R. G. von Reis,[2] J. S. A. von Wrangel,[8,9] W. E. Vossius,[232] L. Vujeva,[140] S. P. Vyatchanin,[108] J. Wack,[11] L. E. Wade,[104] M. Wade,[104] K. J. Wagner,[110] L. Wallace,[11] E. J. Wang,[89] H. Wang,[217] J. Z. Wang,[90] W. H. Wang,[165] Y. F. Wang,[1] G. Waratkar,[194] J. Warner,[2] M. Was,[31] T. Washimi,[25] N. Y. Washington,[11] D. Watarai,[42] B. Weaver,[2] S. A. Webster,[86] N. L. Weickhardt,[97] M. Weinert,[8,9] A. J. Weinstein,[11] R. Weiss,[35] L. Wen,[72] K. Wette,[34] J. T. Whelan,[110] B. F. Whiting,[46] C. Whittle,[11] E. G. Wickens,[73] D. Wilken,[8,9,9] A. T. Wilkin,[211] B. M. Williams,[120] D. Williams,[86] M. J. Williams,[73] N. S. Williams,[1] J. L. Willis,[11] B. Willke,[9,8,9] M. Wils,[109] L. Wilson,[104] C. W. Winborn,[105] J. Winterflood,[72] C. C. Wipf,[11] G. Woan,[86] J. Woehler,[36,37] N. E. Wolfe,[35] H. T. Wong,[141] I. C. F. Wong,[219,109] K. Wong,[189] T. Wouters,[71,37] J. L. Wright,[2] M. Wright,[86,71] B. Wu,[78] C. Wu,[142] D. S. Wu,[8,9] H. Wu,[142] K. Wu,[120] Q. Wu,[53] Y. Wu,[96] Z. Wu,[100] E. Wuchner,[54] D. M. Wysocki,[10] V. A. Xu,[207] Y. Xu,[98] N. Yadav,[28] H. Yamamoto,[11] K. Yamamoto,[152] T. S. Yamamoto,[42] T. Yamamoto,[50] R. Yamazaki,[231] T. Yan,[119] F. Yang,[46] K. Z. Yang,[18] Y. Yang,[146] Z. Yarbrough,[12] J. Yebana,[98] S.-W. Yeh,[142] A. B. Yelikar,[144] X. Yin,[35] J. Yokoyama,[320,42] T. Yokozawa,[50] S. Yuan,[72] H. Yuzurihara,[50] M. Zanolin,[65] M. Zeeshan,[110] T. Zelenova,[62] J.-P. Zendri,[92] M. Zeoli,[15] M. Zerrad,[40] M. Zevin,[96] H. Zhang,[145] L. Zhang,[11] N. Zhang,[57] R. Zhang,[150] T. Zhang,[119] C. Zhao,[72] Yue Zhao,[162] Yuhang Zhao,[20] Z.-C. Zhao,[321] Y. Zheng,[105] H. Zhong,[18] H. Zhou,[78] H. O. Zhu,[72] Z.-H. Zhu,[321,322] A. B. Zimmerman,[148] L. Zimmermann,[56] M. E. Zucker,[35,11] and J. Zweizig[11]

(LIGO Scientific Collaboration, Virgo Collaboration, and KAGRA Collaboration)

[1]*Max Planck Institute for Gravitational Physics (Albert Einstein Institute), D-14476 Potsdam, Germany*
[2]*LIGO Hanford Observatory, Richland, Washington 99352, USA*
[3]*Dipartimento di Farmacia, Università di Salerno, I-84084 Fisciano, Salerno, Italy*
[4]*INFN, Sezione di Napoli, I-80126 Napoli, Italy*
[5]*University of Warwick, Coventry CV4 7AL, United Kingdom*
[6]*OzGrav, School of Physics and Astronomy, Monash University, Clayton 3800, Victoria, Australia*
[7]*The Pennsylvania State University, University Park, Pennsylvania 16802, USA*
[8]*Max Planck Institute for Gravitational Physics (Albert Einstein Institute), D-30167 Hannover, Germany*
[9]*Leibniz Universität Hannover, D-30167 Hannover, Germany*
[10]*University of Wisconsin–Milwaukee, Milwaukee, Wisconsin 53201, USA*
[11]*LIGO Laboratory, California Institute of Technology, Pasadena, California 91125, USA*
[12]*Louisiana State University, Baton Rouge, Louisiana 70803, USA*
[13]*Tata Institute of Fundamental Research, Mumbai 400005, India*
[14]*Centre de Physique Théorique, Aix-Marseille Université, Campus de Luminy, 163 Avenue de Luminy, 13009 Marseille, France*
[15]*Université catholique de Louvain, B-1348 Louvain-la-Neuve, Belgium*
[16]*Queen Mary University of London, London E1 4NS, United Kingdom*
[17]*University of California, Davis, Davis, California 95616, USA*
[18]*University of Minnesota, Minneapolis, Minnesota 55455, USA*
[19]*Instituto Nacional de Pesquisas Espaciais, 12227-010 São José dos Campos, São Paulo, Brazil*
[20]*Université Paris Cité, CNRS, Astroparticule et Cosmologie, F-75013 Paris, France*
[21]*Università di Roma Tor Vergata, I-00133 Roma, Italy*
[22]*INFN, Sezione di Roma Tor Vergata, I-00133 Roma, Italy*

[23]*Universiteit Antwerpen, 2000 Antwerpen, Belgium*
[24]*International Centre for Theoretical Sciences, Tata Institute of Fundamental Research, Bengaluru 560089, India*
[25]*Gravitational Wave Science Project, National Astronomical Observatory of Japan, 2-21-1 Osawa, Mitaka City, Tokyo 181-8588, Japan*
[26]*Advanced Technology Center, National Astronomical Observatory of Japan, 2-21-1 Osawa, Mitaka City, Tokyo 181-8588, Japan*
[27]*Theoretisch-Physikalisches Institut, Friedrich-Schiller-Universität Jena, D-07743 Jena, Germany*
[28]*INFN Sezione di Torino, I-10125 Torino, Italy*
[29]*INFN, Sezione di Genova, I-16146 Genova, Italy*
[30]*Dipartimento di Fisica, Università degli Studi di Genova, I-16146 Genova, Italy*
[31]*Université Savoie Mont Blanc, CNRS, Laboratoire d'Annecy de Physique des Particules - IN2P3, F-74000 Annecy, France*
[32]*Università di Napoli "Federico II," I-80126 Napoli, Italy*
[33]*Cardiff University, Cardiff CF24 3AA, United Kingdom*
[34]*OzGrav, Australian National University, Canberra, Australian Capital Territory 0200, Australia*
[35]*LIGO Laboratory, Massachusetts Institute of Technology, Cambridge, Massachusetts 02139, USA*
[36]*Maastricht University, 6200 MD Maastricht, Netherlands*
[37]*Nikhef, 1098 XG Amsterdam, Netherlands*
[38]*INFN, Sezione di Roma, I-00185 Roma, Italy*
[39]*Università di Roma "La Sapienza," I-00185 Roma, Italy*
[40]*Aix Marseille Université, CNRS, Centrale Med, Institut Fresnel, F-13013 Marseille, France*
[41]*Université Paris-Saclay, CNRS/IN2P3, IJCLab, 91405 Orsay, France*
[42]*University of Tokyo, Tokyo, 113-0033, Japan*
[43]*Institut de Física d'Altes Energies (IFAE), The Barcelona Institute of Science and Technology, Campus UAB, E-08193 Bellaterra (Barcelona), Spain*
[44]*Gran Sasso Science Institute (GSSI), I-67100 L'Aquila, Italy*
[45]*INFN, Laboratori Nazionali del Gran Sasso, I-67100 Assergi, Italy*
[46]*University of Florida, Gainesville, Florida 32611, USA*
[47]*Dipartimento di Scienze Matematiche, Informatiche e Fisiche, Università di Udine, I-33100 Udine, Italy*
[48]*INFN, Sezione di Trieste, I-34127 Trieste, Italy*
[49]*Tecnologico de Monterrey, Escuela de Ingeniería y Ciencias, 64849 Monterrey, Nuevo León, Mexico*
[50]*Institute for Cosmic Ray Research, KAGRA Observatory, The University of Tokyo, 238 Higashi-Mozumi, Kamioka-cho, Hida City, Gifu 506-1205, Japan*
[51]*INFN, Sezione di Perugia, I-06123 Perugia, Italy*
[52]*Università di Camerino, I-62032 Camerino, Italy*
[53]*University of Washington, Seattle, Washington 98195, USA*
[54]*California State University Fullerton, Fullerton, California 92831, USA*
[55]*SUPA, University of Strathclyde, Glasgow G1 1XQ, United Kingdom*
[56]*Université Claude Bernard Lyon 1, CNRS, IP2I Lyon / IN2P3, UMR 5822, F-69622 Villeurbanne, France*
[57]*Georgia Institute of Technology, Atlanta, Georgia 30332, USA*
[58]*Royal Holloway, University of London, London TW20 0EX, United Kingdom*
[59]*Astronomical course, The Graduate University for Advanced Studies (SOKENDAI), 2-21-1 Osawa, Mitaka City, Tokyo 181-8588, Japan*
[60]*Università degli Studi di Urbino "Carlo Bo," I-61029 Urbino, Italy*
[61]*INFN, Sezione di Firenze, I-50019 Sesto Fiorentino, Firenze, Italy*
[62]*European Gravitational Observatory (EGO), I-56021 Cascina, Pisa, Italy*
[63]*LIGO Livingston Observatory, Livingston, Louisiana 70754, USA*
[64]*Université de Strasbourg, CNRS, IPHC UMR 7178, F-67000 Strasbourg, France*
[65]*Embry-Riddle Aeronautical University, Prescott, Arizona 86301, USA*
[66]*Dipartimento di Fisica "E. R. Caianiello," Università di Salerno, I-84084 Fisciano, Salerno, Italy*
[67]*King's College London, University of London, London WC2R 2LS, United Kingdom*
[68]*Korea Institute of Science and Technology Information, Daejeon 34141, Republic of Korea*
[69]*International College, Osaka University, 1-1 Machikaneyama-cho, Toyonaka City, Osaka 560-0043, Japan*
[70]*Accelerator Laboratory, High Energy Accelerator Research Organization (KEK), 1-1 Oho, Tsukuba City, Ibaraki 305-0801, Japan*
[71]*Institute for Gravitational and Subatomic Physics (GRASP), Utrecht University, 3584 CC Utrecht, Netherlands*

[72]OzGrav, University of Western Australia, Crawley, Western Australia 6009, Australia
[73]University of Portsmouth, Portsmouth, PO1 3FX, United Kingdom
[74]Università di Trento, Dipartimento di Fisica, I-38123 Povo, Trento, Italy
[75]INFN, Trento Institute for Fundamental Physics and Applications, I-38123 Povo, Trento, Italy
[76]Università di Perugia, I-06123 Perugia, Italy
[77]University of Oregon, Eugene, Oregon 97403, USA
[78]Syracuse University, Syracuse, New York 13244, USA
[79]Inter-University Centre for Astronomy and Astrophysics, Pune 411007, India
[80]INFN, Sezione di Pisa, I-56127 Pisa, Italy
[81]Università di Pisa, I-56127 Pisa, Italy
[82]Institut de Ciències del Cosmos (ICCUB), Universitat de Barcelona (UB), c. Martí i Franquès, 1, 08028 Barcelona, Spain
[83]Departament de Física Quàntica i Astrofísica (FQA), Universitat de Barcelona (UB), c. Martí i Franqués, 1, 08028 Barcelona, Spain
[84]Institut d'Estudis Espacials de Catalunya, c. Gran Capità, 2-4, 08034 Barcelona, Spain
[85]Dipartimento di Medicina, Chirurgia e Odontoiatria "Scuola Medica Salernitana," Università di Salerno, I-84081 Baronissi, Salerno, Italy
[86]IGR, University of Glasgow, Glasgow G12 8QQ, United Kingdom
[87]HUN-REN Wigner Research Centre for Physics, H-1121 Budapest, Hungary
[88]Concordia University Wisconsin, Mequon, Wisconsin 53097, USA
[89]Stanford University, Stanford, California 94305, USA
[90]University of Michigan, Ann Arbor, Michigan 48109, USA
[91]Università di Padova, Dipartimento di Fisica e Astronomia, I-35131 Padova, Italy
[92]INFN, Sezione di Padova, I-35131 Padova, Italy
[93]Institute for Plasma Research, Bhat, Gandhinagar 382428, India
[94]Universiteit Gent, B-9000 Gent, Belgium
[95]Nicolaus Copernicus Astronomical Center, Polish Academy of Sciences, 00-716, Warsaw, Poland
[96]Northwestern University, Evanston, Illinois 60208, USA
[97]Universität Hamburg, D-22761 Hamburg, Germany
[98]IAC3-IEEC, Universitat de les Illes Balears, E-07122 Palma de Mallorca, Spain
[99]Aix-Marseille Université, Université de Toulon, CNRS, CPT, Marseille, France
[100]Laboratoire des 2 Infinis - Toulouse (L2IT-IN2P3), F-31062 Toulouse Cedex 9, France
[101]Università di Siena, Dipartimento di Scienze Fisiche, della Terra e dell'Ambiente, I-53100 Siena, Italy
[102]Villanova University, Villanova, Pennsylvania 19085, USA
[103]RRCAT, Indore, Madhya Pradesh 452013, India
[104]Kenyon College, Gambier, Ohio 43022, USA
[105]Missouri University of Science and Technology, Rolla, Missouri 65409, USA
[106]Indian Institute of Technology Madras, Chennai 600036, India
[107]Department of Physics and Astronomy, Vrije Universiteit Amsterdam, 1081 HV Amsterdam, Netherlands
[108]Lomonosov Moscow State University, Moscow 119991, Russia
[109]Katholieke Universiteit Leuven, Oude Markt 13, 3000 Leuven, Belgium
[110]Rochester Institute of Technology, Rochester, New York 14623, USA
[111]Université libre de Bruxelles, 1050 Bruxelles, Belgium
[112]Bar-Ilan University, Ramat Gan, 5290002, Israel
[113]Université Côte d'Azur, Observatoire de la Côte d'Azur, CNRS, Artemis, F-06304 Nice, France
[114]University of British Columbia, Vancouver, British Columbia V6T 1Z4, Canada
[115]Vrije Universiteit Brussel, 1050 Brussel, Belgium
[116]OzGrav, University of Adelaide, Adelaide, South Australia 5005, Australia
[117]Centre national de la recherche scientifique, 75016 Paris, France
[118]Université Rennes, CNRS, Institut FOTON - UMR 6082, F-35000 Rennes, France
[119]University of Birmingham, Birmingham B15 2TT, United Kingdom
[120]Washington State University, Pullman, Washington 99164, USA
[121]Cornell University, Ithaca, New York 14850, USA
[122]Laboratoire Kastler Brossel, Sorbonne Université, CNRS, ENS-Université PSL, Collège de France, F-75005 Paris, France
[123]Christopher Newport University, Newport News, Virginia 23606, USA
[124]OzGrav, University of Melbourne, Parkville, Victoria 3010, Australia
[125]Astronomical Observatory Warsaw University, 00-478 Warsaw, Poland
[126]University of Maryland, College Park, Maryland 20742, USA

[127]Università degli Studi di Milano-Bicocca, I-20126 Milano, Italy
[128]INFN, Sezione di Milano-Bicocca, I-20126 Milano, Italy
[129]Université de Lyon, Université Claude Bernard Lyon 1, CNRS, Institut Lumière Matière, F-69622 Villeurbanne, France
[130]University of Chicago, Chicago, Illinois 60637, USA
[131]University of Arizona, Tucson, Arizona 85721, USA
[132]INFN, Sezione di Napoli, Gruppo Collegato di Salerno, I-80126 Napoli, Italy
[133]University of Massachusetts Dartmouth, North Dartmouth, Massachusetts 02747, USA
[134]Niels Bohr Institute, Copenhagen University, 2100 København, Denmark
[135]Universidad de Guadalajara, 44430 Guadalajara, Jalisco, Mexico
[136]Istituto di Astrofisica e Planetologia Spaziali di Roma, 00133 Roma, Italy
[137]Colorado State University, Fort Collins, Colorado 80523, USA
[138]Departamento de Astronomía y Astrofísica, Universitat de València, E-46100 Burjassot, València, Spain
[139]Observatori Astronòmic, Universitat de València, E-46980 Paterna, València, Spain
[140]Niels Bohr Institute, University of Copenhagen, 2100 Kóbenhavn, Denmark
[141]National Central University, Taoyuan City 320317, Taiwan
[142]National Tsing Hua University, Hsinchu City 30013, Taiwan
[143]OzGrav, Charles Sturt University, Wagga Wagga, New South Wales 2678, Australia
[144]Vanderbilt University, Nashville, Tennessee 37235, USA
[145]University of the Chinese Academy of Sciences / International Centre for Theoretical Physics Asia-Pacific, Beijing 100049, China
[146]Department of Electrophysics, National Yang Ming Chiao Tung University, 101 University Street, Hsinchu, Taiwan
[147]Kamioka Branch, National Astronomical Observatory of Japan, 238 Higashi-Mozumi, Kamioka-cho, Hida City, Gifu 506-1205, Japan
[148]University of Texas, Austin, Texas 78712, USA
[149]CaRT, California Institute of Technology, Pasadena, California 91125, USA
[150]Northeastern University, Boston, Massachusetts 02115, USA
[151]Dipartimento di Ingegneria Industriale (DIIN), Università di Salerno, I-84084 Fisciano, Salerno, Italy
[152]Faculty of Science, University of Toyama, 3190 Gofuku, Toyama City, Toyama 930-8555, Japan
[153]Carleton College, Northfield, Minnesota 55057, USA
[154]University of Szeged, Dóm tér 9, Szeged 6720, Hungary
[155]OzGrav, Swinburne University of Technology, Hawthorn, Victoria 3122, Australia
[156]INFN Cagliari, Physics Department, Università degli Studi di Cagliari, Cagliari 09042, Italy
[157]Università degli Studi di Cagliari, Via Università 40, 09124 Cagliari, Italy
[158]Université Libre de Bruxelles, Brussels 1050, Belgium
[159]INAF, Osservatorio Astronomico di Brera sede di Merate, I-23807 Merate, Lecco, Italy
[160]Departamento de Matemáticas, Universitat de València, E-46100 Burjassot, València, Spain
[161]Montana State University, Bozeman, Montana 59717, USA
[162]The University of Utah, Salt Lake City, Utah 84112, USA
[163]Johns Hopkins University, Baltimore, Maryland 21218, USA
[164]University of Rhode Island, Kingston, Rhode Island 02881, USA
[165]The University of Texas Rio Grande Valley, Brownsville, Texas 78520, USA
[166]Université de Liège, B-4000 Liège, Belgium
[167]DIFA- Alma Mater Studiorum Università di Bologna, Via Zamboni, 33–40126 Bologna, Italy
[168]Istituto Nazionale Di Fisica Nucleare - Sezione di Bologna, viale Carlo Berti Pichat 6/2–40127 Bologna, Italy
[169]University of Manitoba, Winnipeg, Manitoba R3T 2N2, Canada
[170]INFN-CNAF - Bologna, Viale Carlo Berti Pichat, 6/2, 40127 Bologna BO, Italy
[171]Università degli Studi di Sassari, I-07100 Sassari, Italy
[172]INFN, Laboratori Nazionali del Sud, I-95125 Catania, Italy
[173]Université de Normandie, ENSICAEN, UNICAEN, CNRS/IN2P3, LPC Caen, F-14000 Caen, France
[174]Laboratoire de Physique Corpusculaire Caen, 6 boulevard du maréchal Juin, F-14050 Caen, France
[175]The University of Sheffield, Sheffield S10 2TN, United Kingdom
[176]Université Claude Bernard Lyon 1, CNRS, Laboratoire des Matériaux Avancés (LMA), IP2I Lyon / IN2P3, UMR 5822, F-69622 Villeurbanne, France
[177]Università di Firenze, Sesto Fiorentino I-50019, Italy
[178]IGFAE, Universidade de Santiago de Compostela, E-15782 Santiago de Compostela, Spain

[179]Dipartimento di Scienze Matematiche, Fisiche e Informatiche, Università di Parma, I-43124 Parma, Italy
[180]INFN, Sezione di Milano Bicocca, Gruppo Collegato di Parma, I-43124 Parma, Italy
[181]California State University, Los Angeles, Los Angeles, California 90032, USA
[182]Marquette University, Milwaukee, Wisconsin 53233, USA
[183]Perimeter Institute, Waterloo, Ontario N2L 2Y5, Canada
[184]Corps des Mines, Mines Paris, Université PSL, 60 Boulevard Saint-Michel, 75272 Paris, France
[185]Dipartimento di Fisica, Università di Trieste, I-34127 Trieste, Italy
[186]Université Côte d'Azur, Observatoire de la Côte d'Azur, CNRS, Lagrange, F-06304 Nice, France
[187]National Center for Nuclear Research, 05-400 Świerk-Otwock, Poland
[188]University of Zurich, Winterthurerstrasse 190, 8057 Zurich, Switzerland
[189]Canadian Institute for Theoretical Astrophysics, University of Toronto, Toronto, Ontario M5S 3H8, Canada
[190]Stony Brook University, Stony Brook, New York 11794, USA
[191]Center for Computational Astrophysics, Flatiron Institute, New York, New York 10010, USA
[192]Montclair State University, Montclair, New Jersey 07043, USA
[193]HUN-REN Institute for Nuclear Research, H-4026 Debrecen, Hungary
[194]Indian Institute of Technology Bombay, Powai, Mumbai 400 076, India
[195]Centro de Física das Universidades do Minho e do Porto, Universidade do Minho, PT-4710-057 Braga, Portugal
[196]Aix Marseille Université, CNRS/IN2P3, CPPM, Marseille, France
[197]CNR-SPIN, I-84084 Fisciano, Salerno, Italy
[198]Scuola di Ingegneria, Università della Basilicata, I-85100 Potenza, Italy
[199]Western Washington University, Bellingham, Washington 98225, USA
[200]SUPA, University of the West of Scotland, Paisley PA1 2BE, United Kingdom
[201]Barry University, Miami Shores, Florida 33168, USA
[202]Eötvös University, Budapest 1117, Hungary
[203]Institute for Cosmic Ray Research, KAGRA Observatory, The University of Tokyo, 5-1-5 Kashiwa-no-Ha, Kashiwa City, Chiba 277-8582, Japan
[204]Department of Physics, Graduate School of Science, Osaka Metropolitan University, 3-3-138 Sugimoto-cho, Sumiyoshi-ku, Osaka City, Osaka 558-8585, Japan
[205]Department of Physics, Ochanomizu University, Bunkyo, Tokyo 112-8610, Japan
[206]University of Sannio at Benevento, I-82100 Benevento, Italy and INFN, Sezione di Napoli, I-80100 Napoli, Italy
[207]University of California, Berkeley, California 94720, USA
[208]Instituto de Fisica Teorica UAM-CSIC, Universidad Autonoma de Madrid, 28049 Madrid, Spain
[209]Laboratoire d'Acoustique de l'Université du Mans, UMR CNRS 6613, F-72085 Le Mans, France
[210]University of Southampton, Southampton SO17 1BJ, United Kingdom
[211]University of California, Riverside, Riverside, California 92521, USA
[212]Dipartimento di Ingegneria Industriale, Elettronica e Meccanica, Università degli Studi Roma Tre, I-00146 Roma, Italy
[213]University of Nevada, Las Vegas, Las Vegas, Nevada 89154, USA
[214]University of Nottingham NG7 2RD, United Kingdom
[215]Ariel University, Ramat HaGolan Street 65, Ari'el, West Bank
[216]The University of Mississippi, University, Mississippi 38677, USA
[217]Graduate School of Science, Institute of Science Tokyo, 2-12-1 Ookayama, Meguro-ku, Tokyo 152-8551, Japan
[218]Institute of Physics, Academia Sinica, 128 Sec. II, Academia Road, Nankang, Taipei 11529, Taiwan
[219]The Chinese University of Hong Kong, Shatin, NT, Hong Kong
[220]American University, Washington, DC 20016, USA
[221]Dipartimento di Fisica, Università degli studi di Milano, Via Celoria 16, I-20133, Milano, Italy
[222]INFN, sezione di Milano, Via Celoria 16, I-20133, Milano, Italy
[223]Department of Applied Physics, Fukuoka University, 8-19-1 Nanakuma, Jonan, Fukuoka City, Fukuoka 814-0180, Japan
[224]University of Cambridge, Cambridge CB2 1TN, United Kingdom
[225]University of Lancaster, Lancaster Louisiana1 4YW, United Kingdom
[226]College of Industrial Technology, Nihon University, 1-2-1 Izumi, Narashino City, Chiba 275-8575, Japan
[227]Gravitational Wave Science Project, National Astronomical Observatory of Japan, 2-21-1 Osawa, Mitaka City, Tokyo 181-8588, Japan

[228]Faculty of Engineering, Niigata University, 8050 Ikarashi-2-no-cho,
Nishi-ku, Niigata City, Niigata 950-2181, Japan
[229]Department of Physics, Tamkang University, No. 151, Yingzhuan Road, Danshui District,
New Taipei City 25137, Taiwan
[230]Rutherford Appleton Laboratory, Didcot OX11 0DE, United Kingdom
[231]Department of Physical Sciences, Aoyama Gakuin University,
5-10-1 Fuchinobe, Sagamihara City, Kanagawa 252-5258, Japan
[232]Helmut Schmidt University, D-22043 Hamburg, Germany
[233]Nambu Yoichiro Institute of Theoretical and Experimental Physics (NITEP),
Osaka Metropolitan University, 3-3-138 Sugimoto-cho, Sumiyoshi-ku,
Osaka City, Osaka 558-8585, Japan
[234]Directorate of Construction, Services and Estate Management, Mumbai 400094, India
[235]Observatoire Astronomique de Strasbourg, 11 Rue de l'Université, 67000 Strasbourg, France
[236]Faculty of Physics, University of Białystok, 15-245 Białystok, Poland
[237]National Astronomical Observatories, Chinese Academic of Sciences,
20A Datun Road, Chaoyang District, Beijing, China
[238]School of Astronomy and Space Science, University of Chinese Academy of Sciences,
20A Datun Road, Chaoyang District, Beijing, China
[239]Sungkyunkwan University, Seoul 03063, Republic of Korea
[240]Department of Physics, Ulsan National Institute of Science and Technology (UNIST),
50 UNIST-gil, Ulju-gun, Ulsan 44919, Republic of Korea
[241]Institute for Cosmic Ray Research, The University of Tokyo,
5-1-5 Kashiwa-no-Ha, Kashiwa City, Chiba 277-8582, Japan
[242]Chung-Ang University, Seoul 06974, Republic of Korea
[243]University of Washington Bothell, Bothell, Washington 98011, USA
[244]Laboratoire de Physique et de Chimie de l'Environnement, Université Joseph KI-ZERBO,
9GH2 + 3V5, Ouagadougou, Burkina Faso
[245]Ewha Womans University, Seoul 03760, Republic of Korea
[246]National Institute for Mathematical Sciences, Daejeon 34047, Republic of Korea
[247]Korea Astronomy and Space Science Institute, Daejeon 34055, Republic of Korea
[248]Department of Astronomy and Space Science, Chungnam National University,
9 Daehak-ro, Yuseong-gu, Daejeon 34134, Republic of Korea
[249]Institute of Particle and Nuclear Studies (IPNS),
High Energy Accelerator Research Organization (KEK),
1-1 Oho, Tsukuba City, Ibaraki 305-0801, Japan
[250]Division of Science, National Astronomical Observatory of Japan,
2-21-1 Osawa, Mitaka City, Tokyo 181-8588, Japan
[251]Nagoya University, Nagoya, 464-8601, Japan
[252]Department of Physics, Aristotle University of Thessaloniki, 54124 Thessaloniki, Greece
[253]Bard College, Annandale-On-Hudson, New York 12504, USA
[254]Technical University of Braunschweig, D-38106 Braunschweig, Germany
[255]Institute of Mathematics, Polish Academy of Sciences, 00656 Warsaw, Poland
[256]Astronomical Observatory, Jagiellonian University, 31-007 Cracow, Poland
[257]Department of Physics and Astronomy, University of Padova, Via Marzolo, 8-35151 Padova, Italy
[258]Sezione di Padova, Istituto Nazionale di Fisica Nucleare (INFN), Via Marzolo, 8-35131 Padova, Italy
[259]Department of Physics, Nagoya University,
ES building, Furocho, Chikusa-ku, Nagoya, Aichi 464-8602, Japan
[260]Université de Montréal/Polytechnique, Montreal, Quebec H3T 1J4, Canada
[261]Indian Institute of Science Education and Research, Kolkata, Mohanpur, West Bengal 741252, India
[262]Seoul National University, Seoul 08826, Republic of Korea
[263]Department of Computer Simulation, Inje University,
197 Inje-ro, Gimhae, Gyeongsangnam-do 50834, Republic of Korea
[264]NAVIER, École des Ponts, Université Gustave Eiffel, CNRS, Marne-la-Vallée, France
[265]Gravitational Wave Science Project, National Astronomical Observatory of Japan (NAOJ), Mitaka City,
Tokyo 181-8588, Japan
[266]Department of Physics, National Cheng Kung University,
No. 1, University Road, Tainan City 701, Taiwan
[267]St. Thomas University, Miami Gardens, Florida 33054, USA
[268]Scuola Normale Superiore, I-56126 Pisa, Italy
[269]Institució Catalana de Recerca i Estudis Avançats, E-08010 Barcelona, Spain

[270]Institut de Física d'Altes Energies, E-08193 Barcelona, Spain
[271]Institut fuer Theoretische Astrophysik, Zentrum fuer Astronomie Heidelberg, Universitaet Heidelberg, Albert Ueberle Strasse 2, 69120 Heidelberg, Germany
[272]Institucio Catalana de Recerca i Estudis Avançats (ICREA), Passeig de Lluís Companys, 23, 08010 Barcelona, Spain
[273]Research Center for Space Science, Advanced Research Laboratories, Tokyo City University, 3-3-1 Ushikubo-Nishi, Tsuzuki-Ku, Yokohama, Kanagawa 224-8551, Japan
[274]Tsinghua University, Beijing 100084, China
[275]Institut des Hautes Etudes Scientifiques, F-91440 Bures-sur-Yvette, France
[276]Faculty of Law, Ryukoku University, 67 Fukakusa Tsukamoto-cho, Fushimi-ku, Kyoto City, Kyoto 612-8577, Japan
[277]Phenikaa Institute for Advanced Study (PIAS), Phenikaa University, Yen Nghia, Ha Dong, Hanoi, Vietnam
[278]University of Stavanger, 4021 Stavanger, Norway
[279]Physics Program, Graduate School of Advanced Science and Engineering, Hiroshima University, 1-3-1 Kagamiyama, Higashihiroshima City, Hiroshima 739-8526, Japan
[280]GRAPPA, Anton Pannekoek Institute for Astronomy and Institute for High-Energy Physics, University of Amsterdam, 1098 XH Amsterdam, Netherlands
[281]University College London, London WC1E 6BT, United Kingdom
[282]Observatoire de Paris, 75014 Paris, France
[283]Laboratoire Univers et Théories, Observatoire de Paris, 92190 Meudon, France
[284]Graduate School of Science and Technology, Niigata University, 8050 Ikarashi-2-no-cho, Nishi-ku, Niigata City, Niigata 950-2181, Japan
[285]University of Maryland, Baltimore County, Baltimore, Maryland 21250, USA
[286]CSIR-Central Glass and Ceramic Research Institute, Kolkata, West Bengal 700032, India
[287]Consiglio Nazionale delle Ricerche - Istituto dei Sistemi Complessi, I-00185 Roma, Italy
[288]Department of Astronomy, Yonsei University, 50 Yonsei-Ro, Seodaemun-Gu, Seoul 03722, Republic of Korea
[289]Department of Physics, University of Guadalajara, Avenida Revolucion 1500, Colonia Olimpica C.P. 44430, Guadalajara, Jalisco, Mexico
[290]Hobart and William Smith Colleges, Geneva, New York 14456, USA
[291]INAF, Osservatorio Astronomico di Padova, I-35122 Padova, Italy
[292]Dipartimento di Ingegneria, Università del Sannio, I-82100 Benevento, Italy
[293]Museo Storico della Fisica e Centro Studi e Ricerche "Enrico Fermi," I-00184 Roma, Italy
[294]Kennesaw State University, Kennesaw, Georgia 30144, USA
[295]Subatech, CNRS/IN2P3 - IMT Atlantique - Nantes Université, 4 rue Alfred Kastler, B.P. 20722, 44307 Nantes CÉDEX 03, France
[296]Universidad de Antioquia, Medellín, Colombia
[297]Departamento de Física - ETSIDI, Universidad Politécnica de Madrid, 28012 Madrid, Spain
[298]Department of Electronic Control Engineering, National Institute of Technology, Nagaoka College, 888 Nishikatakai, Nagaoka City, Niigata 940-8532, Japan
[299]Trinity College, Hartford, Connecticut 06106, USA
[300]Dipartimento di Fisica e Scienze della Terra, Università Degli Studi di Ferrara, Via Saragat, 1, 44121 Ferrara FE, Italy
[301]Faculty of Science, Toho University, 2-2-1 Miyama, Funabashi City, Chiba 274-8510, Japan
[302]Indian Institute of Technology, Palaj, Gandhinagar, Gujarat 382355, India
[303]Kavli Institute for Astronomy and Astrophysics, Peking University, Yiheyuan Road 5, Haidian District, Beijing 100871, China
[304]Laboratoire MSME, Cité Descartes, 5 Boulevard Descartes, Champs-sur-Marne, 77454 Marne-la-Vallée Cedex 2, France
[305]Faculty of Information Science and Technology, Osaka Institute of Technology, 1-79-1 Kitayama, Hirakata City, Osaka 573-0196, Japan
[306]NASA Goddard Space Flight Center, Greenbelt, Maryland 20771, USA
[307]Faculty of Science and Technology, Kochi University, 2-5-1 Akebono-cho, Kochi-shi, Kochi 780-8520, Japan
[308]Laboratoire de Physique de l'École Normale Supérieure, ENS, (CNRS, Université PSL, Sorbonne Université, Université Paris Cité), F-75005 Paris, France
[309]Faculty of Physics, University of Warsaw, Ludwika Pasteura 5, 02-093 Warszawa, Poland

[310]Institute for Cosmic Ray Research, KAGRA Observatory, The University of Tokyo, 5-1-5 Kashiwa-no-Ha, Kashiwa City, Chiba 277-8582, Japan
[311]Research Center for Space Science, Advanced Research Laboratories, Tokyo City University, 3-3-1 Ushikubo-Nishi, Tsuzuki-Ku, Yokohama, Kanagawa 224-8551, Japan
[312]Laser Interferometry and Gravitational Wave Astronomy, Max Planck Institute for Gravitational Physics, Callinstrasse 38, 30167 Hannover, Germany
[313]The Hakubi Center for Advanced Research, Kyoto University, Yoshida-honmachi, Sakyou-ku, Kyoto City, Kyoto 606-8501, Japan
[314]Department of Physics, Kyoto University, Kita-Shirakawa Oiwake-cho, Sakyou-ku, Kyoto City, Kyoto 606-8502, Japan
[315]Yukawa Institute for Theoretical Physics (YITP), Kyoto University, Kita-Shirakawa Oiwake-cho, Sakyou-ku, Kyoto City, Kyoto 606-8502, Japan
[316]University of Catania, Department of Physics and Astronomy, Via S. Sofia, 64, 95123 Catania CT, Italy
[317]National Institute of Technology, Fukui College, Geshi-cho, Sabae-shi, Fukui 916-8507, Japan
[318]Department of Communications Engineering, National Defense Academy of Japan, 1-10-20 Hashirimizu, Yokosuka City, Kanagawa 239-8686, Japan
[319]Eindhoven University of Technology, 5600 MB Eindhoven, Netherlands
[320]Kavli Institute for the Physics and Mathematics of the Universe (Kavli IPMU), WPI, The University of Tokyo, 5-1-5 Kashiwa-no-Ha, Kashiwa City, Chiba 277-8583, Japan
[321]Department of Astronomy, Beijing Normal University, Xinjiekouwai Street 19, Haidian District, Beijing 100875, China
[322]School of Physics and Technology, Wuhan University, Bayi Road 299, Wuchang District, Wuhan, Hubei, 430072, China
[323]Deutsches Elektronen-Synchrotron DESY, Notkestr. 85, 22607 Hamburg, Germany

[†]Deceased.